\documentclass[floatfix,aps,preprintnumbers,nofootinbib,superscriptaddress,showpacs,letterpaper,twocolumn]{revtex4}
\pdfoutput=1
\usepackage{amsmath,amssymb,amsbsy,booktabs,array}
\usepackage{bm,comment}
\usepackage{graphicx,color}
\usepackage[sort&compress]{natbib}
\usepackage[colorlinks=true, linkcolor=blue, filecolor=blue, urlcolor=blue, citecolor=blue, pdftex, plainpages=false]{hyperref}

\newcommand{\mres}{m_{\rm res}}
\newcommand\riken{RIKEN-BNL Research Center, Brookhaven National
Laboratory, Upton, NY 11973, USA}
\newcommand{\nishina}{Theoretical Research Division, Nishina Center, RIKEN, Wako 351-0198, Japan}
\newcommand\bnlaf{Physics Department, Brookhaven
National Laboratory, Upton, NY 11973, USA}

\newcommand\cuaff{Physics Department, Columbia University, New York, NY 10027, USA}
\newcommand\soton{School of Physics and Astronomy, University of
Southampton, Southampton SO17 1BJ, UK}
\newcommand\fermilabaf{Theoretical Physics Department, Fermi National Accelerator Laboratory, Batavia, IL 60510, USA}
\newcommand\buaff{Center for Computational Science, Boston University, Boston, MA 02215, USA}

\def\CO{{\cal O}}
\def\CC{{\cal C}}
\def\bar{\overline}

\def\spose#1{\hbox to 0pt{#1\hss}}
\def\ltapprox{\mathrel{\spose{\lower 3pt\hbox{$\mathchar"218$}}
\raise 2.0pt\hbox{$\mathchar"13C$}}}
\def\gtapprox{\mathrel{\spose{\lower 3pt\hbox{$\mathchar"218$}}
\raise 2.0pt\hbox{$\mathchar"13E$}}}
\def\inapprox{\mathrel{\spose{\lower 3pt\hbox{$\mathchar"218$}}
\raise 2.0pt\hbox{$\mathchar"232$}}}

\begin{document}
\preprint{FERMILAB-PUB-14-100-T}

\title{$B$-meson decay constants from 2+1-flavor lattice QCD with domain-wall light quarks and relativistic heavy quarks}

\author{N.~H.~Christ}\affiliation\cuaff
\author{J.~M.~Flynn}\affiliation\soton
\author{T.~Izubuchi}\affiliation\riken\affiliation\bnlaf
\author{T.~Kawanai}\affiliation\nishina\affiliation\bnlaf
\author{C.~Lehner}\affiliation\bnlaf
\author{A.~Soni}\affiliation\bnlaf
\author{R.~S.~Van~de~Water}\affiliation\fermilabaf
\author{O.~Witzel}\affiliation\buaff
\collaboration{RBC and UKQCD Collaborations}\noaffiliation


\date{\today}

\begin{abstract}
We calculate the $B$-meson decay constants $f_B$, $f_{B_s}$, and their ratio in unquenched lattice QCD using domain-wall light quarks and relativistic $b$-quarks.  We use gauge-field ensembles generated by the RBC and UKQCD collaborations using the domain-wall fermion action and Iwasaki gauge action with three flavors of light dynamical quarks.  We analyze data at two lattice spacings of $a \approx 0.11, 0.086$~fm with unitary pion masses as light as $M_\pi \approx 290$~MeV; this enables us to control the extrapolation to the physical light-quark masses and continuum.  For the $b$-quarks we use the anisotropic clover action with the relativistic heavy-quark interpretation, such that discretization errors from the heavy-quark action are of the same size as from the light-quark sector.  We renormalize the lattice heavy-light axial-vector current using a mostly nonperturbative method in which we compute the bulk of the matching factor nonperturbatively, with a small correction, that is close to unity, in lattice perturbation theory.  We also improve the lattice heavy-light current through ${\mathcal O}(\alpha_s a)$.  We extrapolate our results to the physical light-quark masses and continuum using SU(2) heavy-meson chiral perturbation theory, and provide a complete systematic error budget.  We obtain {$f_{B^0} = 199.5(12.6)$} MeV, {$f_{B^+} = 195.6(14.9)$} MeV, $f_{B_s} = 235.4(12.2)$ MeV,  {$f_{B_s}/f_{B^0} = 1.197(50)$}, and {$f_{B_s}/f_{B^+} = 1.223(71)$}, where the errors are statistical and total systematic added in quadrature.  These results are in good agreement with other published results and provide an important independent cross check of other three-flavor determinations of $B$-meson decay constants using staggered light quarks. 
\end{abstract}

\maketitle


\section{Introduction}
\label{Sec:Intro}

Leptonic decays of bottom mesons probe the quark-flavor-changing transitions $b \to u$ and $b \to s$, and therefore play an important role in constraining and searching for new physics in the flavor sector.  

In the Standard Model, the decay rate for $B^+ \to \ell^+ \nu_{\ell}$ is given by 
\begin{equation}
\Gamma ( B \to \ell \nu_{\ell} ) =  \frac{ m_B}{8 \pi} G_F^2  f_B^2 |V_{ub}|^2 m_{\ell}^2 
           \left(1-\frac{ m_{\ell}^2}{m_B^2} \right)^2 \;,  \label{eq:B_leptonic_rate}
\end{equation}
where $f_B$ is the leptonic decay constant that parameterizes nonperturbative QCD contributions to the electroweak decay process, and we use the convention $f_\pi \sim 130$~MeV.  The decay rate in Eq.~(\ref{eq:B_leptonic_rate}) is suppressed by the small value of the CKM matrix element $|V_{ub}|$, which is of ${\mathcal O}(10^{-3})$, and is further helicity suppressed for light final-state charged leptons.   When combined with an experimental measurement of the decay rate, a lattice-QCD calculation of $f_B$ enables the determination of the CKM matrix element $|V_{ub}|$ within the Standard Model.  This is particularly important given the long-established $\sim 3\sigma$ disagreement between $|V_{ub}|$ obtained from exclusive $B\to\pi\ell\nu$ semileptonic decay and inclusive $B\to X_u\ell\nu$ decay~\cite{CKMfitter,UTfit,Antonelli:2009ws,Laiho:2009eu,Beringer:2012zz,Aoki:2013ldr}.  Thus far only the charged-current decay $B^+ \to \tau^+ \nu_{\tau}$ has been observed experimentally.  The experimental measurements from Belle and BaBar have $\sim 30\%$ errors~\cite{Aubert:2009wt,Hara:2010dk,Lees:2012ju,Adachi:2012mm}, but no individual measurement has $5\sigma$ significance.  The precision will improve, however, with additional data collected by Belle~II, which is expected to begin running in around 2016.   At this point the independent determination of $|V_{ub}|$ from $B^+ \to \tau^+ \nu_{\tau}$ may be sufficiently precise to provide some insight into the current $|V_{ub}|$ puzzle.

The leptonic decays of neutral $B_d^0$ and $B_s^0$ mesons proceed via flavor-changing neutral currents.  Thus they are loop suppressed in the Standard Model, and  potentially more sensitive to new physics than $B^+$ leptonic decays.  The decay rate for these neutral-current processes is given by: 
\begin{widetext}
\begin{equation}
\Gamma ( B_q \to \ell^+ \ell^-) = \frac{G_F^2}{\pi} \, Y \,
\left(  \frac{\alpha}{4 \pi \sin^2 \Theta_W} \right)^2
m_{B_q} f_{B_q}^2 |V_{tb}^*V_{tq}|^2 m_{\ell}^2 
           \sqrt{1- 4 \frac{ m_{\ell}^2}{m_B^2} }\;, \label{eq:B_to_ll_rate}
\end{equation}
\end{widetext}
where $q=d,s$ and the loop function $Y$ includes next-to-leading-order short-distance QCD and electroweak corrections~\cite{Buchalla:1993bv}.  Here lattice-QCD calculations of the decay constants $f_B$ and $f_{B_s}$ are needed to calculate predictions for $B_{d,s} \to \ell^+ \ell^-$ both within the Standard Model and in beyond-the-Standard Model theories (see, e.g. Refs.~\cite{Buras:2012ru,Bobeth:2013uxa}).  Evidence for $B_s \to \mu^+ \mu^-$ decay has been seen at the $\sim 4\sigma$ level by both LHCb~\cite{Aaij:2012nna,Aaij:2013aka} and CMS~\cite{Chatrchyan:2013bka}, while LHCb has also seen $\sim 2 \sigma$ evidence for $B_d^0 \to \mu^+ \mu^-$ decay~\cite{Aaij:2013aka}.  The statistical significance of both of these results will increase in the next few years.  Many uncertainties cancel, or are at least suppressed, in the Standard-Model prediction for ${\mathcal B}( B_s \to \mu^+ \mu^-)/{\mathcal B}( B_d^0 \to \mu^+ \mu^-)$, which is proportional to the (squared) ratio of decay constants.  Therefore, once the experimental measurements are more precise, this $SU(3)$-breaking ratio will provide an especially clean test of the Standard Model given a similarly accurate lattice-QCD calculation of $f_{B_s}/f_{B_d^0}$.

In this work we present a new calculation of the leptonic decay constants $f_B$, $f_{B_s}$, and the ratio $f_{B_s}/f_B$ in (2+1)-flavor lattice QCD.  We use the gauge-field ensembles generated by the RBC and UKQCD collaborations with the domain-wall fermion action and Iwasaki gluon action which include the effects of dynamical $u,d$, and $s$ quarks~\cite{Allton:2008pn,Aoki:2010dy}.  For the bottom quarks, we use the relativistic heavy-quark (RHQ) action introduced by Christ, Li, and Lin in Ref.~\cite{Christ:2006us}, with the parameters of the action that were obtained nonperturbatively in Ref.~\cite{Aoki:2012xaa}.   We improve the lattice heavy-light axial vector current through ${\mathcal O}(\alpha_s a)$, and renormalize the current using the mostly nonperturbative method introduced in Ref.~\cite{ElKhadra:2001rv}.  We analyze data with several values of the light-quark mass (down to $\approx$ 290~MeV) and two lattice spacings of $a \approx$ 0.11 and 0.086~fm.  We then extrapolate our numerical simulation to the physical light-quark mass and continuum limit using next-to-leading order SU(2) heavy-light meson chiral perturbation theory (HM$\chi$PT)~\cite{Goity:1992tp,Arndt:2004bg,Aubin:2005aq,Albertus:2010nm}.   

This work is the first application of the RHQ action to weak-matrix element calculations relevant for phenomenology.  The general relativistic heavy-quark framework was introduced by El~Khadra, Kronfeld, and Mackenzie in Ref.~\cite{ElKhadra:1996mp}, and can be used to simulate systems with both light quarks $am_0 \ll 1$ (where $a$ is the lattice spacing and $m_0$ is the bare quark mass) and heavy quarks with $a m_0 \gtapprox 1$ with controlled discretization errors.  This method takes advantage of the fact that, in the rest frame of the heavy-light bound states, the spatial momentum carried by the heavy quark is smaller than the mass of the heavy quark and of order of $\Lambda_\text{QCD}$
\begin{equation}
\left| \vec p_{hl} \right| \sim \Lambda_\text{QCD}.
\end{equation}
Performing a Symanzik-like expansion in powers of the spatial derivative $D_i$ and keeping all orders of the mass $m_0a$ and the temporal derivative $D_0$, one arrives at an anisotropic action which breaks the axis-interchange symmetry between spatial and temporal directions.  There are several implementations of the relativistic heavy-quark framework.  Here we use the RHQ action introduced by Christ, Li, and Lin in Refs.~\cite{Christ:2006us}.  These authors showed that if the three coefficients in the anisotropic Sheikholeslami-Wohlert (clover) action --- the bare quark mass $m_0a$, anisotropy $\zeta$, and clover coefficient $c_P$ --- are suitably tuned, one can eliminate errors of $O(|\vec p|a)$, $O([m_0a]^n)$, and $O(|\vec p|a[m_0a]^n)$ from on-shell Green's functions. Thus the RHQ action allows us to simulate heavy quarks such as bottom with discretization errors of similar size to those of light-quark systems.   In this work we use the nonperturbatively-determined values of $\{m_0a, c_P, \zeta \}$ on the RBC/UKQCD domain-wall + Iwasaki ensembles corresponding to the physical $b$-quark.  The values of these parameters were fixed using masses in the $B_s$ system, and validated by comparison with the experimentally-measured low-lying masses and mass-splittings in the bottomonium system.

There are several (2+1)-flavor and (2+1+1) calculations of the $B_{(s)}$-meson decay constants and their ratio in the literature using a variety of actions for the bottom and light quarks~\cite{Albertus:2010nm,McNeile:2011ng,Bazavov:2011aa,Na:2012kp,Dowdall:2013tga,Carrasco:2013naa,Bazavov:2013wia}.   Of these, our calculation is most similar to that of the Fermilab Lattice and MILC collaborations, who also use the relativistic heavy quark framework.  Their calculation uses the Fermilab interpretation of the isotropic clover action~\cite{ElKhadra:1996mp} with the tadpole-improved tree-level value of the clover coefficient $c_{SW}$.  They also ${\mathcal O}(a)$-improve the heavy-light axial-vector current at tree level.  Thus, for similar values of the lattice spacing, their calculation suffers from larger heavy-quark discretization errors than ours.   All of the published $N_f \geq 3$ results for $f_B$, $f_{B_s}$, and $f_{B_s}/f_B$ use staggered light quarks; the three $N_f = 2+1$ calculations use the same asqtad-improved ensembles generated by the MILC Collaboration.  Our calculation using domain-wall light quarks therefore provides a valuable independent check for these phenomenologically-important quantities.  

This paper is organized as follows.  In Sec.~\ref{Sec:Actions} we describe the lattice actions and simulation parameters used in this work.  Next we present the determination of the $B_{(s)}$ meson decay amplitudes in Sec.~\ref{Sec:DecayAmplitudes}.  First we discuss the operator renormalization and improvement, followed by the two-point correlator fits, the interpolation to the tuned $b$-quark mass, and finally (for $B_s$ meson quantities) the interpolation to the physical $s$ quark.  In Sec.~\ref{Sec:Extrapolations} we extrapolate the numerical simulation data to the physical light-quark masses and the continuum limit using SU(2) HM$\chi$PT.   Section~\ref{Sec:SysErrors} presents our complete uncertainty budget; for clarity, we discuss each source of systematic uncertainty in a separate subsection.  Finally, we conclude in Sec.~\ref{Sec:Conc} with a comparison of our results with other lattice determinations, and with an outlook for the future.  This paper also has two appendices describing our determination of the heavy-heavy current renormalization factor $Z_V^{bb}$ (App.~\ref{App:Zvbb}) and our estimate of heavy-quark discretization errors (App.~\ref{App:HQdiscErr}).

\section{Lattice actions and parameters}
\label{Sec:Actions}

In this section we describe the setup of our numerical lattice simulations, which is the same in our earlier work on tuning the parameters of the RHQ action~\cite{Aoki:2012xaa}.  Sec.~\ref{Sec:LightQuark} summarizes the parameters of the light-quark and gluon actions, while \ref{sec:HQAct} summarizes those of the heavy $b$-quark action.

\subsection{Light-quark and gluon actions}
\label{Sec:LightQuark}

We use the dynamical ``2+1''-flavor domain-wall Iwasaki ensembles generated by the RBC and UKQCD Collaborations with two lattice spacings of $a \approx 0.11$ fm ($a^{-1}=1.729$ GeV) and $a \approx 0.08$ ($a^{-1}=2.281$ GeV)~\cite{Allton:2008pn,Aoki:2010dy}.  These ensembles were generated with three dynamical quarks: the two lighter sea quarks have equal masses which are denoted by $m_l$, while the heavier sea quark mass is tuned to within 10\% of the physical strange-quark mass and is denoted by $m_h$.  The lattices employ the five-dimensional Shamir domain-wall action \cite{Shamir:1993zy,Furman:1994ky} for the fermions in combination with the Iwasaki gauge action \cite{Iwasaki:1983ck}.  This combination allows for sufficient tunneling between topological sectors \cite{Antonio:2008zz}. For the calculation of the $B_{(s)}$-meson decay constants, we analyze five ensembles with unitary pion masses as light as $\approx$~290~MeV.  All spatial volumes are about $2.5$ fm, such that $M_\pi L \gtapprox 4$.  Table \ref{tab:lattices} summarizes the parameters of the gauge-field ensembles used in this analysis.  Throughout this work, we refer to the coarser ensembles with $a \approx 0.11$ fm as the ``$24^3$'' ensembles and the finer ($a \approx 0.08$ fm) ensembles as the ``$32^3$'' ensembles.

\begin{table*}[tb]
\caption{Lattice ensemble parameters.  The columns list the lattice volume, approximate lattice spacing, light ($m_l$) and strange ($m_h$) sea-quark masses, residual chiral symmetry breaking parameter $m_{\rm res}$, physical $u/d$- and $s$-quark mass, unitary pion mass, and number of configurations analyzed. The tildes over $a\widetilde m_{u/d}$ and $a \widetilde m_s$ denote that these values include the residual quark mass.} 
\vspace{3mm}
\label{tab:lattices}
\begin{tabular}{c@{\hskip 3mm}cc@{\hskip 4mm}ccc@{\hskip 4mm}cc@{\hskip 3mm}c@{\hskip 3mm}crc} \hline\hline
$\left(\frac{L}{a}\right)^3 \times \left(\frac{T}{a}\right)$ &$\approx a$(fm)& $a^{-1}$ [GeV] & ~~$am_l$ & ~~$am_h$&$a\mres$ &$a\widetilde m_{u/d}$ &$ a \widetilde m_s$&  $M_\pi$[MeV]  & \# configs. \\[0.5mm] \hline
$24^3 \times 64$ &  0.11 &  1.729(25) & 0.005 & 0.040 &0.003152& 0.00136(4)&0.0379(11)& 329 & 1636\\
$24^3 \times 64$ &  0.11 &  1.729(25) & 0.010 & 0.040 &0.003152& 0.00136(4)&0.0379(11)& 422 & 1419\\ \hline
$32^3 \times 64$ &  0.086 & 2.281(28) & 0.004 & 0.030 &0.0006664&0.00102(5)&0.0280(7)& 289 & 628\\ 
$32^3 \times 64$ &  0.086 & 2.281(28) & 0.006 & 0.030 &0.0006664&0.00102(5)&0.0280(7)& 345 & 889\\
$32^3 \times 64$ &  0.086 & 2.281(28) & 0.008 & 0.030 &0.0006664&0.00102(5)&0.0280(7)& 394 & 544\\ \hline\hline
\end{tabular}\end{table*}

For the light valence quarks we use the same fermion action and parameters as in the sea sector.  Hence we can use RBC-UKQCD's earlier determinations of the unitary pion masses, residual quark mass $\mres$, and values of the physical $u/d$- and $s$-quark masses from Ref.~\cite{Aoki:2010dy}.   In particular, we use $L_s = 16$ for the extent of the fifth dimension, a domain-wall height of $M_5 =1.8$, and periodic boundary conditions in all directions.  With these choices the size of residual chiral symmetry breaking is small: $am_{\rm res}$ is approximately $3\times 10^{-3}$ or less on all ensembles.   For the calculation of the $B_{(s)}$-meson decay constants, we generated point-source valence quark propagators with six different masses including approximately the physical strange quark and the unitary point; their values are listed in Tab.~\ref{Tab:ValenceLight}. These point-source domain-wall propagators were saved and are available for non-competing projects upon request.

\begin{table}[tb]
\centering
\caption{Partially quenched light-quark masses analyzed. On the $32^3$ ensembles, two propagators were generated on each configuration with sources separated by $T/2a$.}
\label{Tab:ValenceLight}
\begin{tabular}{l@{\hskip 3mm}c@{\hskip 2mm}c}\hline\hline
& $a^{-1}$ [GeV]& $a m_q$ \\
\hline
$24^3$ & 1.729(25) & 0.005, 0.01, 0.02, 0.03, 0.0343, 0.04\\
$32^3$ & 2.281(28) & 0.004, 0.006, 0.008, 0.025, 0.0272, 0.03\\
\hline\hline
\end{tabular}
\end{table}
 
\subsection{Heavy-quark action}
\label{sec:HQAct}

We simulate the heavy $b$-quarks (denoted by $Q(x)$) with the anisotropic Sheikholeslami-Wohlert (clover) action \cite{Sheikholeslami:1985ij}:
\begin{widetext}
\begin{align}
S_{\rm RHQ} &= a^4 \sum_{x,x'} \bar{Q}(x') \left( m_0 + \gamma_0 D_0 + \zeta \vec{\gamma} \cdot \vec{D} - \frac{a}{2} (D^0)^2 - \frac{a}{2} \zeta (\vec{D})^2+ \sum_{\mu,\nu} \frac{ia}{4} c_P \sigma_{\mu\nu} F_{\mu\nu} \right)_{x' x} Q(x) \,,
\label{eq:HQAct}
\intertext{where}
D_\mu Q(x) &= \frac{1}{2a} \left[ U_\mu(x)Q(x+\hat{\mu}) - U_\mu^\dagger(x-\hat{\mu})Q(x-\hat{\mu}) \right] \\
D^2_\mu Q(x) &= \frac{1}{a^2} \left[  U_\mu(x) Q(x+\hat{\mu}) + U_\mu^\dagger(x - \hat{\mu})Q(x-\hat{\mu}) - 2 Q(x)  \right] \\
F_{\mu\nu} Q(x)&= \frac{1}{8 a^2} \sum_{s,s'= \pm 1} s s' \left[ U_{s\mu}(x) U_{s'\nu}(x+s\hat{\mu}) U_{s\mu}^\dagger(x + s'\nu) U_{s'\nu}^\dagger (x)- \textrm{h.c.} \right] Q(x)
\end{align}
\end{widetext}
and  $\gamma_\mu = \gamma_\mu^\dagger$ , $\{\gamma_\mu,\gamma_\nu\} = 2 \delta_{\mu\nu}$ and $\sigma_{\mu\nu} = \frac{i}{2} [ \gamma_\mu , \gamma_\nu ]$. 
In Reference \cite{Aoki:2012xaa} we nonperturbatively determined the values of the three parameters $m_0a$, $c_P$, and $\zeta$ that correspond to the physical $b$-quark mass using the same set of gauge field configurations as in this work.  We follow the same approach for our computation of the decay constants so that we can propagate statistical uncertainties from the tuning procedure directly to the decay constants.  Here we briefly summarize the aspects of the tuning procedure needed to understand the error propagation;  further details can be found in Ref.~\cite{Aoki:2012xaa}.

\begin{figure}[tb]
\centering
\includegraphics[scale=0.5,clip]{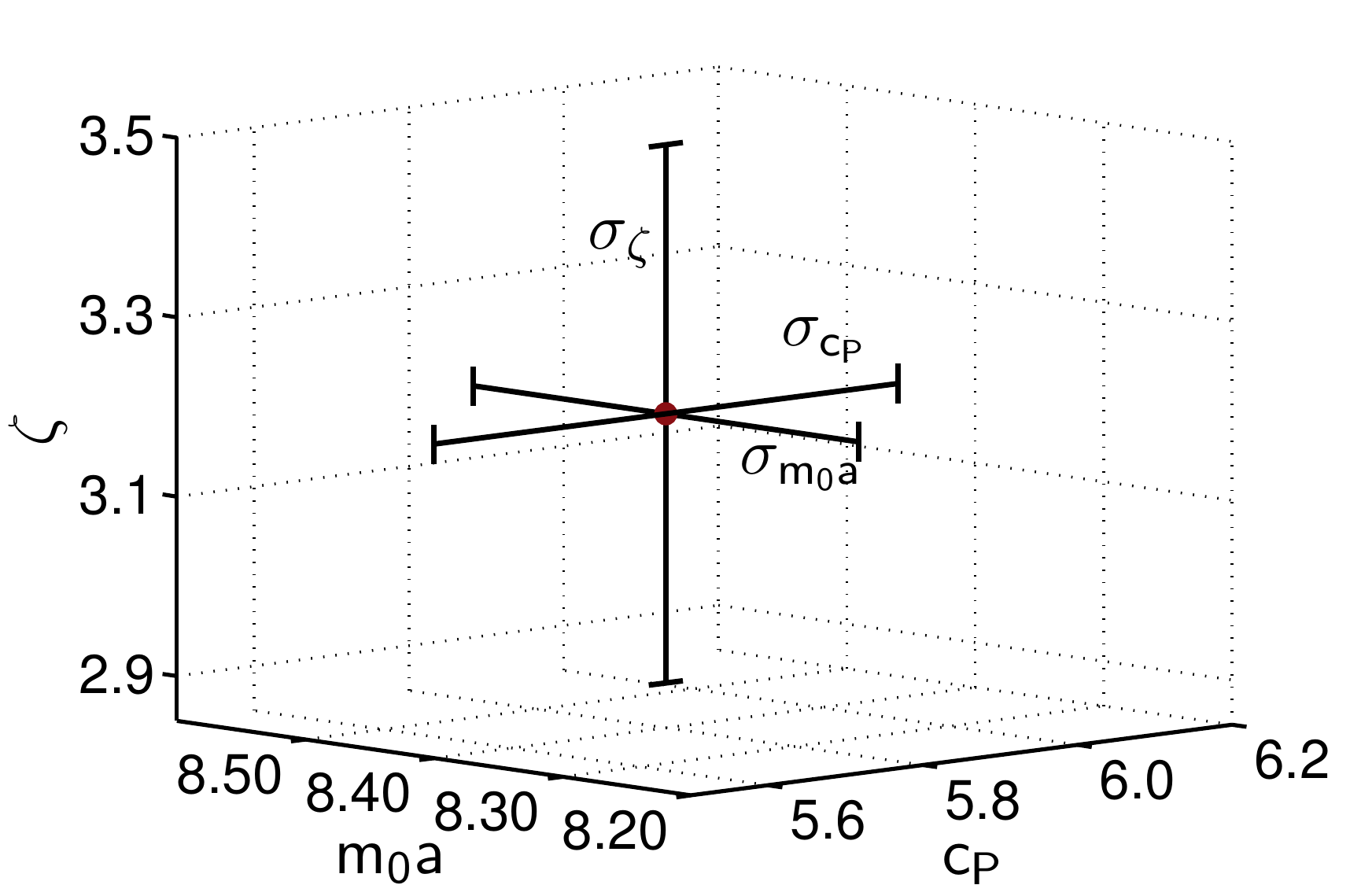}
\caption{Parameter sets used to obtain the tuned coefficients of the RHQ action.  The seven sets of $\{m_0a, c_P, \zeta\}$ are located on a cube at the centers of the six faces and at the midpoint.}
\label{fig:RHQBox}
\end{figure}

The RHQ parameters were tuned using two experimental inputs from the $B_s$-meson system -- the spin-averaged mass $\overline M_{B_s} = (M_{B_s} + 3 M_{B_s^*})/4$ and the hyperfine-splitting $\Delta_{M_{B_s}} =  M_{B_s^*} - M_{B_s}$ -- along with the constraint that the lattice rest mass (measured from the exponential decay of meson correlators) equals the kinetic mass (measured from the meson dispersion relation).  They were obtained nonperturbatively via an iterative procedure as follows.  We began with an initial guess for the tuned values of $\{m_0a$, $c_P$, $\zeta\}$, and computed $B_s$-meson two-point correlation functions for seven sets of parameters centered on these values, as depicted in Fig.~\ref{fig:RHQBox}.  For each of the seven parameter sets we computed $\overline M_{B_s}$, $\Delta_{M_{B_s}}$, and $M_1/M_2$, and then linearly interpolated/extrapolated to the values of $\{m_0a, c_P, \zeta\}$ that reproduced the experimental meson masses from the 2010 PDG~\cite{Nakamura:2010zzi} and $M_1/M_2=1$. We repeated this procedure, re-centering the seven parameter sets each time, until all of the tuned parameter values remained inside the ``box" depicted in Fig.~\ref{fig:RHQBox}, and thus were the result of an interpolation rather than an extrapolation.  We confirmed the assumption that the meson masses depend linearly on $\{m_0a$, $c_P$, $\zeta\}$ with additional simulations using larger box sizes.

The nonperturbatively tuned RHQ parameters determined in \cite{Aoki:2012xaa} and used in this work are presented in Tab.~\ref{tab:RHQ24} and \ref{tab:RHQ32}. These tables list the final choice for the seven parameter sets used in the interpolation and the tuned results on the individual ensembles.  Because we do not observe any statistically significant dependence on the sea-quark mass, we can average the values on the different ensembles.  The tuned RHQ parameters on the $24^3$ and $32^3$ lattice spacings obtained from the weighted averages of different sea-quark ensembles are given in Table~\ref{tab:RHQParamErr}, along with our estimate of the systematic uncertainties in these values as estimated in our earlier work, Ref.~\cite{Aoki:2012xaa}.  These values of $\{m_0a, c_P, \zeta\}$ are used in our calculation of the renormalization factor $Z_V^{bb}$ in Appendix~\ref{App:Zvbb}, our estimation of heavy-quark discretization errors in Appendix~\ref{App:HQdiscErr}, and our companion calculation of the $B\to\pi\ell\nu$ form factor~\cite{Kawanai:2013qxa}.

\begin{table}[tb]
\caption{Tuned RHQ parameters $m_0a$, $c_P$, and $\zeta$ corresponding to the physical $b$-quark obtained on the same $24^3$ gauge field configurations used in this work~\cite{Aoki:2012xaa}. We used the same seven sets of parameters for the final interpolation to the tuned values on both $24^3$ ensembles. Only statistical uncertainties are quoted.} \vspace{1mm} 
\label{tab:RHQ24}
\begin{tabular}{l@{\hskip 4mm}c@{\hskip 4mm}c@{\hskip 2mm}c@{\hskip 2mm}cc}
\hline\hline
&$am_l$ & $m_0a$ & $c_P$ & $\zeta$ \\ \hline
tuning box && $8.40\pm 0.15$&$5.80\pm 0.45$&$3.20 \pm 0.30$\\ \hline
tuned values &0.005 & 8.43(7) & 5.7(2)& 3.11(9) \\
tuned values &0.010 & 8.47(9) & 5.8(2)& 3.1(1) \\ \hline\hline
\end{tabular}
\end{table}

\begin{table}[tb]
\caption{Tuned RHQ parameters $m_0a$, $c_P$, and $\zeta$ corresponding to the physical $b$ quark obtained on the same $32^3$ gauge field configurations used in this work~\cite{Aoki:2012xaa}. We used the same seven sets of parameters for the final interpolation to the tuned values on all $32^3$ ensembles. Only statistical uncertainties are quoted.}
\label{tab:RHQ32}
\begin{tabular}{l@{\hskip 4mm}c@{\hskip 4mm}c@{\hskip 2mm}c@{\hskip 2mm}cc}
\hline\hline
&$am_l$ & $m_0a$ & $c_P$ & $\zeta$ \\ \hline
tuning box &&$3.98\pm 0.10$&$3.60\pm 0.30$&$1.97 \pm 0.15$\\ \hline
tuned values &0.004 & 4.07(6) & 3.7(1)& 1.86(8) \\
tuned values &0.006 & 3.97(5) & 3.5(1)& 1.94(6) \\
tuned values &0.008 & 3.95(6) & 3.6(1)& 1.99(8) \\ \hline\hline
\end{tabular}
\end{table}

\begin{table*}
\caption{Tuned values of the RHQ parameters on the $24^3$ and $32^3$ ensembles~\cite{Aoki:2012xaa}.   The central values and statistical errors are from a weighted average of the results on the individual sea-quark ensembles given in Tables~\ref{tab:RHQ24} and~\ref{tab:RHQ32}.   The errors listed in $m_0a$, $c_P$, and $\zeta$ are from left to right:  statistics, heavy-quark discretization errors, the lattice scale uncertainty, and the uncertainty in the experimental measurement of the $B_s$-meson hyperfine splitting, respectively.  Details on the error estimation can be found in Ref.~\cite{Aoki:2012xaa}.}
\vspace{3mm}
  \begin{tabular}{lr@{.}lr@{.}lr@{.}l} \hline\hline
    & \multicolumn{2}{c}{$m_0a$} &  \multicolumn{2}{c}{$c_P$} & \multicolumn{2}{c}{$\zeta$}  \\[0.5mm] \hline 
    $a \approx 0.11$~fm  &   8&45(6)(13)(50)(7) & 5&8(1)(4)(4)(2) & 3&10(7)(11)(9)(0) \\ 
    $a \approx 0.086$~fm &  3&99(3)(6)(18)(3) & 3&57(7)(22)(19)(14) & 1&93(4)(7)(3)(0) \\ \hline\hline
  \end{tabular}
  \label{tab:RHQParamErr}
\end{table*}

\section{\texorpdfstring{Lattice calculation of $B$-meson decay amplitudes}{Lattice calculation of B-meson decay amplitudes}}
\label{Sec:DecayAmplitudes}

In QCD the $B_q$-meson decay constant is defined by the vacuum-to-meson matrix element of the heavy-light axial-vector current ${\mathcal A}_\mu = \bar b \gamma_\mu\gamma_5 q$:
\begin{align}
\langle 0 | {\mathcal A}_\mu | B_q(p)\rangle = i f_{B_q} p_\mu,
\end{align}
where $q$ denotes the light quark and $p_\mu$ is the $B_q$-meson four-momentum. Because $f_{B_q}$ behaves as $1/\sqrt{M_{B_q}}$ when $M_{B_q}$ is large,  it is advantageous to compute the decay amplitude, 
\begin{align}
\Phi_{B_q} = f_{B_q} \sqrt{M_{B_q}},
\end{align}
which is proportional to $f_{B_q}$.

In this section we describe the numerical computation of the $B$-meson decay amplitudes on the five sea-quark ensembles listed in Table~\ref{tab:lattices}.  We first describe the lattice axial-current operator renormalization and improvement, then the two-point correlator calculations and fits, and finally the {\it a posteriori} interpolation to the physical strange-quark mass.

\subsection{\texorpdfstring{Operator renormalization and improvement}{Operator renormalization and improvement}}

The lattice version of the axial-current operator, $A_\mu$, is related to the continuum current as follows:
\begin{eqnarray}
	 Z_{A_\mu} A_\mu &\doteq&  {\mathcal A}_{\mu} + {\mathcal O}\left(\alpha_s^2 a \Lambda_{\rm QCD} f_i(m_0a, c_P, \zeta)\right) \nonumber\\ &+&  {\mathcal O}\left(a^2 \Lambda_{\rm QCD}^2 f_j(m_0a, c_P, \zeta)\right) \,,
\end{eqnarray}
where $\doteq$ denotes the equality of on-shell matrix elements, and where the ${\mathcal O}(\alpha_s^2a, a^2)$ discretization errors on the right-hand-side are specific to our choice of operator improvement, discussed below.

We calculate the matching factor for the temporal component of the axial current, hereafter called $Z_\Phi$, using the mostly nonperturbative method introduced by El-Khadra et al.~in Reference \cite{ElKhadra:2001rv}. This approach takes advantage of rewriting $Z_\Phi$ as the following product:
\begin{align}
	Z_\Phi = \rho_A^{bl}\sqrt{Z_V^{ll} Z_V^{bb}}.
\label{eq:Zvbl}
\end{align}
Because the flavor-conserving renormalization factors $Z_V^{bb}$ and $Z_V^{ll}$ can be obtained nonperturbatively from standard heavy-light and light-light meson charge normalization conditions, only the residual correction $\rho_A^{bl}$ needs to be computed perturbatively.  The flavor-conserving factors $Z_V^{bb}$ and $Z_V^{ll}$ account for most of the operator renormalization, while $\rho_A^{bl}$ is expected to be close to unity because most of the radiative corrections, including contributions from tadpole graphs, cancel in the ratio $Z_{\Phi}/\sqrt{Z_V^{bb}Z_V^{ll}}$~\cite{Harada:2001fi}.  Therefore $\rho_A^{bl}$ has a more convergent series expansion in $\alpha_s$ than $Z_\Phi$ and can be computed in lattice perturbation theory to few-percent precision.

In practice, we calculate the flavor off-diagonal correction $\rho_A^{bl}$ at 1-loop in tadpole-improved lattice perturbation theory.  The results corresponding to $\alpha_s^{\bar{\rm MS}}(1/a)$ are given in Table~\ref{Tab:Zvbb}.  Details on the calculation will be provided in a forthcoming publication~\cite{CLehnerPT}.  The light-light renormalization factor $Z_V^{ll}$ has already been obtained by the RBC/UKQCD Collaborations (see Ref.~\cite{Aoki:2010dy}), where we use the fact that $Z_A = Z_V$ for domain-wall fermions up to corrections of $\CO(a m_{\rm res})$. We use the determinations in the chiral limit given in Tab.~\ref{Tab:Zvbb}.  We calculate the heavy-heavy renormalization factor $Z_V^{bb}$ as part of this project. Details of the calculation are provided in Appendix \ref{App:Zvbb}; the results are given in Tab.~\ref{Tab:Zvbb}.  As a cross-check of our use of lattice perturbation theory for $\rho_A^{bl}$, we can compare our nonperturbatively determined values of $Z_V^{bb}$ with those computed at one loop in perturbation theory,
\begin{align}
\left(Z_V^{bb}\right)^\textrm{PT}_{24c} =10.72\,, \quad \left(Z_V^{bb}\right)^\textrm{PT}_{32c} = 5.725 \,.
\end{align}
We find agreement to better than 10\% percent, which is consistent with expectations of perturbative errors of ${\mathcal O}(\alpha_s^2)$.

\begin{table}[t]
\centering
\caption{Matching factors and improvement coefficients.  The light-light flavor conserving renormalization factor $Z_V^{ll} = Z_A$ for domain-wall fermions up to corrections of $O(m_\text{res})$~\cite{Aoki:2010dy}; results quoted here are in the chiral limit. Errors shown on $Z_V^{ll}$ and $Z_V^{bb}$ are statistical only. The flavor-diagonal matching factor $\rho_A^{bl}$ and improvement coefficient $c_A$ are both computed at 1-loop in mean-field improved lattice perturbation theory \cite{Lehner:2012bt}.} \vspace{1mm}
\label{Tab:Zvbb}
\begin{tabular}{c@{\hskip 2mm}cc@{\hskip 2mm}ccc}\hline\hline
$a^{-1}$ [GeV]& $Z_V^{ll}$&$Z_V^{bb}$ & $\alpha_s^{\bar{\rm MS}}(a^{-1})$ & $\rho_A^{bl}$ & $c_A$\\
\hline
 1.729(25) & 0.71689(51) & 10.039(25) & 0.23 & 1.02658 & 0.066\\
 2.281(28)& 0.74469(13) &  5.256(8) & 0.22 & 1.01661 & 0.064\\ 
\hline\hline
\end{tabular}
\end{table}

To reduce lattice discretization errors we improve the axial-vector current ${\mathcal O}(a)$ at one-loop in mean field improved lattice perturbation theory. At this order,  only one additional matrix element needs to be computed: 
\begin{align}
\Phi_{B_q}^{(1)} &= \langle 0 | \bar b \gamma_0 \gamma_5 \sum_i \gamma_i \left( 2 \overleftarrow{D}_i \right) q | B_q(p)\rangle / \sqrt{M_{B_q}},
\end{align}
where the symmetric covariant derivative $\overleftarrow{D}_\mu$ acts on fields to the left:
\begin{align}
\bar{b}(x) \overleftarrow{D}_\mu &= \frac{1}{2} \left( \bar{b}(x+\hat{\mu}) U_\mu^\dagger(x) - \bar{b}(x-\hat{\mu}) U_\mu(x-\hat{\mu}) \right) \,. 
\end{align}
The $O(\alpha_s a)$-improved decay amplitude is then given by
\begin{align}
\Phi_{B_q}^\text{imp} = \Phi_{B_q} + c_A \Phi_{B_q}^{(1)},
\label{Eq:PhiBimp}
\end{align}
with values of the coefficient $c_A$ given in Table~\ref{Tab:Zvbb}.  Finally, we obtain the improved, renormalized decay amplitude as follows:
\begin{align}
\Phi_{B_q}^\text{ren} = Z_\Phi \left( \Phi_{B_q} + c_A \Phi_{B_q}^{(1)} \right).
\label{Eq:PhiBren}
\end{align}
Discretization errors in our simulations from the heavy-light axial-vector current are therefore of ${\mathcal O}(\alpha_s^2a, a^2)$.

\subsection{Two-point correlator fits}

To obtain the decay amplitudes we first cacluate the following two-point correlation functions:
\begin{align}
C_{AP}(t,t_0) &=  \sum_{\vec y} \langle {\cal O}_{A}^\dagger(\vec y, t) \widetilde{\cal O}_P(\vec 0, t_0) \rangle \,,
\label{Eq:C_AP} \\
C_{A^{(1)}P}(t,t_0) &=  \sum_{\vec y} \langle {\cal O}_{A^{(1)}}^\dagger(\vec y, t) \widetilde{\cal O}_P(\vec 0, t_0) \rangle \,,
\label{Eq:C_A1P} \\
C_{PP}(t,t_0) &=  \sum_{\vec y} \langle {\cal O}_P^\dagger(\vec y, t) \widetilde{\cal O}_P(\vec 0, t_0) \rangle \,,
\label{Eq:C_PP} \\
\widetilde C_{PP}(t,t_0) &=  \sum_{\vec y} \langle \widetilde{\cal O}_P^\dagger(\vec y, t) \widetilde{\cal O}_P(\vec 0, t_0) \rangle \,.
\label{Eq:Ctilde_PP}
\end{align}
where ${\cal O}_P =  \bar b\gamma_5 q$ is a pseudoscalar interpolating operator, ${\cal O}_{A} =  \bar b\gamma_0\gamma_5 q$ is the leading axial-current operator, and ${\cal O}_{A^{(1)}} = \bar b \gamma_0 \gamma_5 \sum_i \gamma_i \left(D_i^+ + D_i^- \right) q$ is the ${\mathcal O}(a)$ axial-current operator.  We use point sources for the light quarks in the correlation functions and gauge-invariant Gaussian-smeared sources~\cite{Alford:1995dm,Lichtl:2006dt} for the $b$-quark propagators.  The parameters for the Gaussian smearing were optimized in our earlier work to suppress excited-state contamination~\cite{Aoki:2012xaa}.  We use point sinks for the $b$-quarks in order to minimize the statistical errors, except in $\widetilde C_{PP}(t,t_0)$ which is used to obtain the wavefunction renormalization.  The tildes above the operators in Eqs.~(\ref{Eq:C_AP})--(\ref{Eq:Ctilde_PP}) indicate that a smeared source or sink was used for the $b$-quark.

To reduce autocorrelations between results on consecutive configurations, we place the sources of our propagators at the origin of the lattice after translating the gauge field by a random four-vector $(\vec{x},\,t)$.  This is equivalent to selecting a random source position for each configuration, but simplifies the subsequent analysis.  We double our statistics on all ensembles by folding the correlators at the temporal midpoint of the lattice, which allows us to use both forward and backward propagating states.  In the case of the $32^3$ ensembles we also double the statistics by placing a second source on each shifted configuration located at the temporal midpoint of the lattice, $(\vec{x},\,t) = (\vec{0},\, T/2)$.   After folding and averaging the correlators with the two source positions, the entire subsequent analysis chain, including the chiral-continuum extrapolations, is then carried out using a single-elimination jackknife error analysis.

At sufficiently large times, the two-point correlators are dominated by the contribution from the ground-state meson.  We can then extract the masses and renormalized, ${\mathcal O}(\alpha_s a)$-improved decay amplitudes from simple ratios of correlators:
\begin{align}
M_{B_q} &= \lim_{t\gg t_0} \cosh^{-1}\left(\frac{C_{PP}(t,t_0) + C_{PP}(t+2,t_0)}{{2}\, C_{PP}(t+1,t_0)}\right) \,, \label{Eq:EffMass}\\
\Phi_{B_q} &=  \sqrt{2} Z_\Phi \lim_{t\gg t_0} \frac{ \left|C_{AP}(t,t_0) + c_A C_{A^{(1)}P}(t,t_0)\right|}{\sqrt{\widetilde C_{PP}(t,t_0)e^{-M_{B_q}(t-t_0)}}}, \label{Eq:PhiBcorrelator}
\end{align}
where we use the values of the renormalization factor $Z_\Phi = \rho_A^{bl} \left( Z_V^{ll} Z_V^{bb} \right)^{1/2}$ and improvement coefficient $c_A$ given in Table~\ref{Tab:Zvbb}.  

\begin{figure*}[p]
\includegraphics[scale=0.45]{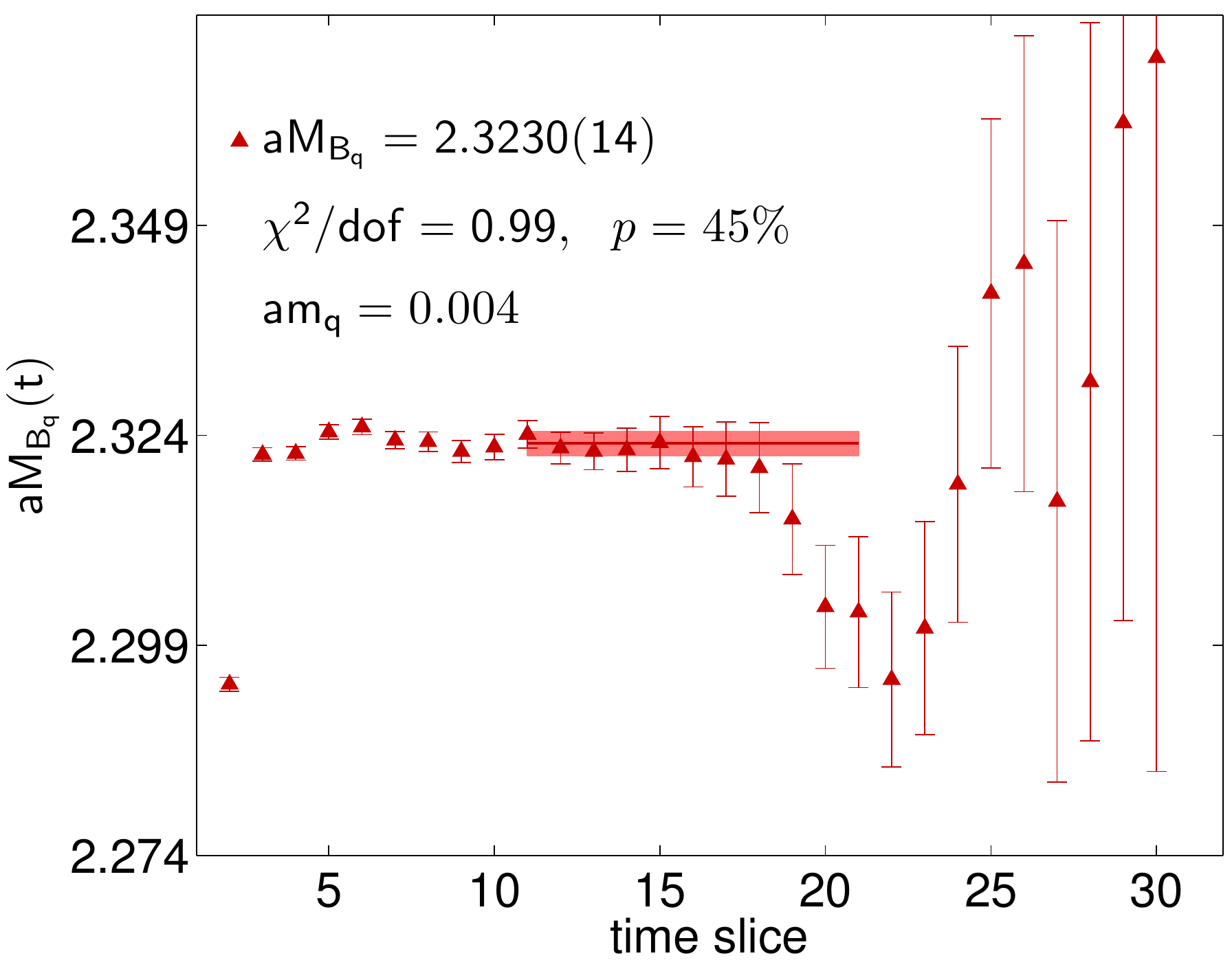}
\includegraphics[scale=0.45]{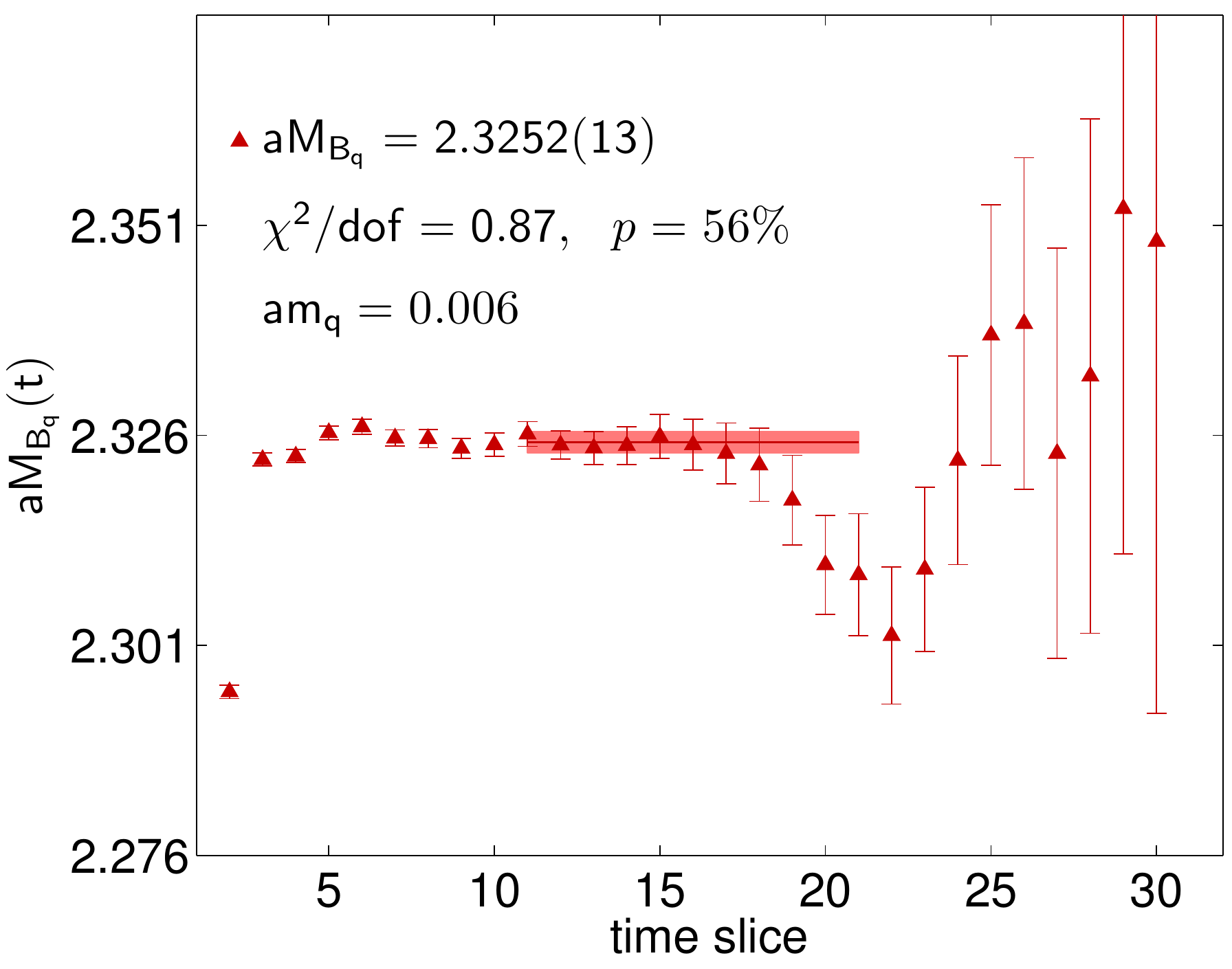}
\includegraphics[scale=0.45]{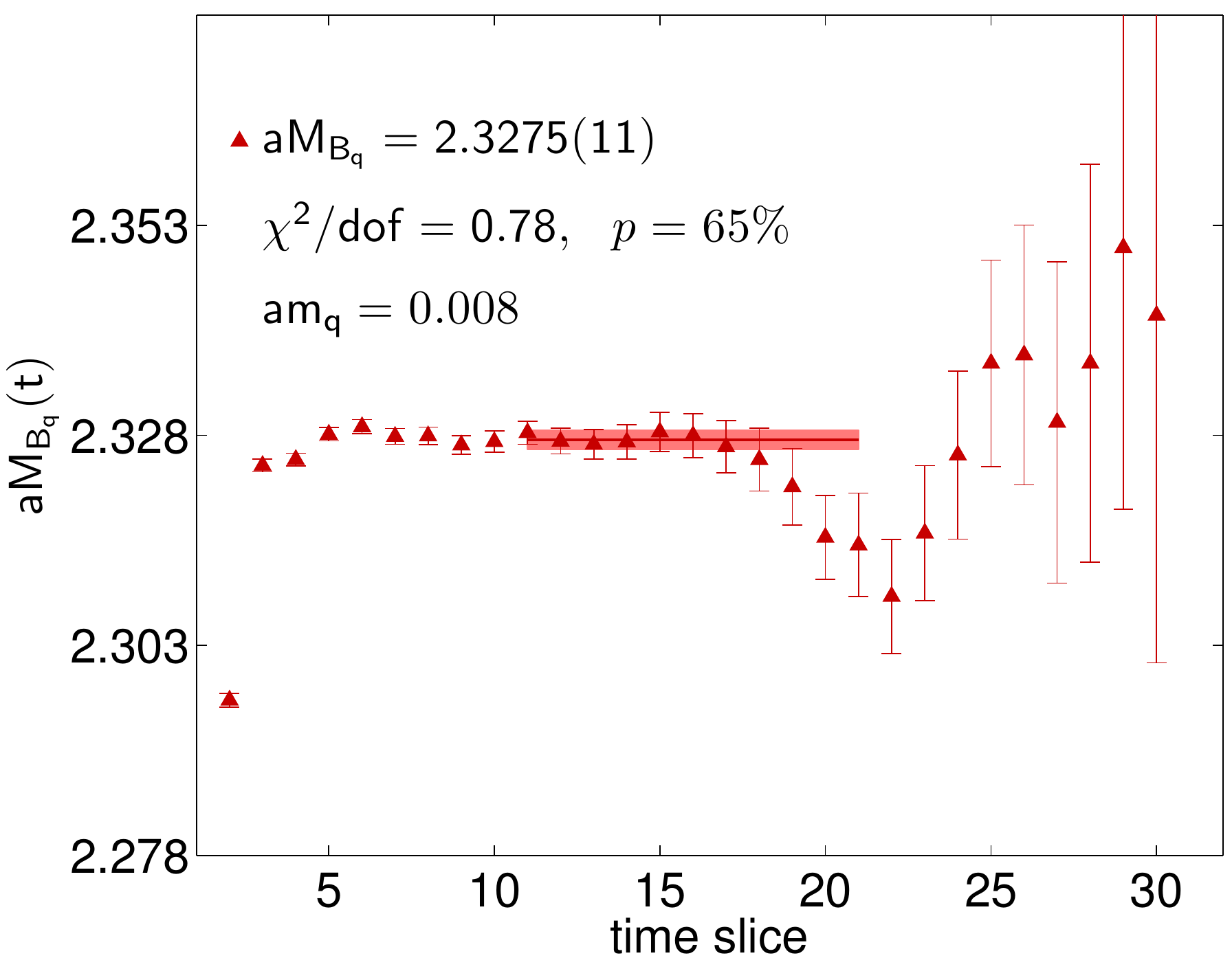}
\includegraphics[scale=0.45]{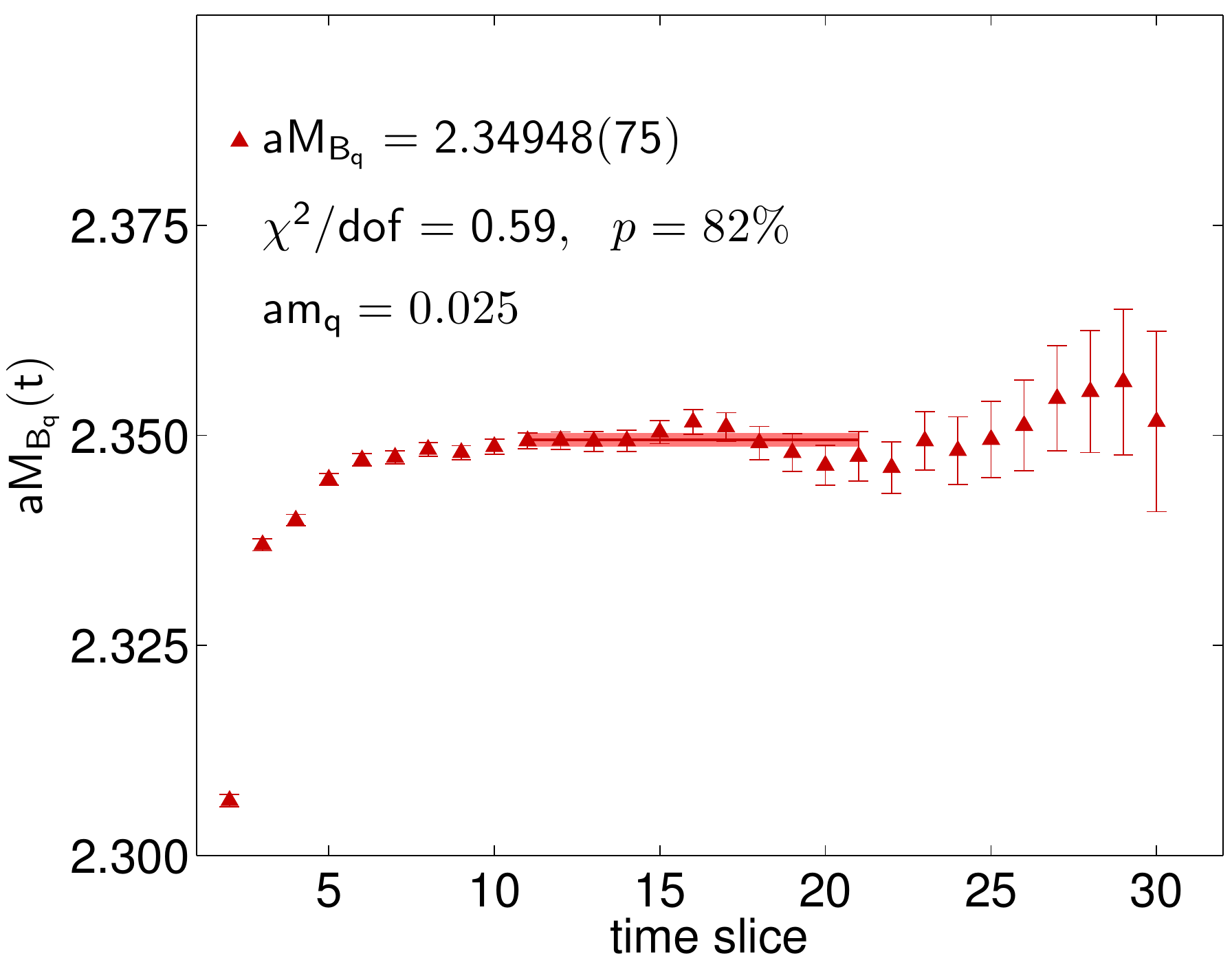}
\includegraphics[scale=0.45]{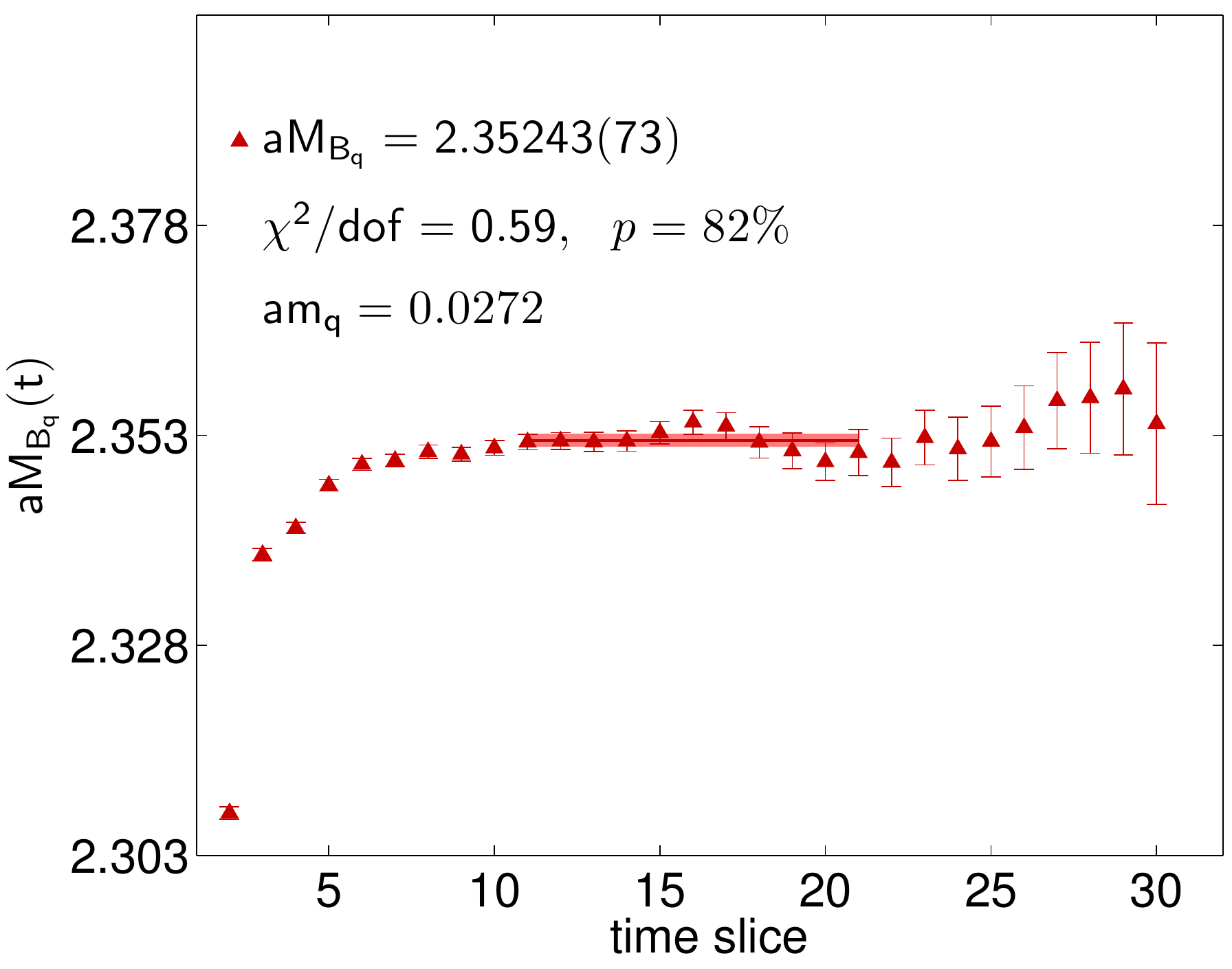}
\includegraphics[scale=0.45]{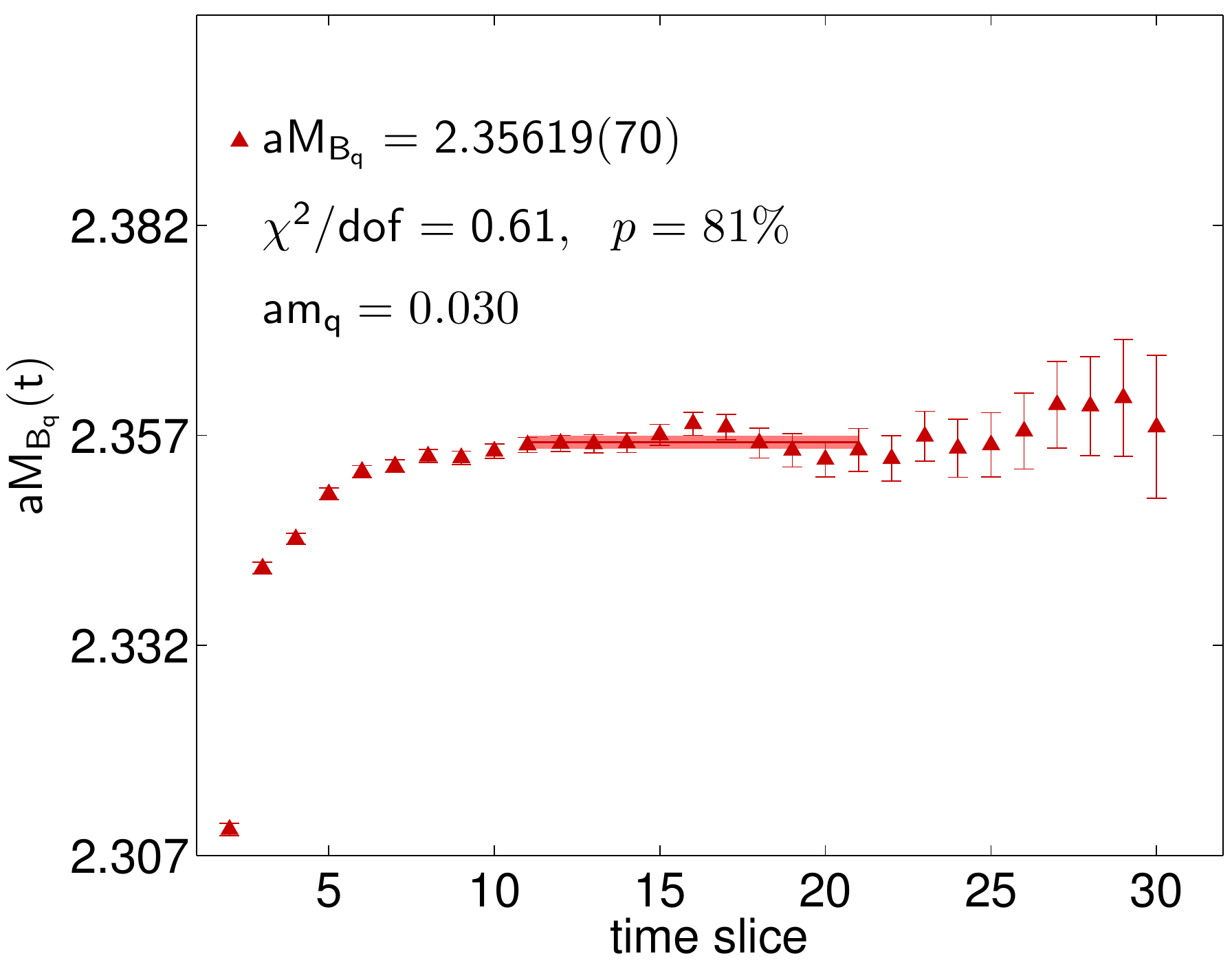}
\caption{Effective masses for all valence-quark masses we use on the $32^3$, $a m_l = 0.006$ ensemble.  The triangles show the data points with jackknife statistical errors, while the horizontal bands show the result of a correlated constant fit to the data on those time slices.}
\label{Fig:EffMasses}
\end{figure*}

\begin{figure*}[p]
\includegraphics[scale=0.45]{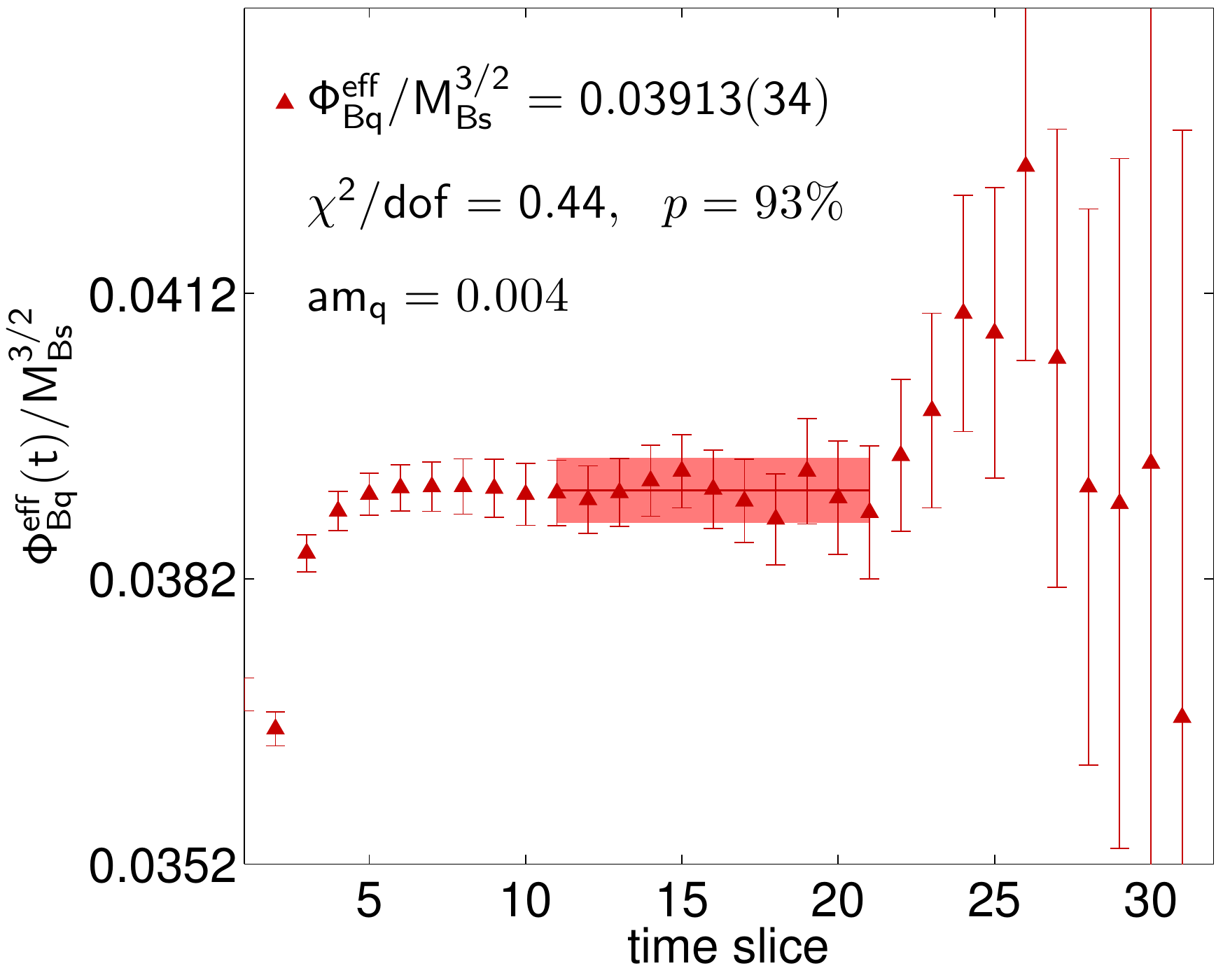}
\includegraphics[scale=0.45]{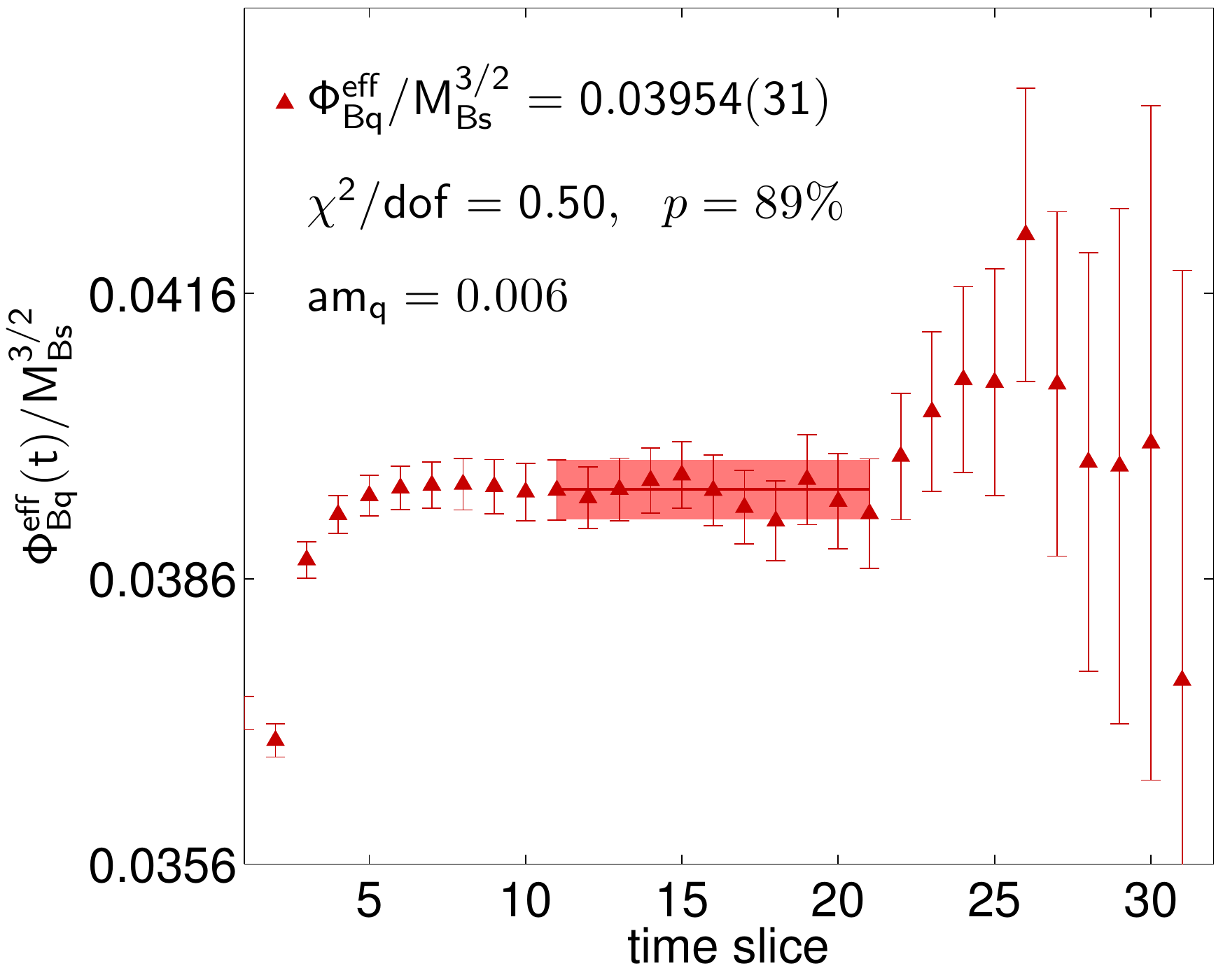}
\includegraphics[scale=0.45]{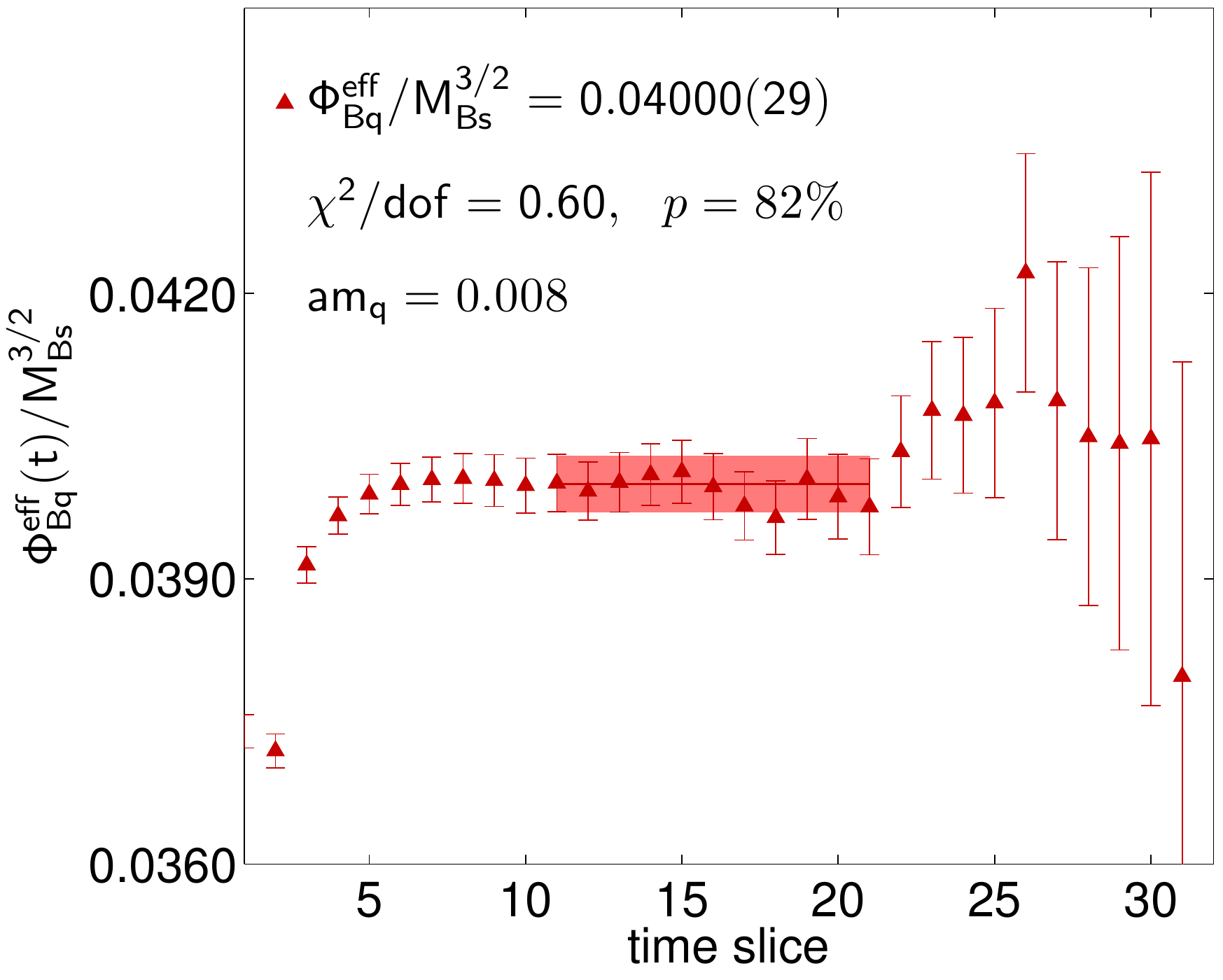}
\includegraphics[scale=0.45]{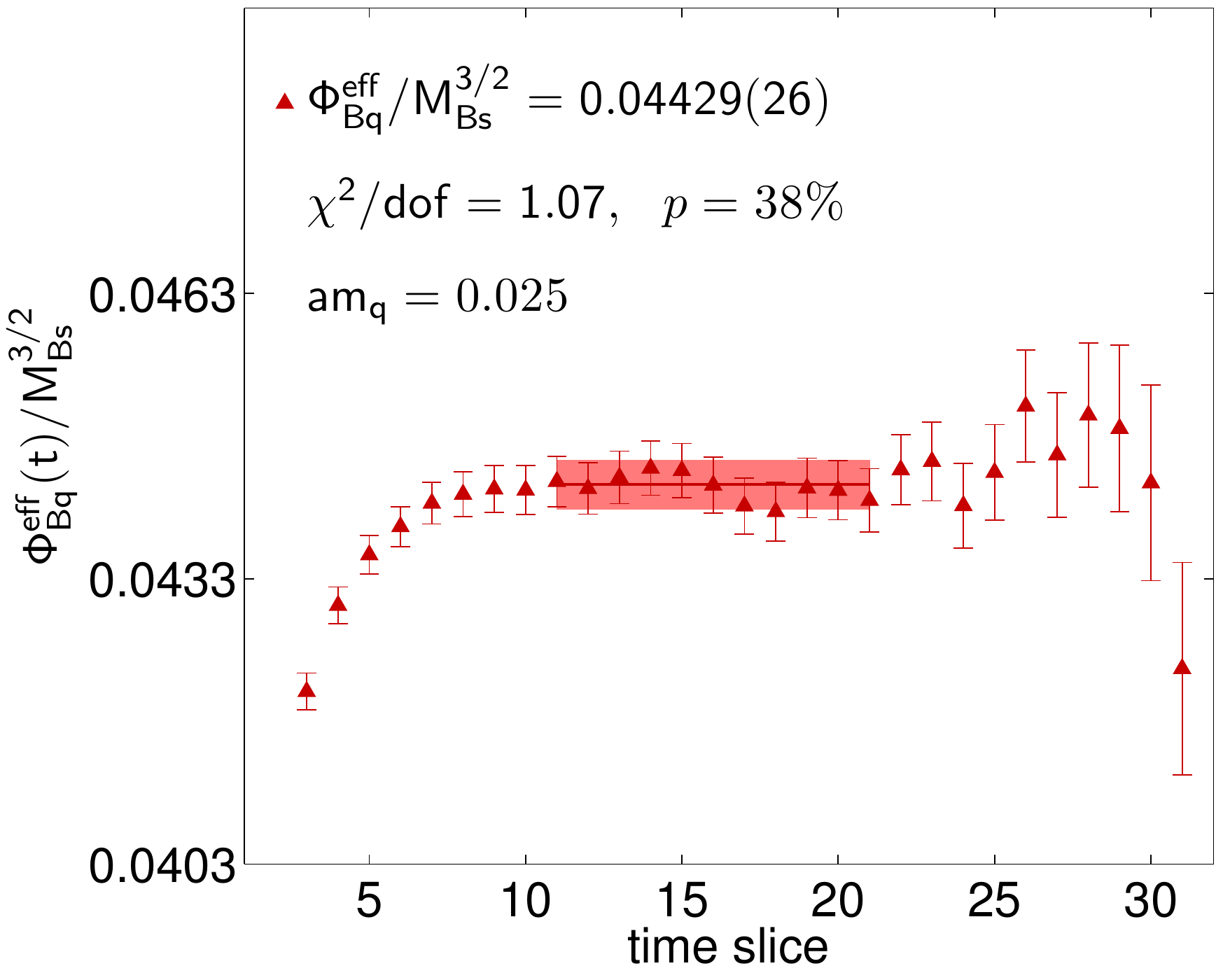}
\includegraphics[scale=0.45]{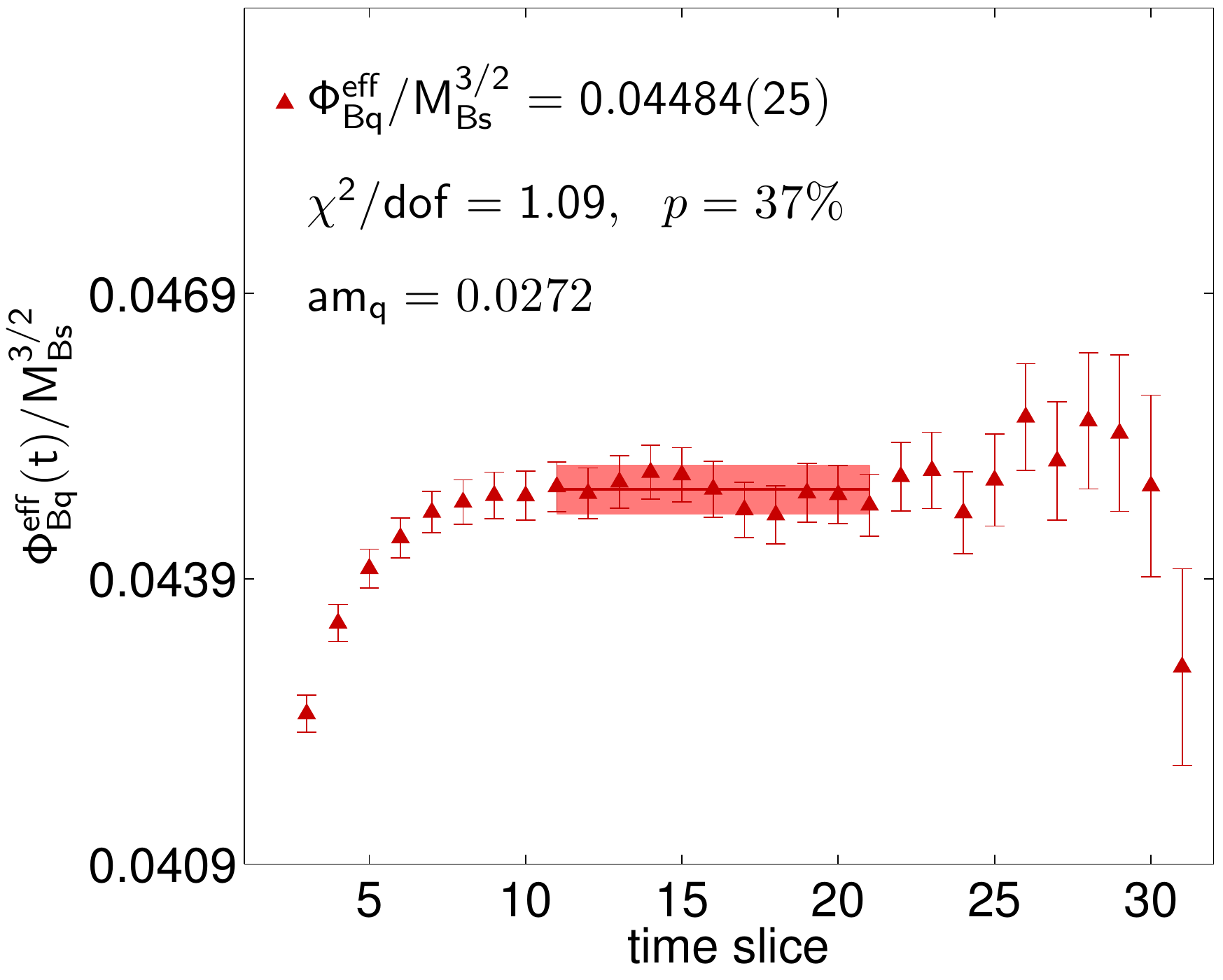}
\includegraphics[scale=0.45]{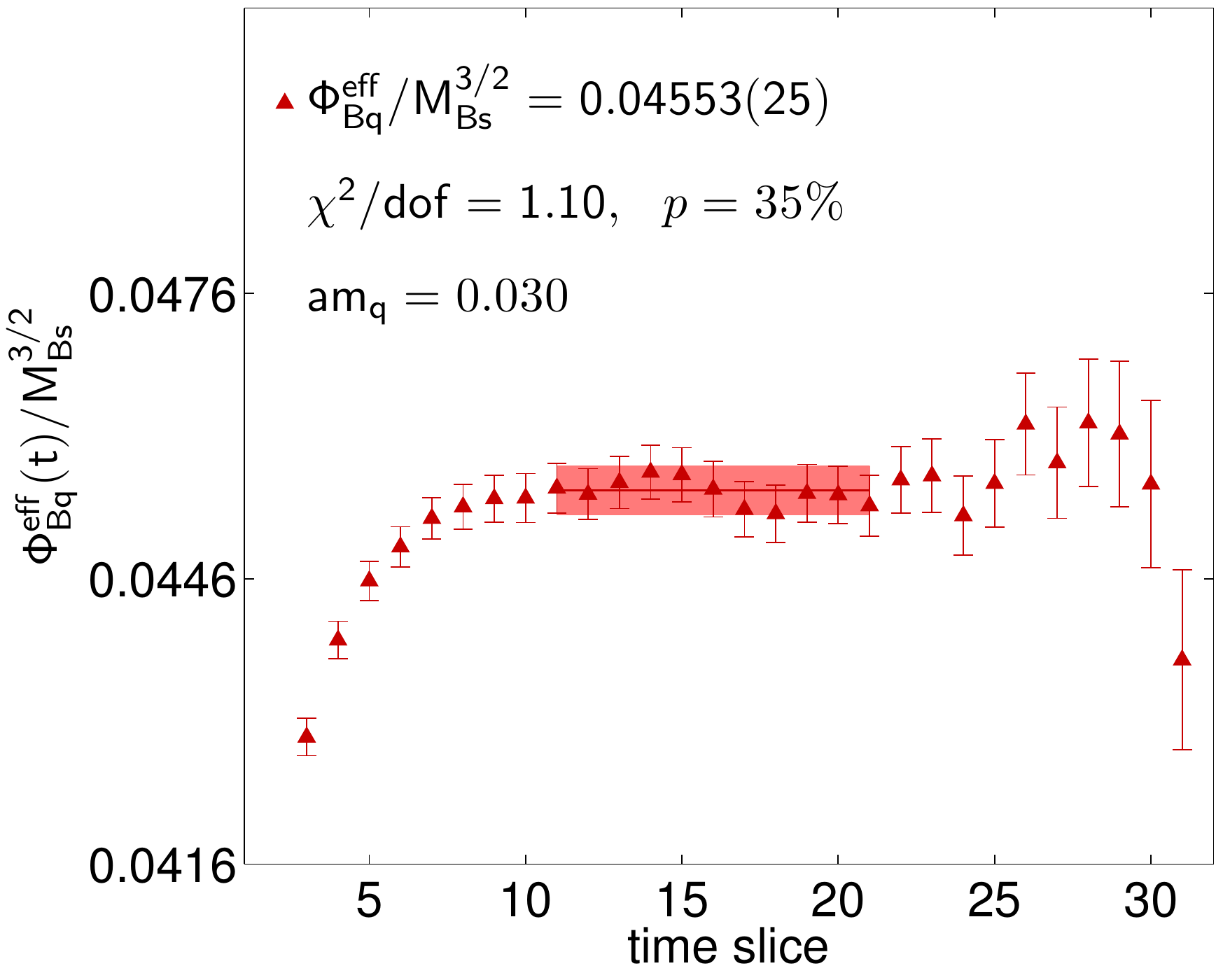}
\caption{Effective decay amplitudes for all valence-quark masses we use on the $32^3$, $a m_l = 0.006$ ensemble. The triangles show the data points with jackknife statistical errors, while the horizontal bands show the result of a correlated constant fit to the data on those time slices.} 
\label{Fig:ZPhis}
\end{figure*}

On each ensemble, for each of the seven sets of RHQ parameters $\{m_0a, c_P, \zeta\}$ listed in in the first rows of Tables~\ref{tab:RHQ24} and \ref{tab:RHQ32} and six valence-quark masses listed in Table~\ref{Tab:ValenceLight}, we obtain the meson masses and decay amplitudes from correlated plateau fits to the above ratios. We use the same range of time slices as in our tuning procedure \cite{Aoki:2012xaa}: $[t_{\rm min},t_{\rm max}] = [10,25]$ on the $24^3$ ensembles and $[t_{\rm min},t_{\rm max}] = [11,21]$ on the $32^3$ ensembles. Figures~\ref{Fig:EffMasses} and~\ref{Fig:ZPhis} show effective mass and decay amplitude plots for the six different light valence-quark masses and the central RHQ parameter set on the $32^3$ ensemble with $m_l = 0.006$ in units of $M_{B_s}$. The plateaus for other ensembles and sets of RHQ parameters look similar. 

{To check for residual autocorrelations, on each ensemble we compare the fit results for the masses and decay amplitudes of $B_q$ mesons with unitary and close-to-strange valence quarks after blocking the configurations.  We consider binning the data in groups of 2 to 8 configurations, and find no significant change in the statistical errors with bin size.  We therefore use the unbinned results in our subsequent analysis.}

Our computation is carried out using the Chroma software library~\cite{Edwards:2004sx} supplemented by our own code for measuring matrix elements for the $O(a)$-improvement and the three-point correlation functions needed for the determination of $Z_V^{bb}$.

\subsection{Interpolation to the tuned RHQ parameters}

As mentioned above, the extraction of $B$-meson decay amplitudes is performed for each of the seven sets of RHQ parameters.  We must then interpolate these results to the tuned values of $\{m_0a, c_P, \zeta\}$ that correspond to the physical $b$-quark.

We first interpolate the seven different masses $M_{B_q}^r$, where the index $r$ runs over the seven parameter sets, to the mass of the $B_q$-meson via a jackknife procedure, in which we utilize the jackknife blocks for the RHQ parameters created as part of our tuning procedure~\cite{Aoki:2012xaa}.  We assume that the masses depend linearly on $\{m_0a, c_P, \zeta\}$:
\begin{align}
M_{B_q}^\text{RHQ} &= J_M \times \left[ \begin{matrix} m_0 a\\c_P\\\zeta \end{matrix}\right]^\text{RHQ} +A_M,
\label{Eq:RHQinterpolate}
\end{align}
where $J_M$ is a three component vector and $A_M$ a constant for each jackknife block,
\begin{align}
J_M &= \left[\frac{M_{B_q}^3-M_{B_q}^2}{2\sigma_{m_0a}},\frac{M_{B_q}^5-M_{B_q}^4}{2\sigma_{c_P}},\frac{M_{B_q}^7-M_{B_q}^6}{2\sigma_{\zeta}}\right],\label{Eq:RHQ_JM}\\
A_M &= M_{B_q}^1 - J_M \times\left[m_0a, c_P, \zeta\right]^T \,,
\label{Eq:RHQ_AM}
\end{align}
and the $\sigma$'s are the variations of the parameters listed in Tables~\ref{tab:RHQ24} and~\ref{tab:RHQ32}.  This procedure allows us to directly propagate statistical uncertainties from the tuning procedure to the meson masses and later also into the decay amplitudes.  We list the values for all meson masses interpolated to the physical $b$-quark in Table~\ref{Tab:Masses+ZPhis}.  We follow the same procedure for the decay amplitudes, but with $M_{B_q} \to \Phi_{B_q}^\text{ren}$ in Eqs.~(\ref{Eq:RHQinterpolate})--(\ref{Eq:RHQ_AM}).  

The renormalized decay amplitudes for all valence-quark masses and ensembles are also listed in Table~\ref{Tab:Masses+ZPhis}. We present the results as dimensionless ratios in units of the $B_s$-meson mass, and perform the subsequent chiral-continuum extrapolation using these ratios. Because the RHQ parameters are tuned such that $M_{B_s}$ reproduces the experimental value, this enables us to avoid two potential sources of uncertainty associated with the lattice-scale determination: (1) in the joint chiral-continuum fits to the data on both lattice spacings we do not need to know the ratio $\left(a_{24}/a_{32}\right)$ to relate the overall normalizations of the decay amplitudes $\Phi_{B_q}$ on the different ensembles, and (2) when we convert the final results for the decay constants to physical units we can simply multiply by the experimental value of $M_{B_s}$.

\begin{table}[tb]
\caption{Masses and renormalized decay amplitudes on the $24^3$ ensembles (upper two panels) and $32^3$ ensembles (lower three panels) with statistical errors.} \vspace{1mm}
\label{Tab:Masses+ZPhis}
\begin{tabular}{c@{~~~}l@{~~~}l@{~~~}l@{~~~}c}
\hline \hline
$a^{-1}$ [GeV]&$am_l$ & $am_{q}$ & \hspace{4mm}$aM_{B_q}$& $\Phi_{B_q}^\text{ren}/M_{B_s}^{3/2}$ \\
\hline
1.729(25) &0.005 & 0.005\phantom{0} & 3.0644(16) &  0.03999(64)\\ 
1.729(25) &0.005 & 0.010\phantom{0} & 3.0715(11) &  0.04107(60)\\ 
1.729(25) &0.005 & 0.020\phantom{0} & 3.0849(5)  &  0.04323(58)\\ 
1.729(25) &0.005 & 0.030\phantom{0} & 3.0978(2)  &  0.04532(58) \\ 
1.729(25) &0.005 & 0.0343 & 3.1034(2) &             0.04619(58)\\ 
1.729(25) &0.005 & 0.040\phantom{0} & 3.1106(3)   & 0.04733(58) \\ 
\hline
1.729(25) &0.010 & 0.005\phantom{0} & 3.0656(19) &  0.04001(75)  \\ 
1.729(25) &0.010 & 0.010\phantom{0} & 3.0723(12) &  0.04105(70)  \\ 
1.729(25) &0.010 & 0.020\phantom{0} & 3.0854(6)  &  0.04315(67)  \\ 
1.729(25) &0.010 & 0.030\phantom{0} & 3.0983(3)  &  0.04520(67)  \\ 
1.729(25) &0.010 & 0.0343 & 3.1038(3) &             0.04607(67)  \\ 
1.729(25) &0.010 & 0.040\phantom{0} & 3.1111(3)  &  0.04718(68)  \\ 
\hline
2.281(28) &0.004 & 0.004\phantom{0} & 2.3231(13) &  0.03961(61) \\ 
2.281(28) &0.004 & 0.006\phantom{0} & 2.3252(10) & 0.04005(59) \\ 
2.281(28) &0.004 & 0.008\phantom{0} & 2.3275(8) & 0.04054(57) \\ 
2.281(28) &0.004 & 0.025\phantom{0} & 2.3497(2) & 0.04504(60) \\ 
2.281(28) &0.004 & 0.0272 & 2.3526(2) &           0.04560(60) \\ 
2.281(28) &0.004 & 0.030\phantom{0} & 2.3564(2) & 0.04632(61) \\ 
\hline
2.281(28) &0.006 & 0.004\phantom{0} & 2.3233(10) &  0.03930(51) \\ 
2.281(28) &0.006 & 0.006\phantom{0} & 2.3254(8) & 0.03971(49) \\ 
2.281(28) &0.006 & 0.008\phantom{0} & 2.3277(6) & 0.04016(48) \\ 
2.281(28) &0.006 & 0.025\phantom{0} & 2.3496(1) & 0.04447(49) \\ 
2.281(28) &0.006 & 0.0272 & 2.3526(1) &           0.04502(50) \\ 
2.281(28) &0.006 & 0.030\phantom{0} & 2.3563(1) & 0.04572(50) \\ 
\hline
2.281(28) &0.008 & 0.004\phantom{0} & 2.3236(14) &  0.03961(67) \\ 
2.281(28) &0.008 & 0.006\phantom{0} & 2.3257(11) &  0.03997(65) \\ 
2.281(28) &0.008 & 0.008\phantom{0} & 2.3281(9) & 0.04041(63) \\ 
2.281(28) &0.008 & 0.025\phantom{0} & 2.3495(1) & 0.04448(64) \\ 
2.281(28) &0.008 & 0.0272 & 2.3523(1) &           0.04500(64) \\ 
2.281(28) &0.008 & 0.030\phantom{0} & 2.3560(1) & 0.04565(65) \\ 
\hline\hline
\end{tabular}
\end{table}

\subsection{Interpolation to the physical strange-quark mass}
\label{sec:InterpMs}

\begin{figure}[tb]
\includegraphics[scale=0.45]{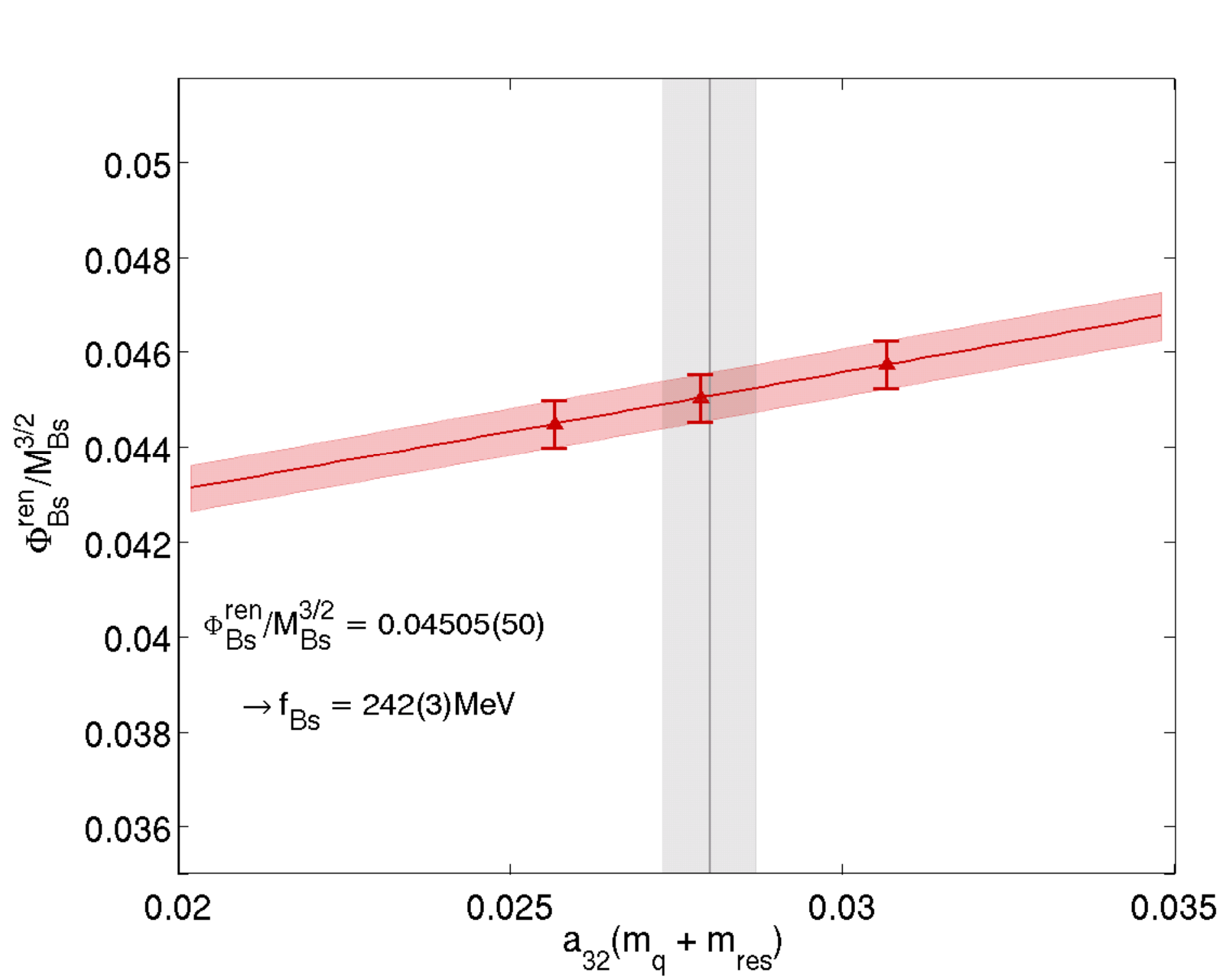}
\caption{Linear interpolation to determine $\Phi_{B_s}$ at the physical strange-quark mass on the $32^3$, $am_l = 0.006$ ensemble.  The black vertical line with error band shows the physical strange-quark mass with errors from Ref.~\cite{Aoki:2010dy}.  The red sloped line with error band shows the interpolation of the three strange-ish data points with jackknife statistical errors from the fit.}
\label{Fig:fBsInterpolation}
\end{figure}

For the determination of $f_{B_s}$ and the ratio $f_{B_s}/f_B$ we must slightly interpolate our data with close-to-strange valence-quark masses to the physical strange quark mass as determined in \cite{Aoki:2010dy}.  We perform a linear, uncorrelated fit to interpolate the three heaviest masses on each ensemble: $am_q = 0.03,\, 0.0343,\,\text{and}\, 0.04$ on the $24^3$ ensembles and $am_q = 0.025,\,0.0272,\,\text{and}\, 0.03$ on the $32^3$ ensembles. Figure~\ref{Fig:fBsInterpolation} shows an example determination of $\Phi_{B_s}^\text{ren}$ on the $32^3$ ensemble with $a m_l = 0.006$. Table~\ref{Tab:PhiBs} lists the $\Phi_{B_s}^\text{ren}$ values for all five ensembles. We then use the interpolated values for $\Phi_{B_s}^\text{ren}$ to obtain the ratio of the SU(3)-breaking ratio $\Phi_{B_s}^\text{ren}/\Phi_{B_q}^\text{ren}$ for all six light valence-quark masses on each ensemble.  We include statistical correlations between the numerator and denominator via a jackknife, and list the results in Tab.~\ref{Tab:ratios}.

\begin{table}[tb]
\caption{Interpolated decay amplitudes $\Phi_{B_s}^\text{ren}$ with statistical errors.}\vspace{1mm}
\label{Tab:PhiBs}
\begin{tabular}{c@{~~~}c@{~~~}c}
\hline \hline
$a^{-1}$ [GeV] &$am_l$ & $\Phi_{B_s}^\text{ren}/M_{B_s}^{3/2}$\\
\hline
1.729(25) & 0.005 & 0.04627(58) \\ 
1.729(25) & 0.010 & 0.04615(67) \\ \hline
2.281(28) & 0.004 & 0.04563(61) \\ 
2.281(28) & 0.006 & 0.04505(50)\\ 
2.281(28) & 0.008 & 0.04503(64)\\ 
\hline\hline
\end{tabular}
\end{table}

\begin{table}[tb]
\caption{Decay-amplitude ratios $\Phi_{B_s}^\text{ren}/\Phi_{B_q}^\text{ren}$ at the physical strange-quark mass with statistical errors.}
\label{Tab:ratios}
\begin{tabular}{c@{~~~}l@{~~~}c@{~~~}l}
\hline \hline
$a^{-1}$ [GeV] &$am_l$ & $am_q$ & $\Phi_{B_s}^\text{ren}/\Phi_{B_q}^\text{ren}$\\
\hline
1.729(25) &0.005& 0.005\phantom{0} & 1.1573(94)\\
1.729(25) &0.005& 0.010\phantom{0} & 1.1266(60) \\
1.729(25) &0.005& 0.020\phantom{0} & 1.0705(25) \\
1.729(25) &0.005& 0.030\phantom{0} & 1.02113(61) \\
1.729(25) &0.005& 0.0343           & 1.001807(41) \\
1.729(25) &0.005& 0.040\phantom{0} & 0.97777(59) \\ \hline
1.729(25) &0.010& 0.005\phantom{0} & 1.153(11) \\
1.729(25) &0.010& 0.010\phantom{0} & 1.1242(67) \\
1.729(25) &0.010& 0.020\phantom{0} & 1.0694(27) \\
1.729(25) &0.010& 0.030\phantom{0} & 1.02086(65) \\
1.729(25) &0.010& 0.0343           & 1.001789(44) \\
1.729(25) &0.010& 0.040\phantom{0} & 0.97803(63) \\\hline
2.281(28) & 0.004& 0.004\phantom{0} & 1.1522(81)\\
2.281(28) & 0.004& 0.006\phantom{0} & 1.1393(62)\\
2.281(28) & 0.004& 0.008\phantom{0} & 1.1255(49)\\
2.281(28) & 0.004& 0.025\phantom{0} & 1.01329(28)\\
2.281(28) & 0.004& 0.0272           & 1.000694(18)\\
2.281(28) & 0.004& 0.030\phantom{0} & 0.98529(30)\\ \hline
2.281(28) & 0.006& 0.004\phantom{0} & 1.1465(64)\\
2.281(28) & 0.006& 0.006\phantom{0} & 1.1347(49)\\
2.281(28) & 0.006& 0.008\phantom{0} & 1.1218(38)\\
2.281(28) & 0.006& 0.025\phantom{0} & 1.01308(22)\\
2.281(28) & 0.006& 0.0272           & 1.000691(11)\\
2.281(28) & 0.006& 0.030\phantom{0} & 0.98551(23)\\ \hline
2.281(28) & 0.008& 0.004\phantom{0} & 1.1367(85)\\
2.281(28) & 0.008& 0.006\phantom{0} & 1.1264(66)\\
2.281(28) & 0.008& 0.008\phantom{0} & 1.1143(52)\\
2.281(28) & 0.008& 0.025\phantom{0} & 1.01235(30)\\
2.281(28) & 0.008& 0.0272           & 1.000666(19)\\
2.281(28) & 0.008& 0.030\phantom{0} & 0.98629(32)\\
\hline\hline
\end{tabular}
\end{table}

\section{Chiral and continuum extrapolations}
\label{Sec:Extrapolations}

In this section we present the extrapolation to the physical light-quark masses and to the continuum limit of the numerical lattice data presented in the previous section and summarized in Tabs.~\ref{Tab:Masses+ZPhis}--\ref{Tab:ratios}.  All extrapolations are performed for dimensionless ratios of decay amplitudes in units of the $B_s$-meson mass; we obtain the physical decay constants in GeV after the chiral-continuum extrapolation by multiplying by the appropriate power of $M_{B_s}$.

\subsection{\texorpdfstring{Chiral-continuum extrapolations of $f_{B}$ and $f_{B_s}/f_B$}{fB and fBs/fB}}
\label{Sec:f_B}

We obtain the decay constant $f_B$ and the ratio $f_{B_s}/f_{B}$ {at the physical pion mass and in the continuum limit using theoretical knowledge of the light-quark-mass and lattice-spacing dependence from chiral perturbation theory for heavy-light mesons (HM$\chi$PT) to guide the extrapolation.  Chiral perturbation theory provides a model-independent, low-energy effective description of QCD in terms of the light-pseudoscalar-meson degrees-of-freedom, provided that the mesons are sufficiently light.  The RBC and UKQCD collaborations find that next-to-leading order (NLO) SU(3) $\chi$PT does not describe their data for light pseudoscalar-meson masses and decay constants near the physical strange-quark mass, but that NLO SU(2) $\chi$PT can be applied to their heavier data and leads to reasonable estimates for the NNLO corrections~\cite{Allton:2008pn}.  Other collaborations also find that,  within its range of validity, SU(2) $\chi$PT converges more quickly than SU(3) $\chi$PT~\cite{Aoki:2008sm,Kadoh:2008sq,Bazavov:2009ir}.  Thus, in this work, we perform the combined chiral- and continuum extrapolation using NLO SU(2) HM$\chi$PT.} 

In the SU(2) theory, the strange-quark mass is integrated out, and only the light-quarks' degrees-of-freedom are included.  The SU(2) low-energy constants therefore depend upon the value of $m_s$, as well as on the value of $m_b$ for heavy-light quantities.  In Ref.~\cite{Albertus:2010nm} we derived the HM$\chi$PT expressions for $B_{(s)}$-meson decay constants in the context of our calculation using domain-wall light quarks and static heavy quarks, which we quote here:
\begin{align}
\label{FitFuncSU2}
\Phi_{B_x} &= \Phi_0 \Bigg\{ 
 1 - \frac{1+3g_b^2}{(4 \pi f_\pi)^2} \cdot M_{xl}^2 \ln(M_{xl}^2/\Lambda_\chi^2)\nonumber \\
   &- \frac{1+3g_b^2}{(4 \pi f_\pi)^2} \frac{1}{4}\cdot \Big[(M_{ll}^2-M_{xx}^2)\cdot (\ln(M_{xx}^2/\Lambda_\chi^2) +1)\nonumber \\
 &\qquad-  M_{xx}^2\ln(M_{xx}^2/\Lambda_\chi^2) \Big]\nonumber \\  
&+ c_\text{sea} \cdot \frac{2B m_l}{(4 \pi f_\pi)^2}  
+ c_\text{val} \cdot \frac{2B m_x}{(4 \pi f_\pi)^2} \nonumber \\
&+ c_\text{a} \cdot \frac{a^2}{(4 \pi f_\pi)^2a_{32}^4}\Bigg\},
\end{align}
\begin{align}
\label{FitRatioSU2}
\frac{\Phi_{B_s}}{\Phi_{B_x}} & =
 R_{\Phi^{(2)}} \Bigg\{1 + \frac{1+3g_b^2}{(4 \pi f_\pi)^2} \cdot M_{xl}^2 \ln(M_{xl}^2/\Lambda_\chi^2)\nonumber \\
   &{+} \frac{1+3g_b^2}{(4 \pi f_\pi)^2} \frac{1}{4}\cdot \Big[(M_{ll}^2-M_{xx}^2)\cdot (\ln(M_{xx}^2/\Lambda_\chi^2) +1)\nonumber \\
 &\qquad-  M_{xx}^2\ln(M_{xx}^2/\Lambda_\chi^2) \Big]\Bigg\}\nonumber \\  
&+ d_\text{sea}^{(2)}\cdot\frac{2B }{(4\pi f)^2} m_l + d_\text{val}^{(2)}\cdot\frac{2B m_x}{(4\pi f)^2} \nonumber\\
&+ d_\text{a}^{(2)}\cdot\frac{ a^2}{(4\pi f)^2 a_{32}^4}.
\end{align}
where $x$ denotes the light valence quark in the $B_x$ meson, $l$ the light sea quark, and $M_{xy}$ denotes a ``pion" composed of two domain-wall valence quarks with flavors $x$ and $y$.  At tree level, the light pseudoscalar pion masses are given in terms of the constituent quark masses $m_x$ and $m_y$ by
\begin{align}
\label{eq:MPiSq}
M_{xy}^2 = B(m_x +m_y + 2\mres).
\end{align}
The fit functions Eq.~(\ref{FitFuncSU2}) and~(\ref{FitRatioSU2}) incorporate discretization errors due to the light-quark and gluon actions via the residual-quark mass in Eq.~(\ref{eq:MPiSq}) and the analytic term in $a^2$.  

From simple power-counting, we estimate that discretization errors in the decay amplitudes on the $32^3$ ensembles from the light-quark and gluon actions are of ${\mathcal O}\left(a \Lambda_{\rm QCD} \right)^2 \sim 5\%$, using $\Lambda_{\rm QCD} = 500$~MeV.   There are also light-quark and gluon discretization errors in the heavy-light current, and heavy-quark discretization errors from both the action and current.  In Secs.~\ref{Sec:HeavyQuarkDiscErrors}--\ref{Sec:LightQuarkDiscErrors} and App.~\ref{App:HQdiscErr} we estimate the size of these other discretization errors  to be below 2\%.  Thus we expect light-quark and gluon discretization errors from the action to dominate the scaling behavior of the decay amplitudes, and that including an $a^2$ term in the fit will largely remove these contributions.  Heavy-quark discretization errors as well as light-quark and gluon discretization errors in the current will be estimated using power-counting and added {\it a posteriori} to the systematic error budget.

Several parameters enter the expressions in Eq.~(\ref{FitFuncSU2}) and~(\ref{FitRatioSU2}).  We take the values of the lattice spacings and low-energy constant $B$ from the RBC/UKQCD analysis of light pseudoscalar meson masses and decay constants in Ref.~\cite{Aoki:2010dy}.  We use the experimental value of $f_\pi = 130.4$ MeV from the PDG~\cite{Beringer:2012zz}, and use $\Lambda_\chi = 1$~GeV for the scale in the chiral logarithms.   We take the $B^\ast B \pi$-coupling constant, $g_b = 0.57(8)$ from our companion analysis~\cite{Flynn:2013kwa} using the same actions and ensembles.  The constant parameters used in our chiral fits are compiled in Table~\ref{Tab:ChiPTconst}.

\begin{table}[tb]
\caption{Constants used in the chiral and continuum extrapolations of $\Phi_{B}^\text{ren}$, $\Phi_{B_s}^\text{ren}$ and $\Phi_{B_s}^\text{ren}/\Phi_{B}^\text{ren}$~\cite{Aoki:2010dy,Beringer:2012zz,Flynn:2013kwa}.}
\label{Tab:ChiPTconst}
\begin{tabular}{ccc}
\hline\hline
& $24^3$& $32^3$\\ \hline
$a^{-1}$& 1.729~GeV& 2.28~GeV \\
$aB$ & 2.348 & 1.826\\
$f_\pi$ &\multicolumn{2}{c}{130.4 MeV}\\
$g_b$ &\multicolumn{2}{c}{0.57}\\
$\Lambda_\chi$ &\multicolumn{2}{c}{1~GeV}\\
\hline\hline
\end{tabular}
\end{table}

We cannot obtain a good fit (as measured by the $\chi^2/{\rm dof}$ or $p$-value) to our entire data set using the NLO SU(2) HM$\chi$PT expressions above. This is not surprising given that our heaviest pseudoscalar mesons, in which the valence-quark masses are close to that of the physical strange quark, have masses around 600~MeV.  We therefore tried removing the heaviest points from our fits.  We find that we can obtain good fits {of $\Phi_B$} with NLO HM$\chi$PT while including as much of our data as possible when we impose the following cut: $M_\pi^\text{sea} {\ltapprox} 425$~MeV and $M_\pi^\text{val}  {\ltapprox} 350$~MeV. {To obtain an acceptable fit for the decay-constant ratio $\Phi_{B_s}/\Phi_B$, however, we must make an equally-stringent cut on the sea-pion masses: $M_\pi^\text{val}, M_\pi^\text{sea} \ltapprox 350$~MeV.} {We also tried adding higher-order terms analytic in the pion mass in order to extend the reach of the HM$\chi$PT expressions.  We find, however, that multiple NNLO analytic terms are needed to improve the $p$-value, at which point the errors on the extrapolated values of $f_B$ and $f_{B_s}/f_B$ become uncontrolled because we cannot sufficiently constrain the coefficients with data at only two lattice spacings and a narrow range of light-quark masses.  Finally, we tried NLO SU(2) fits to only the five unitary data points with $M_\pi^\text{val} = M_\pi^\text{sea}$, all of which satisfy $M_\pi \ltapprox 425$~MeV.  Without the partially-quenched data, which are strongly correlated on a given ensemble, we obtain good $p$-values for fits of both $\Phi_B$ and $\Phi_{B_s}/\Phi_B$.  Thus we take the unitary NLO SU(2) HM$\chi$PT fit results as our central values for $f_B$ and $f_{B_s}/f_B$.}  

\begin{figure*}[tb]
\includegraphics[scale=0.45]{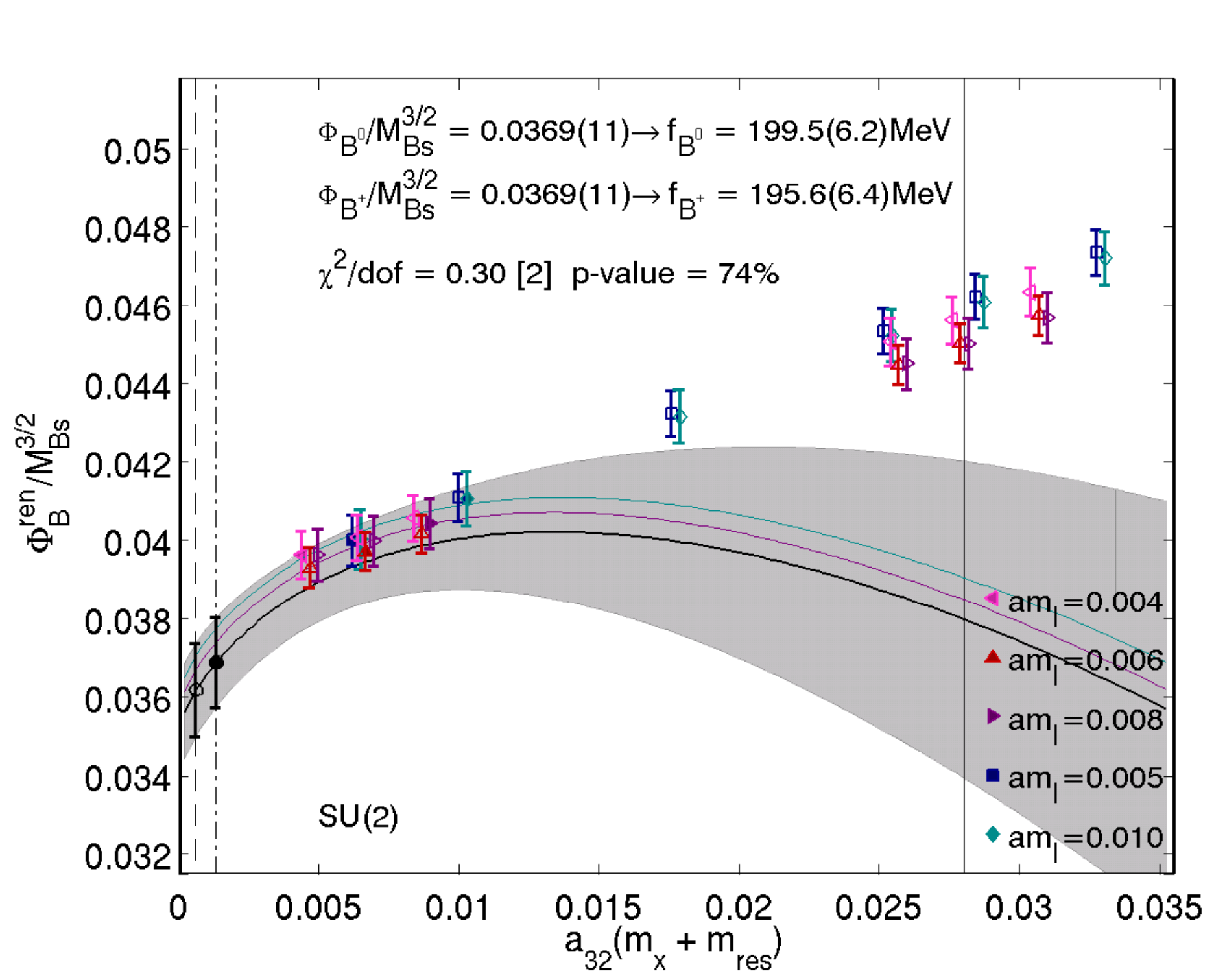}
\hfill \includegraphics[scale=0.45]{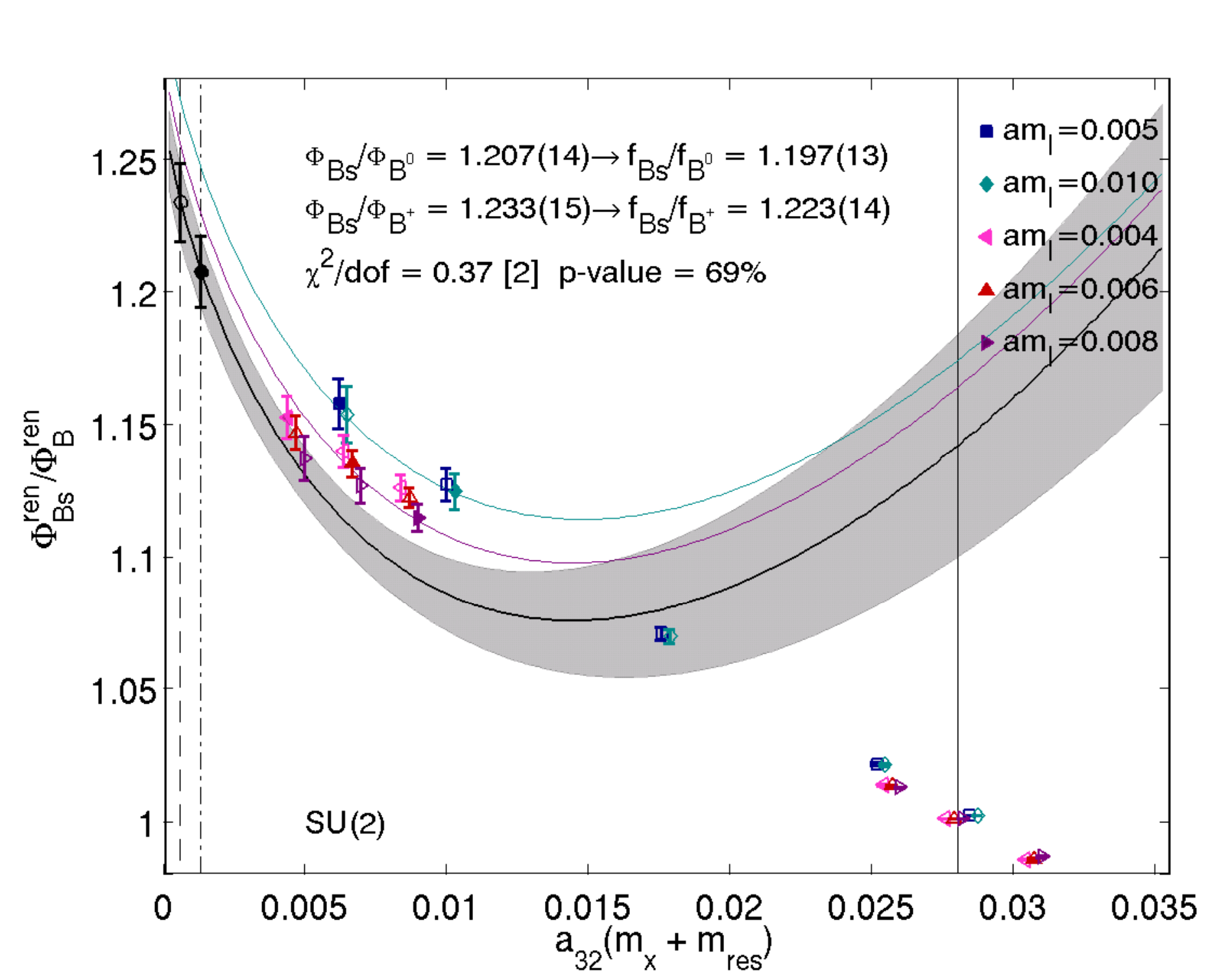}
\caption{Chiral and continuum extrapolation of $\Phi_{B_q}$ (left) and  $\Phi_{B_s}/\Phi_{B_q}$ (right) from a correlated fit using NLO SU(2) HM$\chi$PT.  The different colors/symbols distinguish our data points on the five different ensembles.  For better visibility data points on the $a m_l=0.004,\,0.008,\,0.01$ ensembles are plotted with a small horizontal offset.   {We plot all partially-quenched data, but the fit only includes the five unitary points (filled).}   The colored fit curves show the extrapolation in {light-quark mass}:  the fit function is evaluated {in full QCD with $m_x= m_l$ at the} nonzero lattice spacings  on the different ensembles, such that the curves should approximately go through the {filled} data points of {similar} color.  The chiral extrapolation in full QCD and the continuum is shown by the black line with grey error band.  The physical values of $\Phi_{B^+}$ ($\Phi_{B^0}$) and $\Phi_{B_s}/\Phi_{B^+}$ ($\Phi_{B_s}/\Phi_{B^0}$) correspond to the intersection of this curve with the dashed (dot-dashed) vertical line on the left-hand side indicating the physical $u$-quark ($d$-quark) mass.  The right-hand solid, vertical line indicates the $s$-quark mass.  Only statistical errors are shown.}
\label{Fig:FitfB+FitRatio}
\end{figure*}

{Our} preferred fits are shown in Fig.~\ref{Fig:FitfB+FitRatio}. The decay amplitudes of the neutral and charged $B$ mesons at the physical light-quark masses and in the continuum are obtained by setting the lattice spacing to zero and the {light-quark mass} to $m_d$ and $m_u$, respectively, in {the chiral-continum fit function}.\footnote{{Technically, the light sea-quark mass should be fixed to $(m_u + m_d)/2$, but we cannot change the sea- and valence-quark masses independently in our unitary chiral-continuum extrapolation.  We expect the sea-quark mass dependence of the decay constants, however, to be much smaller than the valence-quark mass dependence from the partially-quenched HM$\chi$PT expressions in Eqs.~(\ref{FitFuncSU2})--(\ref{FitRatioSU2}).  Further, we do not observe any statistically-significant sea-quark mass dependence in our data.  Thus fixing both the valence- and sea-quark masses to either $m_u$ or $m_d$ provides a good approximation.}}  Our results for the decay constants are {$f_{B^0} = 199.5(6.2)$~MeV and $f_{B^+} = 195.6(6.4)$ MeV, and for the ratios are $f_{B_s}/f_{B^0} = 1.197(13)$ and $f_{B_s}/f_{B^+} = 1.223(14)$}, where all errors are statistical only.  {We find about a 1.5\% difference between $f_{B^0}$ and $f_{B^+}$, which is consistent with the recent four-flavor lattice calculation of this splitting by HPQCD using NRQCD $b$ quarks~\cite{Dowdall:2013tga}.} 

{In Fig.~\ref{Fig:FitfB+FitRatio}, the colored fit curves show the extrapolation in light-quark mass at fixed lattice spacing.  They go approximately through the unitary data points included in the fit, and curve downward for $\Phi_B$ (upward for $\Phi_{B_s}/\Phi_B$) as they approach chiral limit due to the chiral logarithms in the SU(2) HM$\chi$PT fit functions, Eqs.~(\ref{FitFuncSU2})--(\ref{FitRatioSU2}).}  
Because the coefficients of the NLO chiral logarithms are fixed in terms of $g_b$ and $f_\pi$, the fit yields large chiral logarithms at pion masses below $\sim 200$~MeV despite the fact that our data is too heavy for us to observe their onset.  {Towards the right-hand sides of the plots in Fig.~~\ref{Fig:FitfB+FitRatio}, the valence quarks become appreciably heavier than the light sea quarks ($m_x \gg m_l$) and the full-QCD fit curves deviate substantially from the partially-quenched data points.}    Our decay-constant data displays no significant dependence on either the light sea-quark mass or the lattice spacing.  We find that the coefficients of the $a^2$ terms are $\sim 0.01$ or smaller for both fits in Fig.~\ref{Fig:FitfB+FitRatio}, and that $c_a$ is in fact statistically consistent with zero in the left-hand fit.  {In alternate fits that include some partially-quenched data,} the coefficients of the sea-quark mass terms are closer to ${\mathcal O}(1)$, but with $\sim 50$\% or larger errors.  

We also tried fits using SU(3) HM$\chi$PT, in which the strange-quark mass is explicit in the fit functions.  This in principle has the advantage of building in the constraint that $\Phi_{B_s}/\Phi_{B_q} = 1$ in the SU(3) limit $m_q = m_s$.  {Although we can obtain reasonable NLO SU(3) fits of $\Phi_B$ when $M_\pi^\text{sea} \ltapprox 425$~MeV and $M_\pi^\text{val} \ltapprox 400$~MeV, we were unable to obtain any acceptable NLO SU(3) fit of the ratio $\Phi_{B_s}^\text{ren}/\Phi_{B_q}^\text{ren}$ because the data are so precise.}   This is consistent with observations by the RBC and UKQCD collaborations in earlier analyses of light pseudoscalar meson masses and decay constants~\cite{Allton:2008pn}, in which they concluded that NLO SU(3) $\chi$PT was incompatible with their heaviest data.  {We can only successfully describe the data for $\Phi_{B_s}^\text{ren}/\Phi_{B_q}^\text{ren}$ with SU(3) HM$\chi$PT after adding several NNLO analytic terms and restricting the masses to $M_\pi^\text{sea} < 425$ MeV and $M_\pi^\text{val} < 400$~MeV.  Although we obtain results that are consistent with our preferred NLO SU(2) fits, the statistical errors from the NNLO SU(3) fits are larger because of the increased number of fit parameters, several of which are not well-constrained by the data.}

We use the alternative SU(3) fits {with good $p$-values}, as well as fits with only analytic dependence on the quark masses and lattice spacings, to estimate the systematic uncertainty due to the chiral-continuum extrapolation in Sec.~\ref{Sec:ChiPT}.

\subsection{\texorpdfstring{Decay constant $f_{B_s}$}{Decay constant fBs}}
\label{Sec:f_Bs}

\begin{figure}
\includegraphics[scale=0.45]{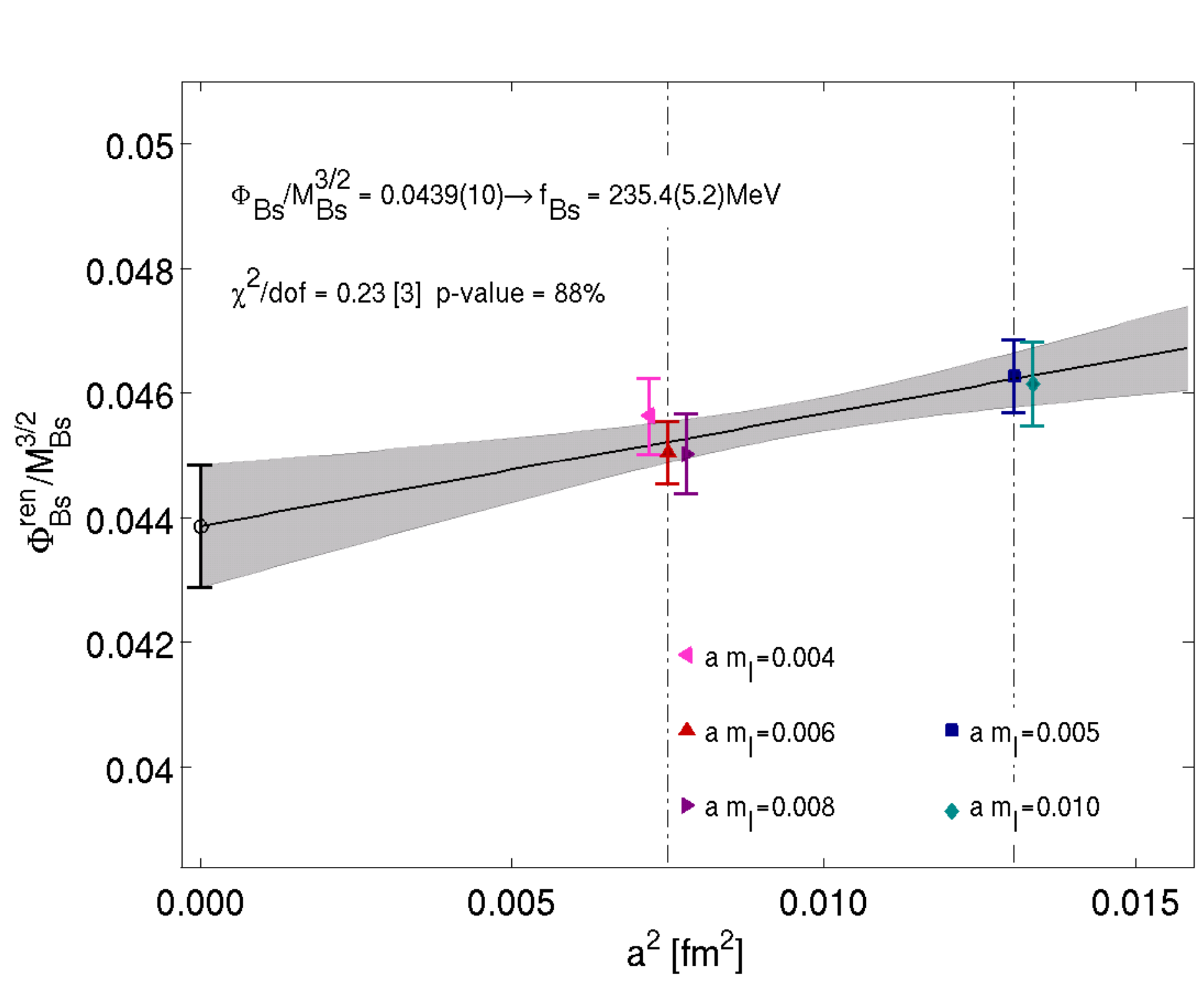}
\caption{Continuum extrapolation of $\Phi_{B_s}/M_{B_s}^{3/2}$ from a linear fit in $a^2$. We plot the five data points for $\Phi_{B_s}/M_{B_s}^{3/2}$ interpolated to the physical strange-quark mass on each ensemble using the same colors/symbols as in Fig.~\ref{Fig:FitfB+FitRatio}. The extrapolation is shown by the black line with gray error band. For better visibility data points on the $a m_l=0.004,\,0.008,\,0.01$ ensembles are plotted with a small horizontal offset. Only statistical errors are shown.}
\label{Fig:ExtfBs}
\end{figure}

After interpolating the decay amplitude $\Phi_{B_s}^\text{ren}$ to the physical strange-quark mass, we only need to extrapolate to the continuum. We do not observe any sea-quark mass dependence in our data, so we use a simple linear function in $a^2$, 
\begin{align}
\Phi_{B_s} = \varrho \, a^2 + \varphi,
\end{align}
which captures the leading scaling behavior from the light-quark and gluon actions.  Again, discretization errors from the heavy-quark action will be estimated via heavy-quark power-counting in Sec.~\ref{Sec:HeavyQuarkDiscErrors} and added to the systematic error budget.  We show the continuum extrapolation of $\Phi_{B_s}$ in Fig.~\ref{Fig:ExtfBs};  our result is $\Phi_{B_s} = 0.158(3)$, which corresponds to {$f_{B_s} = 235.4(5.2)$ MeV} (statistical errors only). 

\section{Estimation of systematic errors}
\label{Sec:SysErrors}

We now discuss the sources of systematic uncertainties in our determinations of the $B_{(s)}$-meson decay constants and their ratio.  Each uncertainty is discussed in a separate subsection and the total error budgets are provided in Table~\ref{Tab:SysErrors}.

\begin{table*}[t]
  \caption{Error budgets for the $B_{(s)}$-meson decay constants and their ratios.  Errors that were considered but found to be negligible are listed as ``0.0."  
Errors are given in \%.  The total error is obtained by adding the individual errors in quadrature.}
\label{Tab:SysErrors}
\begin{tabular}{l@{~}c@{~}c@{~~}c@{~~}c@{~~}c}
\hline\hline
&$f_{B^0}$(\%)&$f_{B^+}$(\%) &$f_{B_s}$(\%) &$f_{B_s}/f_{B^0}$(\%)& $f_{B_s}/f_{B^+}$(\%)\\\hline
statistics & {3.1} & {3.3} & 2.2 & {1.1} & {1.1} \\ 
chiral-continuum extrapolation&  {4.4}&{5.9} &3.1&{3.9} &{5.5}\\
lattice-scale uncertainty & 1.5 & 1.5 & 1.5 & 0.1 &0.1\\
light- and strange-quark mass uncertainty & 0.1 & {0.1} & 0.9 & {0.8} & {0.9} \\
RHQ parameter tuning & 1.2&1.2 &1.2& 0.1&0.1\\
HQ discretization errors & 1.7& 1.7& 1.7&0.3&0.3\\
LQ and gluon discretization errors & 1.1 & 1.1 & 1.2 & 0.6 & 0.6 \\
renormalization factor & 1.7& 1.7& 1.7& 0.0 & 0.0 \\
finite volume & 0.4 & {0.5} & 0.0 & {0.5}& {0.5} \\
isospin-breaking and EM & 0.7 & 0.7 & 0.7 & 0.1 & 0.7  \\\hline
total & {6.3} & {7.6} & 5.2 & {4.2} & {5.8} \\
\hline\hline
\end{tabular}
\end{table*}

\subsection{Chiral- and continuum extrapolation}
\label{Sec:ChiPT}

We estimate the systematic uncertainty due to the chiral- and continuum extrapolation of the decay constants by varying the chiral-continuum extrapolation fit {inputs and} Ans{\"a}tze.  {From the alternative fits tried, we take the largest difference of the central value to be the chiral-continuum extrapolation error.  We do not include fits with poor $p$-values in looking for the largest difference because such fits are not compatible with the data.}  

For $f_B$ and the ratio $f_{B_s}/f_B$, we {vary the inputs to our preferred NLO SU(2) HM$\chi$PT fits in the following ways}:
\begin{itemize}
	\item {excluding the heaviest unitary data point with $M_\pi \sim $~425~MeV,}
	\item {including some or all partially-quenched data with $M_\pi^\text{val,sea} \ltapprox $~425~MeV,}
	\item varying the value of $f_\pi$ in the coefficients of the chiral logarithms from $f_0$ in the chiral limit (112~MeV~\cite{Aoki:2010dy}) to $f_K = 155.5$~MeV~\cite{Beringer:2012zz},
\item varying the $B^*$-$B$-$\pi$ coupling in the coefficient of the chiral logarithms $g_b = 0.57(8)$ by plus or minus one standard deviation \cite{Flynn:2013kwa}.
\end{itemize}
{For the partially-quenched fits, we obtain good $p$-values for $f_B$ when $M_\pi^{\rm val}$ is below $\sim$~350~MeV, and for $f_{B_s}/f_B$ when both $M_\pi^{\rm sea}$ and $M_\pi^{\rm val}$ are below $\sim$~350~MeV.}  We also consider the following {alternate chiral-continuum extrapolation fit functions:}
\begin{itemize}
	\item analytic fits in which we omit the chiral logarithms in Eqs.~(\ref{FitFuncSU2}) and (\ref{FitRatioSU2}),
	\item {analytic fits in which we remove the chiral logarithms {\it and} also remove either the terms linear in the sea-quark mass $m_l$, squared lattice-spacing $a^2$, or both},
	\item NLO SU(3) HM$\chi$PT,
	\item SU(2) or  SU(3) {HM$\chi$PT} with NNLO analytic terms.
\end{itemize}
{For these fits we include partially-quenched data with $M_\pi^\text{val,sea} \ltapprox $~425~MeV.  The additional constraints are needed for the NNLO fits because  they have more free parameters.}  

\begin{figure}
\includegraphics[scale=0.45]{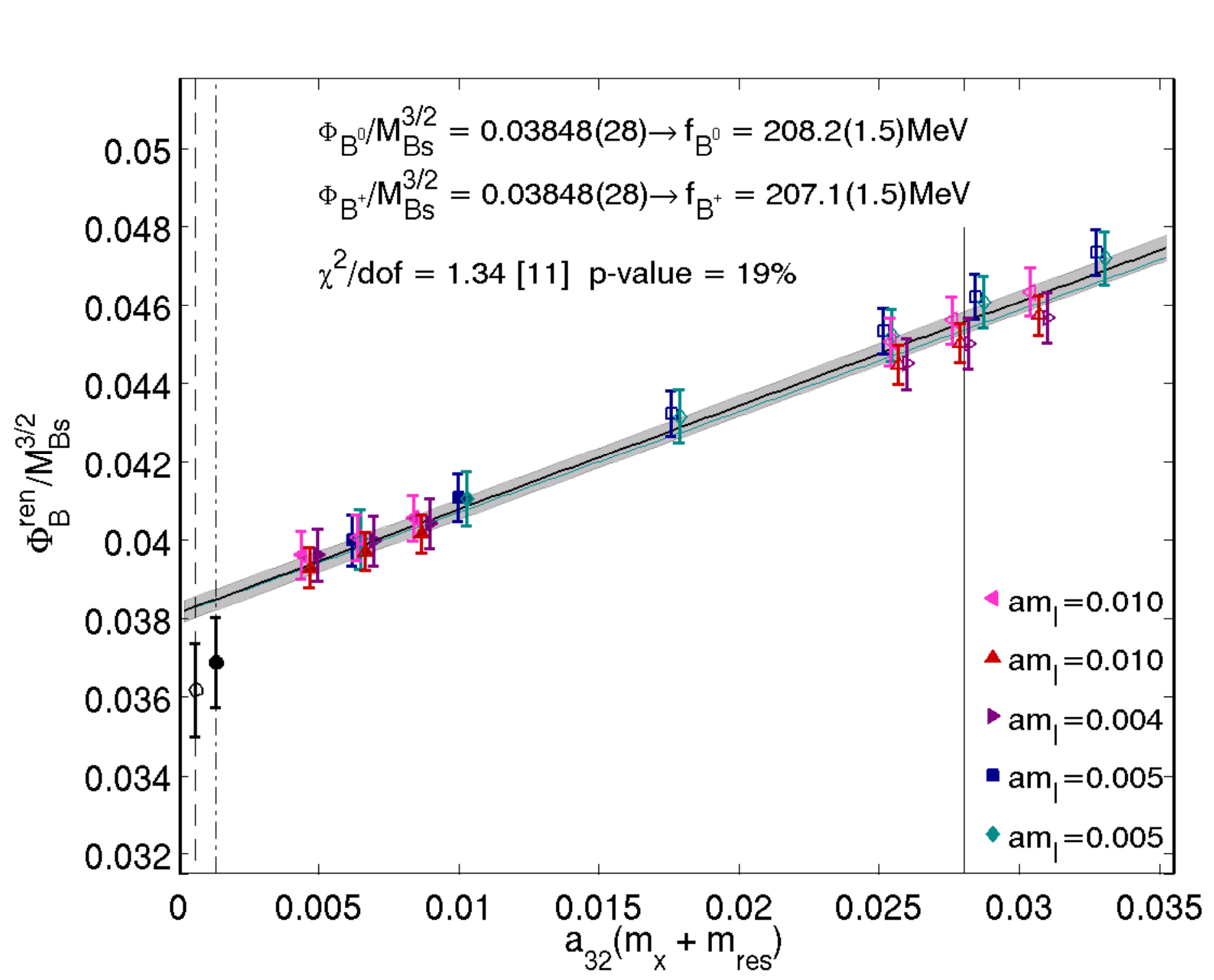}\\
\includegraphics[scale=0.45]{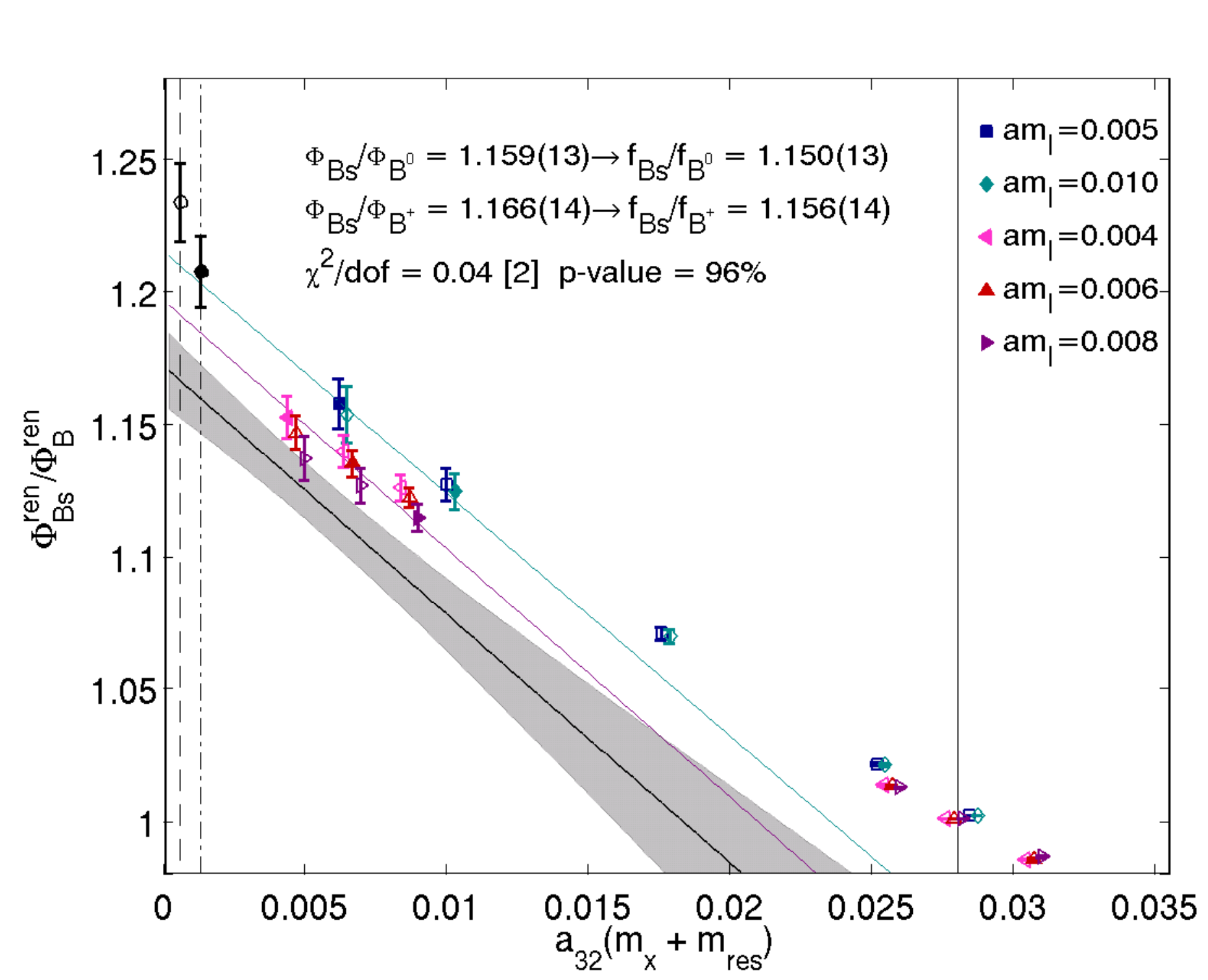}
\caption{Alternate chiral-continuum extrapolations for $\Phi_{B}$ {(upper)} and $\Phi_{B_s}/\Phi_B$ {(lower) used to obtain our chiral-continuum extrapolation error estimate.  Both fits use}  analytic fit {Ans{\"a}tze.  The $\Phi_{B}$ fit includes all partially-quenched data with $M_\pi^\text{val,sea} \ltapprox $~425~MeV and a single analytic term linear in the valence-quark mass $m_x$.  The $\Phi_{B_s}/\Phi_B$ fit uses only the unitary data with $M_\pi \ltapprox $~425~MeV and includes terms proportional to both $m_x$ and $a^2$.}   Colors/symbols are the same as in Fig.~\ref{Fig:FitfB+FitRatio}, and for better visibility data points on the $a m_l=0.004,\,0.008,\,0.01$ ensembles are plotted with a small horizontal offset. Only statistical errors are shown.  For comparison, the results for $f_{B^+}$ ($f_{B^0}$) from our preferred {unitary NLO SU(2) HM$\chi$PT} fits are shown as open (filled) black circles.}
\label{Fig:SysfB}
\end{figure}

For both, $f_B$ and $f_{B_s}/f_B$, {we find that analytic fits produce the largest changes with respect to the preferred NLO SU(2) HM$\chi$PT fit results.  Specifically, for $f_B$ the largest difference is obtained from a fit to all partially-quenched data with $M_\pi^\text{val,sea} \ltapprox $~425~MeV and a single analytic term proportional to the light valence-quark mass $m_x$.  For $f_{B_s}/f_B$, a fit to only the unitary data with  $M_\pi \ltapprox $~425~MeV and terms proportional to both the light-quark mass $m_x$ and squared lattice spacing $a^2$}  leads to the largest shift in the central value{. We}  show these fits in Fig.~\ref{Fig:SysfB}.  {Our data shows no evidence of curvature and the analytic fits have excellent $p$-values.  Once our pion masses are sufficiently light, however, HM$\chi$PT predicts the onset of chiral logarithms that will reduce the value of $f_{B}$ (and increase $f_{B_s}/f_B$) relative to the results of the analytic fits.  Thus we expect the true value of $f_B$ to be lower than that from the analytic fit (and of $f_{B_s}/f_B$ to be higher).  Nevertheless, for both quantities we conservatively take the full difference between the SU(2) HM$\chi$PT and analytic fits as the systematic error due to the chiral extrapolation, so our error estimates cover the results for $f_B$ and $f_{B_s}/f_B$ from the analytic fits.}

For $f_{B_s}$, the preferred continuum extrapolation is from a fit linear in $a^2$.  Our ability to perform alternate fits, however, is limited by the fact that we only have two values of the lattice spacing and therefore do not have sufficient data to add quadratic or even higher-order terms. Because the lattice-spacing dependence of our data is quite mild (the results on of all our five ensembles are statistically consistent),  we take as an alternative the weighted average of the finer $32^3$ data points.

\subsection{Lattice-scale uncertainty}
\label{Sec:Scale}

We exploit the procedure that we use to tune the parameters of the RHQ action to minimize the uncertainty due to the input lattice scale in our final results for the decay constants.  By construction, at the tuned point the $B_s$-meson mass is fixed to the experimentally-measured value.  We therefore choose to perform our decay-constant analysis in terms of dimensionless ratios over $M_{B_s}$.  We can then obtain the decay constants in GeV by multiplying the ratio by $M_{B_s} = 5.366$~GeV from the PDG~\cite{Nakamura:2010zzi}.   Hence our decay-constant results have no explicit dependence on the lattice-scale; we do, however, still need to consider the implicit dependence on the lattice spacing through the RHQ parameters.

\begin{figure*}[p]
\includegraphics[scale=0.45]{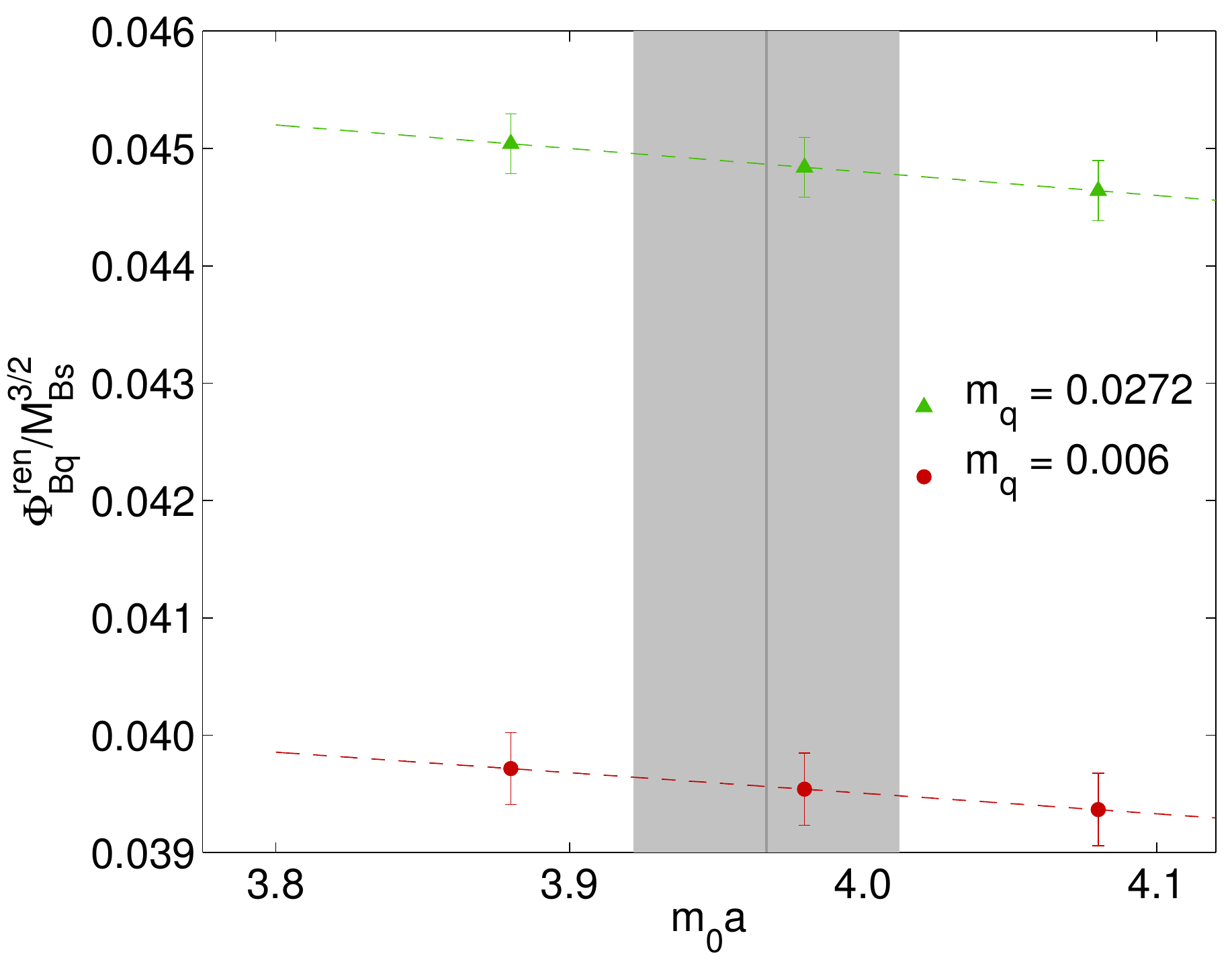}
\includegraphics[scale=0.45]{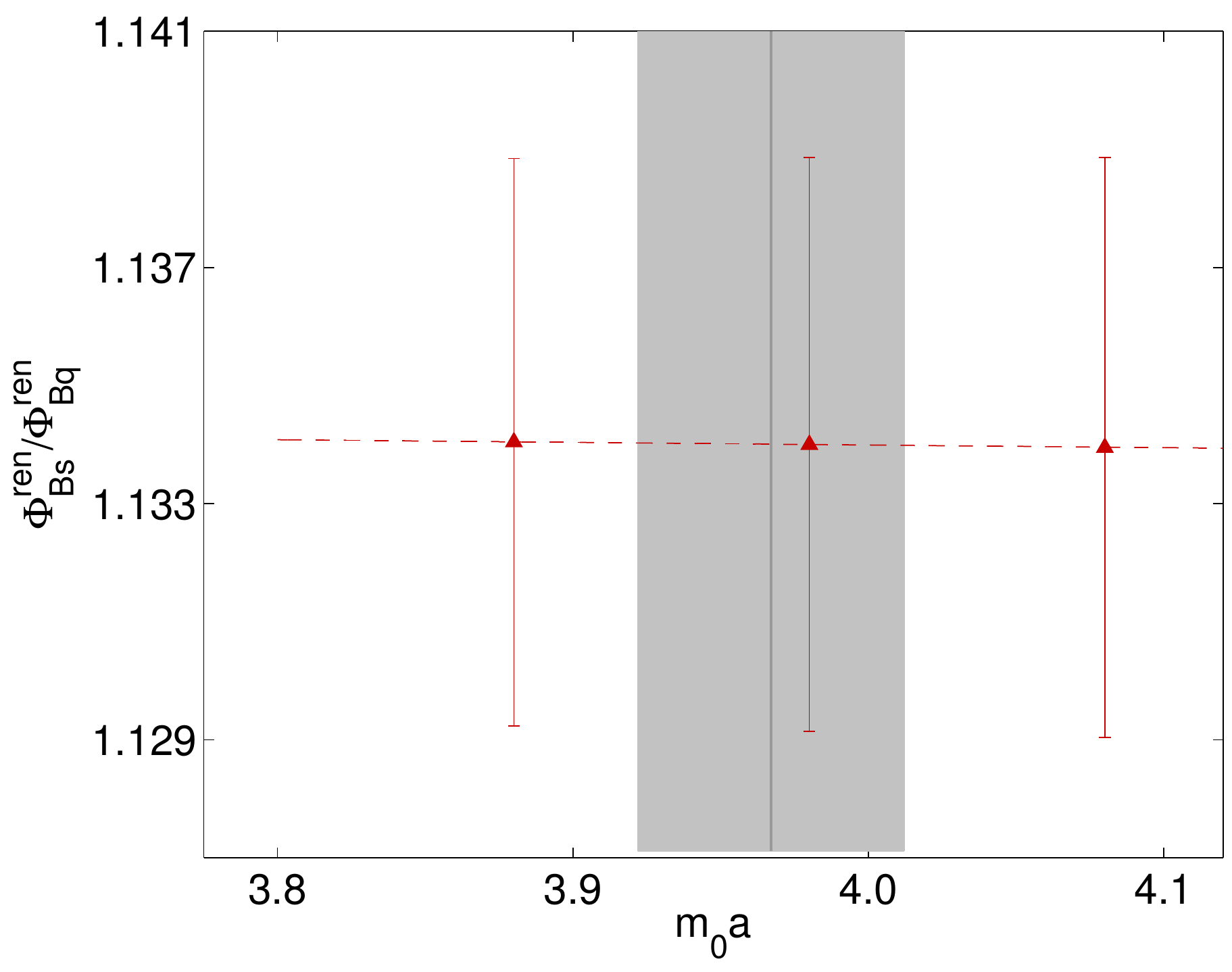}
\includegraphics[scale=0.45]{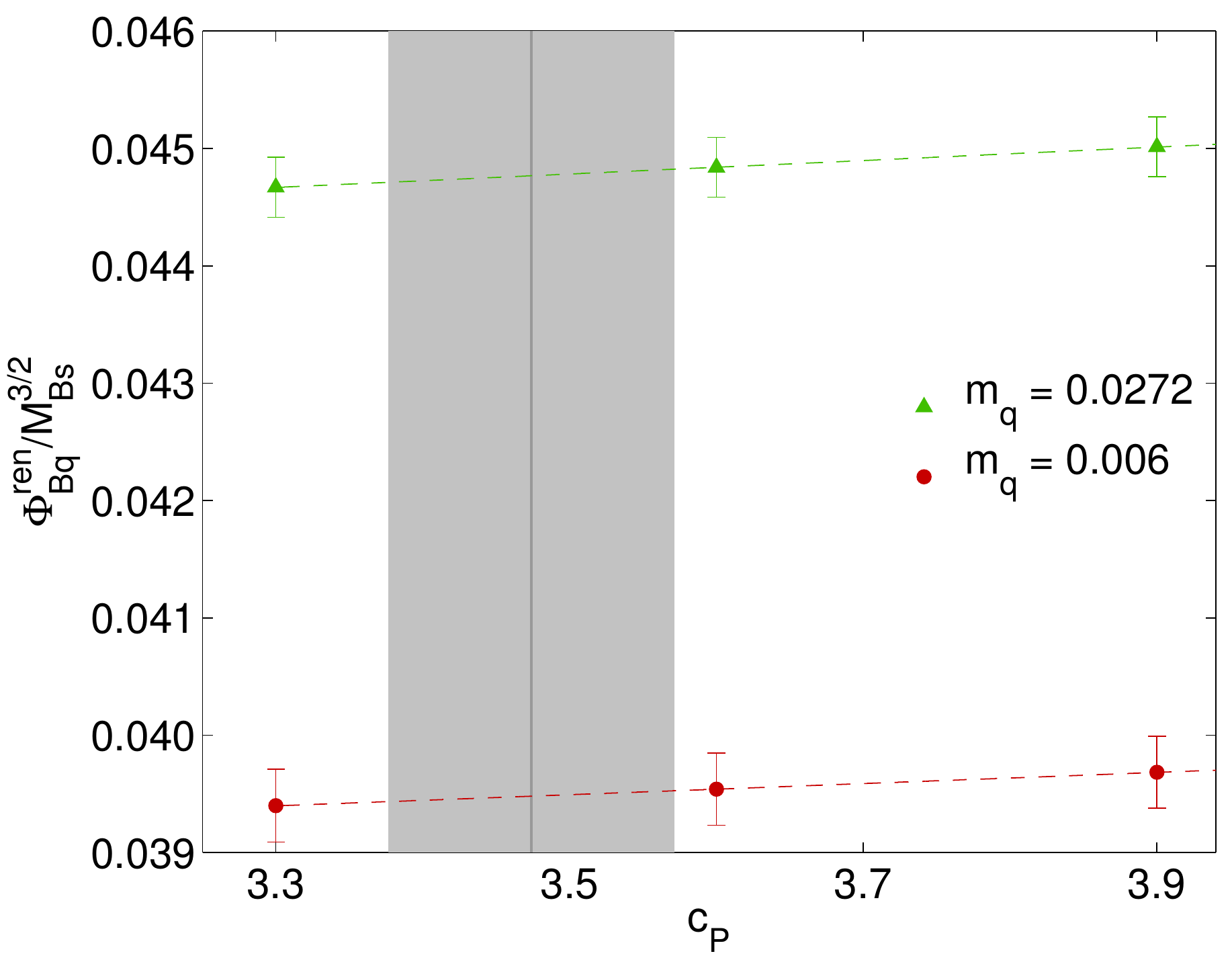}
\includegraphics[scale=0.45]{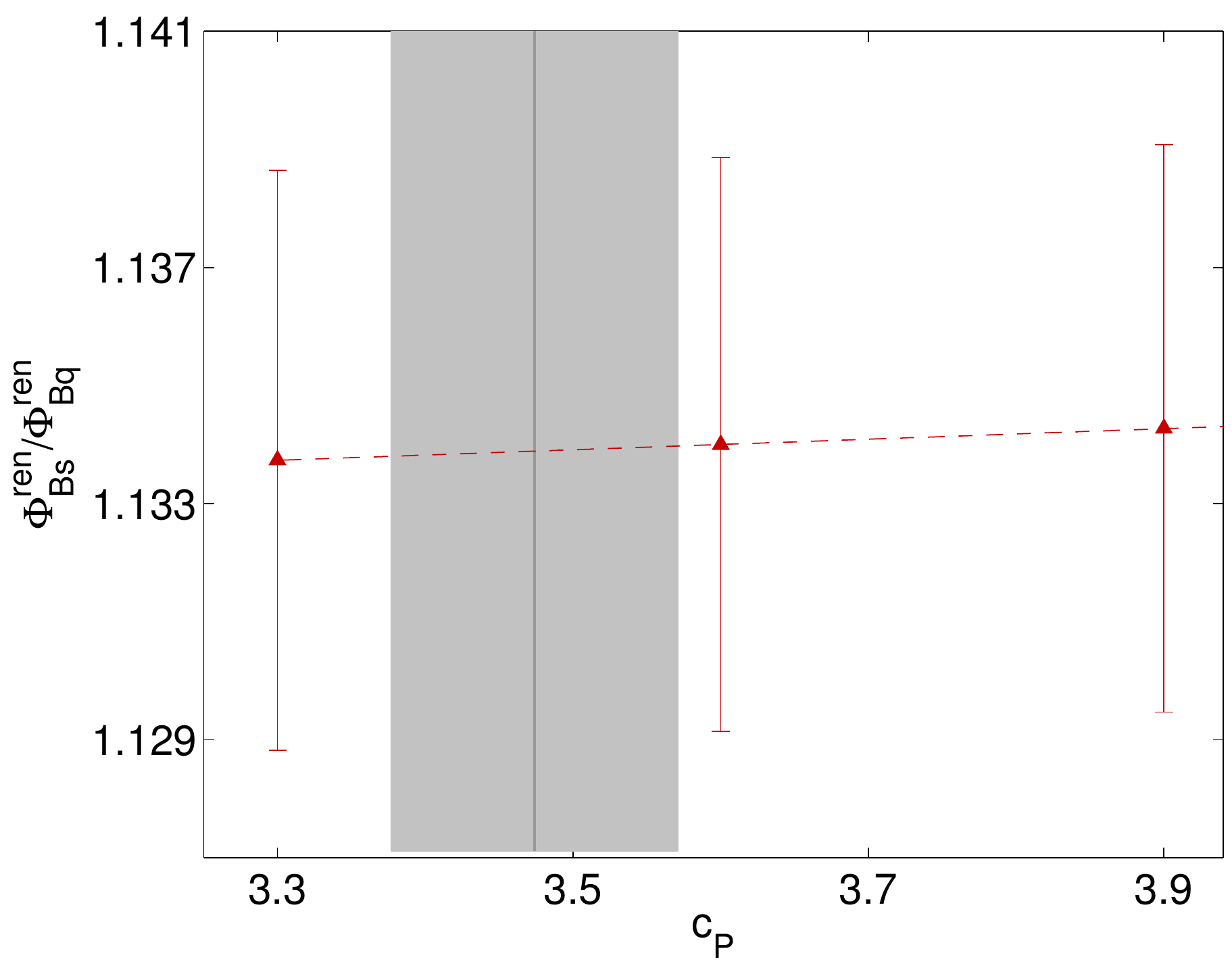}
\includegraphics[scale=0.45]{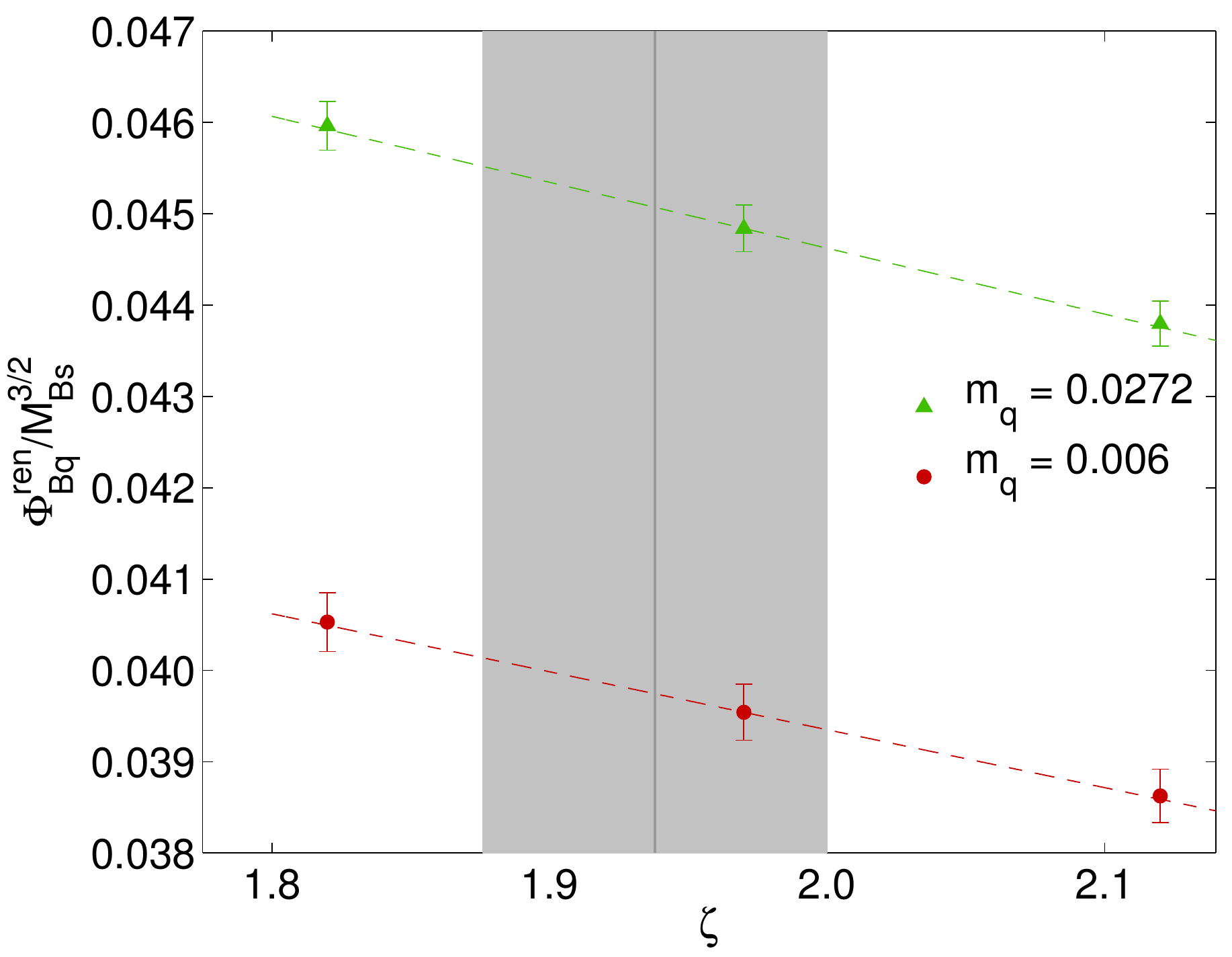}
\includegraphics[scale=0.45]{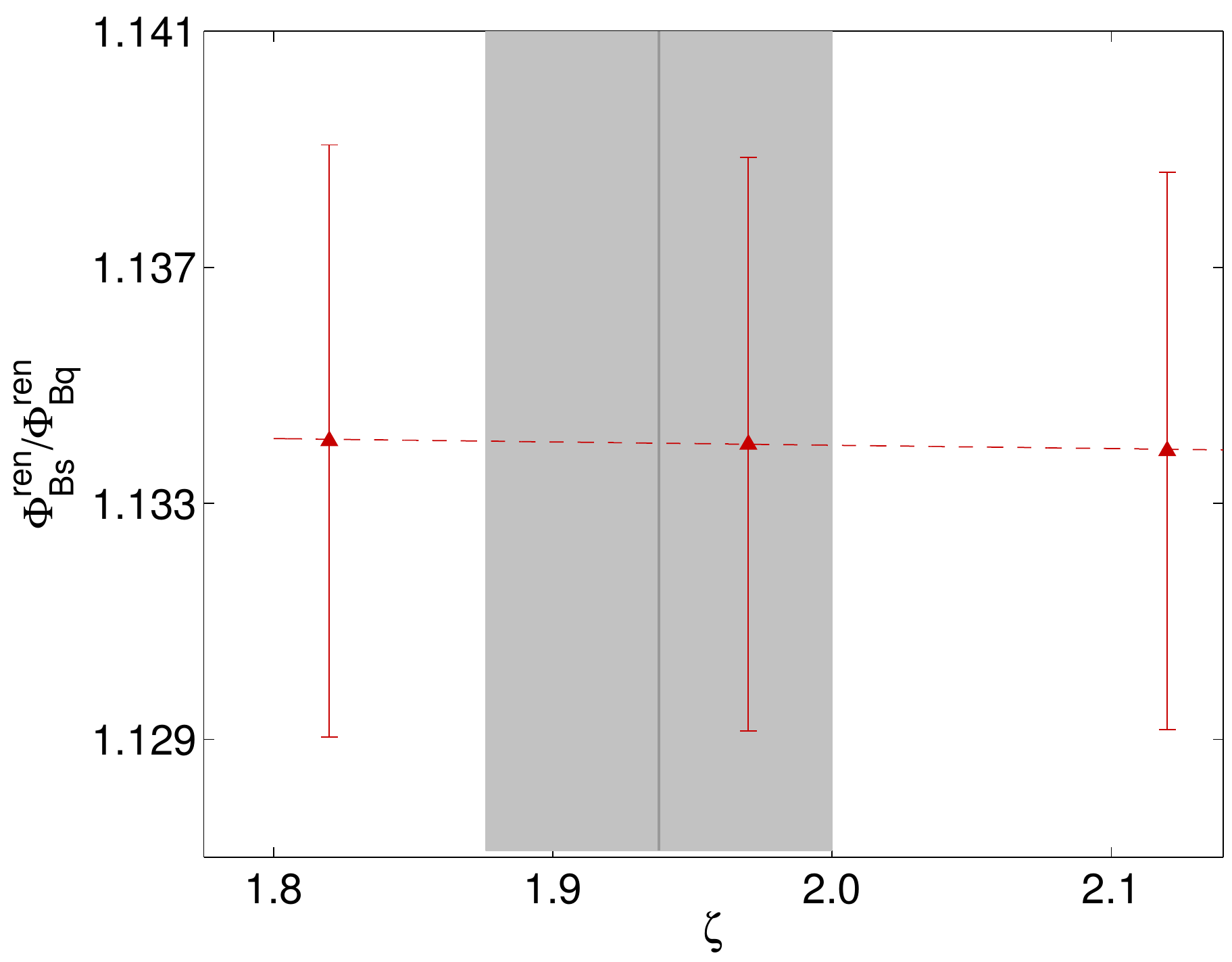}
\caption{Dependence of the decay amplitudes $\Phi_{B_q}/M_{Bs}^{3/2}$ and the ratio $\Phi_{B_s}/\Phi_{B_q}$ on our RHQ parameters. The plots on the left show the decay amplitude for $am_q = 0.006$ (red triangles) and $am_q = 0.0272$ (green circles) vs.~the RHQ parameters $m_0a$, $c_P$, and $\zeta$ (from top to bottom). The plots on the right show dependence of the ratio $\Phi_{Bs}/\Phi_B$ obtained for $am_s=0.0272$ and $am_q =0.006$ on $m_0a$, $c_P$ and $\zeta$. Only statistical errors are shown.}
\label{Fig:RHQdependence}
\end{figure*}

We estimate this source of scale uncertainty by measuring the slope of the decay constants and ratios with respect to the RHQ parameters \{$m_0a$, $c_P$,  $\zeta$\}. These are shown for the $32^3$, $am_l = 0.006$ ensemble in Fig.~\ref{Fig:RHQdependence}.  We then multiply each of these slopes by the uncertainty in the corresponding RHQ parameter due to the lattice scale as provided in Tab.~\ref{tab:RHQParamErr}, {\it e.g.} $\Delta( \Phi) / \Delta(c_P) \times \sigma(c_P)_{a}$.  Finally, for each of our data points on the $24^3$ and $32^3$ ensembles, we add the three contributions for each data point in quadrature.   For each physical quantity, we take the largest estimated total as the uncertainty due to the lattice scale, which gives 1.5\% for the decay constants and 0.1\% for the ratios.

\subsection{Light- and strange-quark mass uncertainties}
\label{Sec:Quark}

Here we estimate the uncertainties due to the input quark masses in the chiral-continuum extrapolations, as well as the mis-tuning of the strange sea quark.  We discuss each source of uncertainty in a separate subsection for clarity.  Because most of the individual uncertainty estimates turn out to be small relative to other errors, we add the numbers from the {three} subsections in quadrature and quote a single error due to to the light- and strange-quark mass uncertainties in Table~\ref{Tab:SysErrors}.

\subsubsection{\texorpdfstring{{Light $u$- and $d$-quark mass uncertainties}}{Light u- and d-quark masses}}

In the chiral-continuum extrapolations of $\Phi_{B_x}$ and $\Phi_{B_s}/\Phi_{B_x}$, we set the {light-quark mass} to the physical $d$-quark mass to obtain the neutral-meson decay constant $f_{B^0}$ and the corresponding ratio $f_{B_s}/f_{B^0}$, and to the physical $u$-quark mass to obtain $f_{B^+}$ and $f_{B_s}/f_{B^+}$.  We use the preliminary values of the quark masses $a_{32}m_{d} = 1.327(13) \times 10^{-3}$ and $a_{32}m_{u}=6.06(24)\times10^{-4}$ from simulations by the RBC/UKQCD collaborations including both QCD and QED. We estimate the uncertainty due to the determination of the {light-quark} masses by repeating the chiral-continuum extrapolation with $m_{d}$($m_u$) shifted by plus and minus one sigma. We observe small changes in the central values between {0.0--0.1\%}.

For $f_{B_s}$, we study the dependence of our three (two) $32^3$ ($24^3$) data points on the light sea-quark mass.  Because we cannot resolve any sea-quark mass dependence within our statistical uncertainties, we take this error to be negligible{.} 

\subsubsection{\texorpdfstring{Valence $s$-quark mass uncertainty}{s quark mass}}

We estimate the errors in $f_{B_s}$ and $f_{B_s}/f_B$ due to the uncertainty in the valence strange quark mass by repeating the interpolation to $m_s$ described in Sec.~\ref{sec:InterpMs} and then using these new values as inputs to the chiral and continuum extrapolations.  We vary independently the values for $a_{32}m_{s} = 0.0280(7)$ and $a_{24}m_{s} = 0.0369(11)$ on the $32^3$ and $24^3$ ensembles again by one sigma~\cite{Aoki:2010dy}. We find shifts in the central values due to varying $m_s$ of {0.8--0.9\%} for $f_{B_s}${, $f_{B_s}/f_{B^0}$, and $f_{B_s}/f_{B^+}$.}

\subsubsection{Strange sea-quark mistuning}
\label{Sec:strangeSeaQuark}

Our ensembles were generated with the heavy sea-quark mass $m_h$ approximately 10\% heavier than that of the physical strange quark, and with only a single value of $m_h$ at each lattice spacing.  Thus we cannot directly study the strange sea-quark mass dependence with our data, and must use the light sea-quark mass dependence as a proxy.  Because, however, we use SU(2) HM$\chi$PT for the chiral-continuum extrapolations of $f_B$ and $f_{B_s}/f_B$, and a linear-in-$a^2$ continuum extrapolation for $f_{B_s},$  the fit functions do not have any explicit strange-quark dependence, so we cannot interpolate to the correct strange sea-quark mass {\it a posteriori}.

We therefore study the data for the decay constants at fixed valence-quark mass on the three $32^3$ ensembles, and on the two $24^3$ ensembles.  In the case of the decay amplitudes $\Phi_B$ and $\Phi_{B_s}$, the statistical uncertainties are too large to resolve any sea-quark mass dependence; hence we quote for these quantities an error of 0.0\% in Tab.~\ref{Tab:SysErrors}.  The statistical errors in the ratio $\Phi_{B}/\Phi_{B_s}$ are sufficiently small that we can resolve the light sea-quark mass dependence.  We therefore perform a linear fit in $m_l$ to the three (two) data points to obtain the slope with respect to $m_l$.  Because the leading sea-quark mass dependence enters as $(2m_l + m_h)$ in SU(3) $\chi$PT, we expect the slope with respect to $m_h$ to be roughly half this value.  Using this slope to correct the central value of $f_{B_s}/f_B$ leads to a change of 0.5\%.  We take this entire shift to be the error in $f_{B_s}/f_B$ due to strange sea-quark mass mistuning, but it is clearly a conservative upper bound.

\subsection{RHQ parameter uncertainty}

Although we tune the values of the RHQ parameters to correspond to the physical $b$-quark, the tuned parameter values have both statistical and systematic uncertainties.

As described in Sec.~\ref{sec:HQAct}, we use seven sets of RHQ parameters and then interpolate to the physical $b$-quark mass using the jackknife blocks from our tuning analysis. The advantage of this method is that the statistical uncertainties in the RHQ parameters are automatically included in the statistical errors of the decay amplitudes.

In Ref.~\cite{Aoki:2012xaa} we presented a systematic error budget for each of the parameters $m_0a$, $c_P$, and $\zeta$; the results are shown in Table~\ref{tab:RHQParamErr}.  We already estimated the uncertainty in the decay constants due to the scale uncertainty in the RHQ parameters in Sec.~\ref{Sec:Scale}.  Hence we need only consider the errors in the RHQ parameters due to the other two sources:  heavy-quark discretization errors and experimental inputs used in the tuning procedure.  As in the case of the lattice-scale uncertainty, we use the slopes of the decay constants with respect to $\{m_0a, c_P, \zeta\}$.  For each of the three RHQ parameters  $\{m_0a, c_P, \zeta \}$ and two sources of uncertainty \{HQ, exp.\}, we estimate the error in the decay amplitudes as, {\it e.g.}, $\Delta (\Phi_B) / \Delta(m_0a) \times \sigma_{\rm HQ}$.  We then add these six individual contributions in quadrature.  We do this for all five sea-quark ensembles, and take the largest total to be the error in the decay constants due to the systematic errors associated with the RHQ tuning procedure.  For the decay constants $f_B$ and $f_{B_s}$, we obtain about 1.2\%, and for the ratio $f_{B_s}/f_B$  about 0.1\%. 

\subsection{Heavy-quark discretization errors}
\label{Sec:HeavyQuarkDiscErrors}

The RHQ action gives rise to nontrivial lattice-spacing dependence in the decay constants in the region $m_0 a \sim 1$.  Thus, instead of including additional functions of $m_0 a$ in the combined chiral-continuum extrapolation, we estimate the size of heavy-quark discretization errors using power-counting.  We follow the approach developed by El~Khadra, Kronfeld, and Mackenzie~\cite{ElKhadra:1996mp} for lattice calculations using the anisotropic Clover action for heavy quarks, and later extended to include dimension~6 and~7 operators in Oktay and Kronfeld~\cite{Oktay:2008ex}.  

Heavy-quark discretization errors in our decay-constant calculation arise from two sources: operators in the heavy-quark action and in the heavy-light axial-vector current.  We tune the parameters of the dimension-5 RHQ action nonperturbatively; therefore the leading discretization errors from the action are of ${\mathcal O}(a^2)$.  We use an $\CO(a)$-improved current operator with the improvement coefficient computed at 1-loop in $\alpha_s$; therefore the leading discretization errors from the current are of ${\mathcal O}(\alpha_s^2a)$ and ${\mathcal O}(a^2)$.\footnote{This general approach for error estimation is also used in similar calculations of heavy-light meson decay constants and form factors using the Fermilab action for $b$-quarks by the Fermilab Lattice and MILC collaborations~\cite{Bailey:2008wp,Bazavov:2011aa}.  The primary differences stem the fact that (1) we remove ${\mathcal O}(a)$ errors from the action to all orders in $\alpha_s$ by tuning the parameters of the anisotropic Clover action nonperturbatively, and (2) we calculate the ${\mathcal O}(a)$-improvement coefficient of the heavy-light axial-vector current to one higher order in $\alpha_s$.  Thus the heavy-quark discretization errors in our calculation are smaller than those in the Fermilab/MILC work.}

To obtain the numerical error estimates we first consider a  nonrelativistic description of the heavy-quark action.  Both the lattice and continuum theories can be described by effective Lagrangians built from the same operators, and discretization errors arise due to mismatches between the short-distance coefficients of higher-dimension operators in the two theories.  For each operator $\CO_i $ in the heavy-quark effective Lagrangian or current, the associated discretization error is given by
\begin{eqnarray}
\textrm{error}_i = \left( \CC_i^\textrm{lat} - \CC_i^\textrm{cont} \right)  \langle \CO_i \rangle \,.
\end{eqnarray}
For heavy-light meson systems, the sizes of matrix elements can be estimated using Heavy-Quark Effective Theory (HQET) power-counting.  Continuum HQET is an expansion in the spatial momentum of the heavy quark, $\vec{p}/m_b$.  The $b$-quarks in $B$ hadrons typically carry a spatial momentum $\vec{p} \approx \Lambda_{\rm QCD}$, the scale of the strong interactions.  The lattice introduces an additional scale, $a$. The relative error contribution to a physical quantity such as the $B$-meson decay constant from an operator with dimension $d$ can then be estimated as $\langle \CO_i \rangle \sim (a \Lambda_{\rm QCD})^{d - 4 + n_\Gamma}$, where $n_\Gamma = 0\,(1)$ if the operator commutes (anticommutes) with $\gamma_4$.\footnote{Operators in the Symanzik effective Lagrangian that anticommute with $\gamma_4$ are suppressed because they connect large upper spinor components with small lower spinor components.  Hence they are promoted to operators of one dimension higher in the heavy-quark effective Lagrangian.}  

The details of our numerical error estimation are provided in Appendix~\ref{App:HQdiscErr}.   We compute the sizes of the mismatch coefficients using the tuned parameters of the RHQ action; their values on the $24^3$ and $32^3$ ensembles are given in Table~\ref{tab:Mismatch_Fcns}.  We take $\Lambda_{\rm QCD} = 500$~MeV as suggested by fits to moments of inclusive $B$-decays~\cite{Buchmuller:2005zv}.  We add the contributions from the individual operators in quadrature to obtain the total uncertainty.  Finally, we take the size of heavy-quark discretization errors in our calculation of the $B_{(s)}$-meson leptonic decay constants to be the estimate on our finer $a^{-1} = 2.281$~GeV lattices (see Table~\ref{tab:HQDiscErrs_fB}), which is 1.7\%.  For the ratios we estimate that discretization errors will be suppressed by the SU(3)-breaking factor $\left(m_s - m_d\right)/\Lambda_\text{QCD}$.  Using the quark-mass determinations from FLAG \cite{Aoki:2013ldr} we estimate the uncertainty in $f_{B_s}/f_B$ from heavy-quark discretization errors to be about 0.3\%. 

\subsection{Light-quark and gluon discretization errors}
\label{Sec:LightQuarkDiscErrors}

The dominant discretization errors from the light-quark and gluon sectors are of ${\mathcal O}\left(a \Lambda_{\rm QCD} \right)^2$ from the action, which we estimate to be $\sim 5\%$ on the finer $32^3$ ensembles.  We remove these errors in our chiral-continuum extrapolation by including an analytic term proportional to $a^2$ in the fit function.  We estimate the light-quark and gluon discretization errors in the heavy-light axial-vector current, which are subleading, with power-counting and add them in quadrature with the other uncertainties in the error budget.

The leading light-quark and gluon discretization errors in the current are of ${\mathcal O}\left( \alpha_s a\widetilde{m}_q,\, (a\widetilde{m}_q)^2,\, \alpha_s^2 a \Lambda_{\rm QCD} \right)$,  where $a\widetilde{m}_q$ denotes the bare lattice mass.  On the $32^3$ ensembles, the first term leads to an $\sim 0.6\%$ uncertainty in $f_{B_s}$ (using $a\widetilde{m}_s$) and uncertainties below $0.1\%$ in $f_{B^+}$ and $f_{B^0}$ (using $a\widetilde{m}_{ud}$).  The second term is significantly smaller, about $\sim 0.1\%$ in $f_{B_s}$ and negligible in $f_{B^+}$ and $f_{B^0}$.  The third term leads to uncertainties of $\sim 1.1\%$ in all three decay constants.  Adding these three contributions in quadrature, we estimate the total uncertainty from light-quark and gluon discretization errors in the heavy-light current to be about $1.2\%$ in $f_{B_s}$ and about $1.1\%$ in $f_{B^+}$ and $f_{B^0}$.  For the decay-constant ratios, we estimate the error from the larger quark-mass dependent term to be of ${\mathcal O}\left( \alpha_s (a\widetilde{m}_s - a\widetilde{m}_{ud}) \right) \sim 0.6\%$.  This contribution is not suppressed because the strange-quark mass is so much larger than the light up- and down-quark masses.  We estimate the error from the lattice-spacing dependent term to be of ${\mathcal O} \left( \alpha_s^2 a (m_s - m_{ud}) \right) \sim 0.2\%$, which is about five times smaller than in the individual decay constants.  Again, adding the contributions in quadrature, we estimate the total uncertainty from light-quark and gluon discretization errors in the heavy-light current to be about $0.6\%$ in all three decay-constant ratios.

\subsection{Renormalization factor}
\label{Sec:rho}

In our computation we divide the heavy-light current renormalization factor into three contributions $Z_\Phi = \rho_A^{bl}\sqrt{Z_V^{ll} Z_V^{bb}}$;  we consider the errors from each of these factors in turn.   For $\rho_A^{bl}$, we must estimate the uncertainty due to truncating the perturbative series in $\alpha_s$.  We conservatively take the full size of the 1-loop correction on the fine lattice, 1.7\%, as the estimate.  For $Z_V^{ll}$ we use the nonperturbative determination of the axial-current renormalization factor $Z_A$ in the chiral limit from Ref.~\cite{Aoki:2010dy}.  The statistical uncertainty in $Z_A$ on the finer ensemble is 0.02\%, which is negligible compared to our other sources of error.  The renormalization factors $Z_V$ and $Z_A$ for domain-wall fermions differ, however, by ${\mathcal O}(am_{\rm res})$, which is about $3 \times 10^{-3}$ on our finer ensembles.  Thus we take 0.3\% to be the systematic uncertainty in $Z_V^{ll}$ due to chiral symmetry breaking.  Further, because we use the values $Z_V^{ll}$ and $\rho_A^{bl}$ in the chiral limit, we must consider the errors due to the nonzero physical up, down, and strange-quark masses.  The leading quark-mass dependent errors in  $Z_V^{ll}$ are of ${\mathcal O}\left((a\widetilde{m}_q)^2 \right)$, and in $\rho_A^{bl}$ are of ${\mathcal O}\left( \alpha_s a\widetilde{m}_q \right)$.  These contributions are already included in the estimate of light-quark and gluon discretization errors in Sec.~\ref{Sec:LightQuarkDiscErrors} above, so we do not count them again here.   For $Z_V^{bb}$, we use the weighted average of the two (three) determinations on the $24^3$ ($32^3$) ensembles.  The statistical uncertainty in $Z_V^{bb}$ on the finer ensemble is again small, 0.15\%.  Adding the contributions from $\rho_A^{bl}$, $Z_V^{ll}$, and $Z_V^{bb}$ in quadrature, we estimate the total error in the decay constants from the renormalization factor to be 1.7\%.  Because we use the values $Z_V^{ll}$ and $\rho_A^{bl}$ in the chiral limit, the renormalization factor $Z_\Phi$ cancels exactly in our computation of the ratio $f_{B_s}/f_B$, so we quote an error of ``0.0\%" in the error budget.

\subsection{Finite volume errors}
\label{Sec:FVerr}

We estimate the error due to the finite spatial lattice volume using one-loop finite-volume SU(2) HM$\chi$PT.  In the finite-volume theory, the loop integrals become sums over lattice sites, such that the chiral logarithms in the fit functions in Eqs.~(\ref{FitFuncSU2}) and~(\ref{FitRatioSU2}) become bessel functions.  For $f_B$ and $f_{B_s}/f_B$ we repeat the combined chiral- and continuum extrapolation using the finite volume SU(2) HM$\chi$PT expressions.  {These lead to shifts in the central values of 0.4--0.5\% for $f_B$ and 0.5\% for $f_{B_s}/f_B$.}   For $f_{B_s}$ we do not perform a chiral extrapolation, but we can still calculate the size of the corrections to the data points using finite-volume HM$\chi$PT.  The use of SU(2) $\chi$PT at the strange-quark mass may in general be questionable, but we expect it to be good enough to obtain a rough estimate of the systematic error.  We find that the finite-volume corrections to $f_{B_s}$ are below $0.05\%$ on all of our ensembles, and hence quote an uncertainty of 0.0\% in the error budget. 

\subsection{Isospin breaking and electromagnetism}
\label{Sec:Isospin}

The decay constants of the charged and neutral $B$ mesons $f_{B^+}$ and $f_{B^0}$ differ due to both the masses and the charges of the constituent light $u$ and $d$ quarks.  The quark-mass contribution to this difference comes from the valence-quark masses, and the leading term is of ${\mathcal O}\left( \Delta m_{ud} / \Lambda_{\rm QCD}\right)$, where $\Delta m_{ud} \equiv (m_d - m_u)$.  We account for this effect by extrapolating the light {valence} quark to either the physical $u$- or $d$-quark mass in the chiral-continuum extrapolation, and find that this leads to a difference of {1.5}\% between $f_{B^+}$ and $f_{B^0}$.  This observed size is {somewhat larger than } the power-counting estimate of 0.5\% obtained using the determination of the quark masses from FLAG~\cite{Aoki:2013ldr} and $\Lambda_{\rm QCD} = 500$~MeV{, but is close to the 2\% difference observed by HPQCD in Ref.~\cite{Dowdall:2013tga}}.   The electromagnetic contribution to the difference between $f_{B^+}$ and $f_{B^0}$ is expected to be the typical size of 1-loop QED corrections, or ${\mathcal O}(\alpha_{QED}) \sim 0.7\%$.  Thus we estimate that the uncertainties in $f_{B^+}$ and $f_{B^0}$ due to isospin breaking and electromagnetism are $\sim 0.7\%$.  Because only the omission of electromagnetic effects contributes significantly to the error, this estimate also applies to $f_{B_s}$.  For the ratio of neutral-meson decay constants $f_{B_s} /f_{B^0}$, the electromagnetic contribution is suppressed due to the equal charges of the down and strange valence quarks.  We estimate its size to be of ${\mathcal O}\left( \alpha_{\rm EM} (m_s - m_d)/\Lambda_{\rm QCD} \right) \sim 0.1\%$.  This cancellation does not occur when the valence-quark charges are different, so electromagnetic effects in $f_{B_s} /f_{B^+}$ are still of ${\mathcal O}(\alpha_{QED}) \sim 0.7\%$.

We note that the difference between the $u$- and $d$-quark masses in the sea sector cannot lead to a difference between $f_{B^+}$ and $f_{B^0}$  because the sea quarks couple to the valence quark in the $B$ meson through $I=0$ gluon exchange.  The use of degenerate $u$ and $d$ sea quarks does lead to identical shifts in $f_{B^+}$, $f_{B^0}$, and $f_{B_s}$.  Such  contributions, however, are negligible because they must be symmetric under the interchange $m_u \leftrightarrow m_d$ and are of of ${\mathcal O}\left( (\Delta m_{ud} / \Lambda_{\rm QCD})^2 \right) \sim 0.003\%$.

\section{Results and conclusions}
\label{Sec:Conc}

After adding the systematic error estimates from Table~\ref{Tab:SysErrors} in quadrature, our final results for the $B_{(s)}$-meson decay constants and their ratios are: 
\begin{align}
	f_{B^0} & = {199.5(6.2)(12.6)} \;\text{MeV}\label{eq:fB0} \\
	f_{B^+} & = {195.6(6.4)(14.9)} \;\text{MeV}  \\
	f_{B_s} & =  235.4(5.2)(11.1) \;\text{MeV} \\
	f_{B_s}/f_{B^0} & = {1.197(13)(49)} \\
	f_{B_s}/f_{B^+} & = {1.223(14)(70)}  \label{eq:fBs/fB+} \,,	
\end{align}
where the errors are statistical and total systematic, respectively.

\begin{figure*}[t]
\includegraphics[width=\textwidth]{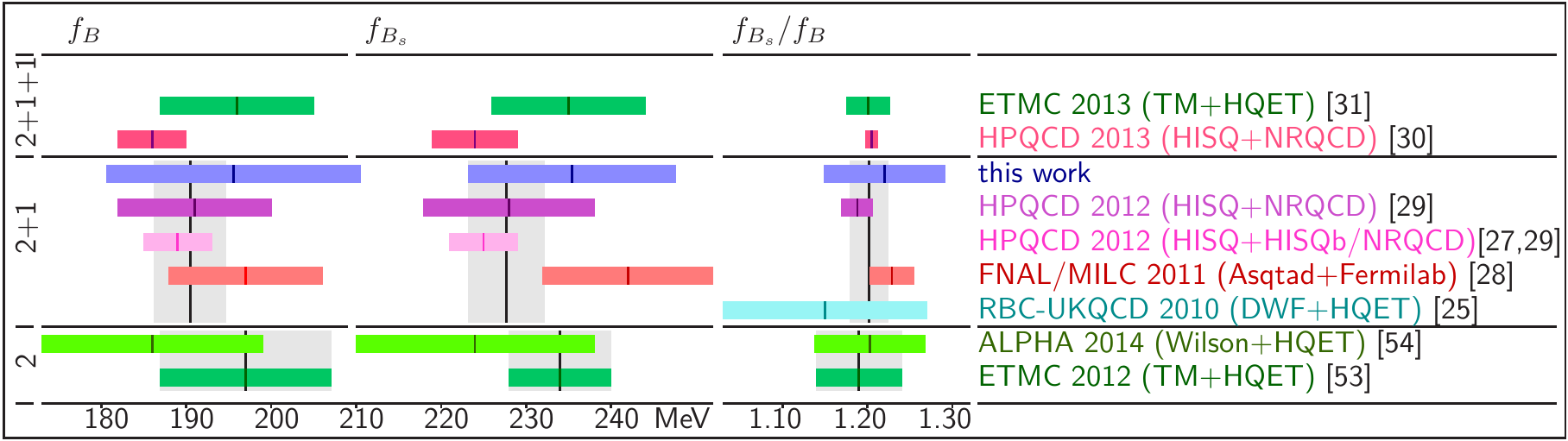}
\caption{Lattice determinations of $f_B$, $f_{B_s}$, and $f_{B_s}/f_B$ using 2, 2+1, and 2+1+1 dynamical sea-quarks~\cite{Albertus:2010nm,McNeile:2011ng,Bazavov:2011aa,Na:2012kp,Carrasco:2012de,Dowdall:2013tga,Carrasco:2013naa,Bernardoni:2014fva}. The gray error bands show the FLAG averages~\cite{Aoki:2013ldr}, which were obtained from the FNAL/MILC and HPQCD determinations for $N_f = 2+1$ and equal the ETMC result for $N_f=2$.  No FLAG average was presented for the 2+1+1-flavor data.  Most of the results shown are for the decay constants in the isospin limit $f_B$ and $f_{B_s}/f_B$ except for the determination by FNAL/MILC who reported results for $f_{B^+}$ and $f_{B_s}/f_{B^+}$; our results refer to the determination of $f_{B^+}$ and  $f_{B_s}/f_{B^+}$. The description in parentheses show the light- and heavy-quark actions used, with the exception of the $N_f=2+1$ HPQCD calculations. The HPQCD works use Asqtad sea quarks, and the determination of $f_B$ labeled ``HISQb/NRQCD'' is obtained by combining $f_{B_s}$ using HISQ $b$ quarks from Ref.~\cite{McNeile:2011ng} with the ratio $f_{B_s}/f_B$ using NRQCD $b$-quarks from \cite{Na:2012kp}.} \label{fig:fBCompare}
\end{figure*}

Figure~\ref{fig:fBCompare} compares our results with other unquenched determinations.  For all quantities, they agree well with the other $N_f > 2$ determinations in the literature.  Our result for $f_{B_s}/f_B$ is more precise than the published $N_f = 2+1$ RBC/UKQCD result~\cite{Albertus:2010nm} using static $b$-quarks because we include domain-wall ensembles with much lighter pions and a finer lattice spacing.  For both the decay constants and their ratio, our uncertainties are comparable to the results of ETM~\cite{Carrasco:2012de,Carrasco:2013naa} ALPHA~\cite{Bernardoni:2014fva}, as well as the similar calculation by the Fermilab Lattice and MILC collaborations~\cite{Bazavov:2011aa} using the Fermilab relativistic heavy-quark interpretation.  Our results are not as precise as those by HPQCD using HISQ $b$-quarks on the MILC asqtad staggered ensembles~\cite{McNeile:2011ng}, which include ensembles as fine as $a \approx 0.045$~fm, or using NRQCD $b$-quarks and HISQ sea quarks~\cite{Dowdall:2013tga}, which include ensembles with the physical pion mass.

The largest source of uncertainty in our calculations of the $B_{(s)}$-meson decay constants is from the chiral extrapolation to the physical light-quark masses and the extrapolation to the continuum limit.  We are currently generating light and strange-quark propagators on the RBC/UKQCD M{\"o}bius domain-wall + Iwasaki ensembles~\cite{Brower:2004xi,Brower:2012vk,MawhinneyLat13Talk} with the same lattice spacing as the $24^3$ ensembles used in our current analysis, but with $m_\pi \approx 140$~MeV.   The inclusion of data at the physical pion mass will significantly reduce our chiral-continuum extrapolation errors.  Statistical errors are the next-largest source of uncertainty in our current analysis.   Shortening the distance of the chiral extrapolation will reduce the statistical errors at the physical point.  We are also investigating the use of all-mode averaging~\cite{Blum:2012uh,Shintani:2014vja} to reduce the statistical errors on the individual numerical data points.  All of the other systematic uncertainties are estimated in Table~\ref{Tab:SysErrors} to be much smaller;  thus we anticipate obtaining significantly smaller errors  in the future.  There has been no difference observed between calculations of the $B$-meson decay constants from two-, three-, and four-flavor lattice simulations (see Fig.~\ref{fig:fBCompare}).  Nevertheless, the inclusion of the dynamical charm quark will be important once calculations reach even higher precision.  The RBC and UKQCD collaborations are currently generating $N_f = 2+1+1$ domain-wall ensembles which will allow a direct study of the effects of  charm-quark loops on the $B$- (and $D$-) meson decay constants.

Our results for the $B_{(s)}$-meson decay constants in Eqs.~(\ref{eq:fB0})--(\ref{eq:fBs/fB+}) are the first from simulations with domain-wall light quarks and relativistic heavy quarks, and also the first application of the RHQ action to weak-matrix elements relevant for phenomenology.  They provide valuable independent cross-checks of the published unquenched calculations using staggered sea quarks.  The good agreement with other works bolsters confidence in lattice-QCD calculations of the $B_{(s)}$-meson decay constants, and provides further support that the RHQ action can be used to obtain accurate results for bottom systems with competitive and reliable uncertainties.  We are also undertaking companion calculations of $B$-meson semileptonic form factors~\cite{Kawanai:2013qxa}, $B^0$-$\bar{B^0}$ mixing matrix elements~\cite{Witzel:2013sla}, and $B^*$-$B$-$\pi$ coupling~\cite{Flynn:2013kwa} using the same lattice actions and ensembles.  These will enable determinations of $|V_{ub}|$ from both leptonic and semileptonic decays and place an important constraint on the apex of the CKM unitarity triangle.  Finally, we note that rare decays such as $B \to K \ell^+ \ell^-$ and $B \to \pi \ell^+ \ell^-$ provide potentially sensitive probes of new physics, and are therefore future possible applications of the RHQ action.

\section*{Acknowledgments}

We thank the referee for invaluable comments on the manuscript that led to improvements in both the presentation and analysis.

Computations for this work were carried out in part on facilities of the USQCD Collaboration, which are funded by the Office of Science of the U.S. Department of Energy.  We thank BNL, Columbia University, Fermilab, RIKEN, and the U.S. DOE for providing the facilities essential for the completion of this work. 

This work was supported in part by the U.S. Department of Energy under grant No. DE-FG02-92ER40699 (N.H.C.), by the UK Science and Technology Facilities Council (STFC) grant ST/J000396/1 (J.M.F.),  and by the Grant-in-Aid of the Ministry of Education, Culture, Sports, Science and Technology, Japan (MEXT Grant) Nos.~26400261, 22540301, and 23105715 (T.I).   T.K. is supported by the JSPS Strategic Young Researcher Overseas Visits Program for Accelerating Brain Circulation (No. R2411).  O.W.~ackknowledges support at Boston University by the U.S.~DOE grant DE-SC0008814.  This manuscript has been authored by employees of Brookhaven Science Associates, LLC under Contract No. DE-AC02-98CH10886 with the U.S. Department of Energy. 

\appendix

\section{\texorpdfstring{Determination of $Z_V^{bb}$}{Determination of ZVbb}}
\label{App:Zvbb}

\begin{figure}[tb]
\includegraphics[scale=0.5]{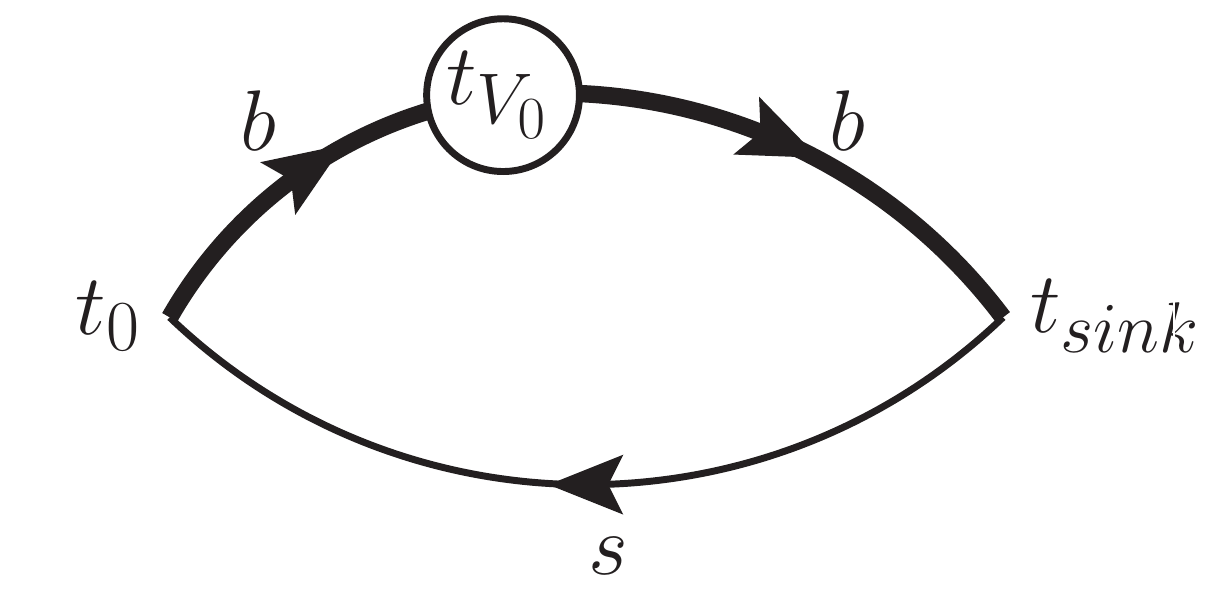}
\caption{Three-point correlation function used to compute the flavor-conserving renormalization factor $Z_V^{bb}$.  The locations of the $B_s$ mesons are fixed and the location of the vector current is varied over all time slices in between.}
\label{Fig:Zvbb}
\end{figure}

The flavor conserving heavy-heavy renormalization factor $Z_V^{bb}$ is obtained from the matrix element of the $b \to b$ vector current between two $B_q$-mesons:
\begin{align}
Z_V^{bb} \times \langle B_q | V^{bb,0} | B_q \rangle = 2 M_{B_q} .
\end{align}
We compute the three-point correlation function shown in Fig.~\ref{Fig:Zvbb} by fixing the locations of the two $B_q$ mesons at $t_0$ and $t_{\rm sink}$ and varying the location $t$ of the operator over all time slices in between:
\begin{align}
C_{PVP}(t_0, t, t_\text{sink}) = \sum_{\vec x, \vec y}\langle \widetilde{{\cal O}}_P(\vec x, t_\text{sink}) {\cal O}_{V_0}(\vec y, t)  \widetilde{{\cal O}}_P(\vec 0, t_0)\rangle \,,
\end{align}
where ${\widetilde{\cal O}}_P$ are pseudoscalar interpolating operators for the $B_q$ mesons and ${\cal O}_{V_0} = \bar{b} \gamma_0 b$ is the leading temporal vector-current operator.   The ${\mathcal O}(a)$-improvement of $Z_V^{bb}$ does not require the computation of any additional matrix elements because we are only interested in the temporal component of the vector-current operator without momentum injected.  In this situation, the equations of motion can be used to parameterize the ${\mathcal O}(a)$ improvement as an overall multiplicative factor that depends upon the $b$-quark mass.  We can then absorb this correction into the values of the perturbative contribution to the renormalization factor $\rho_{A}^{bl}$ given in Table~\ref{Tab:Zvbb}.

We use a point-source propagator for the light spectator quark, and Gaussian smeared sources for the $b$-quarks in the two $B_q$ mesons.
The desired renormalization factor is then given by the ratio
\begin{align}
Z_V^{bb}(t_0, t, t_\text{sink})  = \lim_{t_0\ll t \ll t_\text{sink}} \frac{\widetilde C_{PP}(t,t_0)}{C_{PVP}(t_0,t,t_\text{sink})} \,,
\label{Eq:Zvbb}
\end{align}
where $\widetilde C_{PP}$ is the pseudoscalar-pseudoscalar correlator with a Gaussian smeared heavy quark source and sink [see Eq.~(\ref{Eq:Ctilde_PP})].  For the computation of $Z_V^{bb}$ we double the statistics by computing the sequential $b$-quark propagator for both the forward- and the backward-propagating light spectator quark.  Similarly, we fold the 2-point correlation function about the temporal midpoint of the lattice.  After testing several source-sink separations $ \Delta_t \equiv \left(t_\text{sink} - t_0 \right) =  \{18, 20, 22\}$ on the $24^3$ ensemble with $a m_l=0.005$, we found that $\Delta_t = 20$ led to the best signal-to-noise.  We scaled this value by $a^{-1}_{32c} / a^{-1}_{24c}$ to obtain $\Delta_t = 26$ on the $32^3$ ensembles.

\begin{figure}[t]
\centering
\includegraphics[width=\columnwidth]{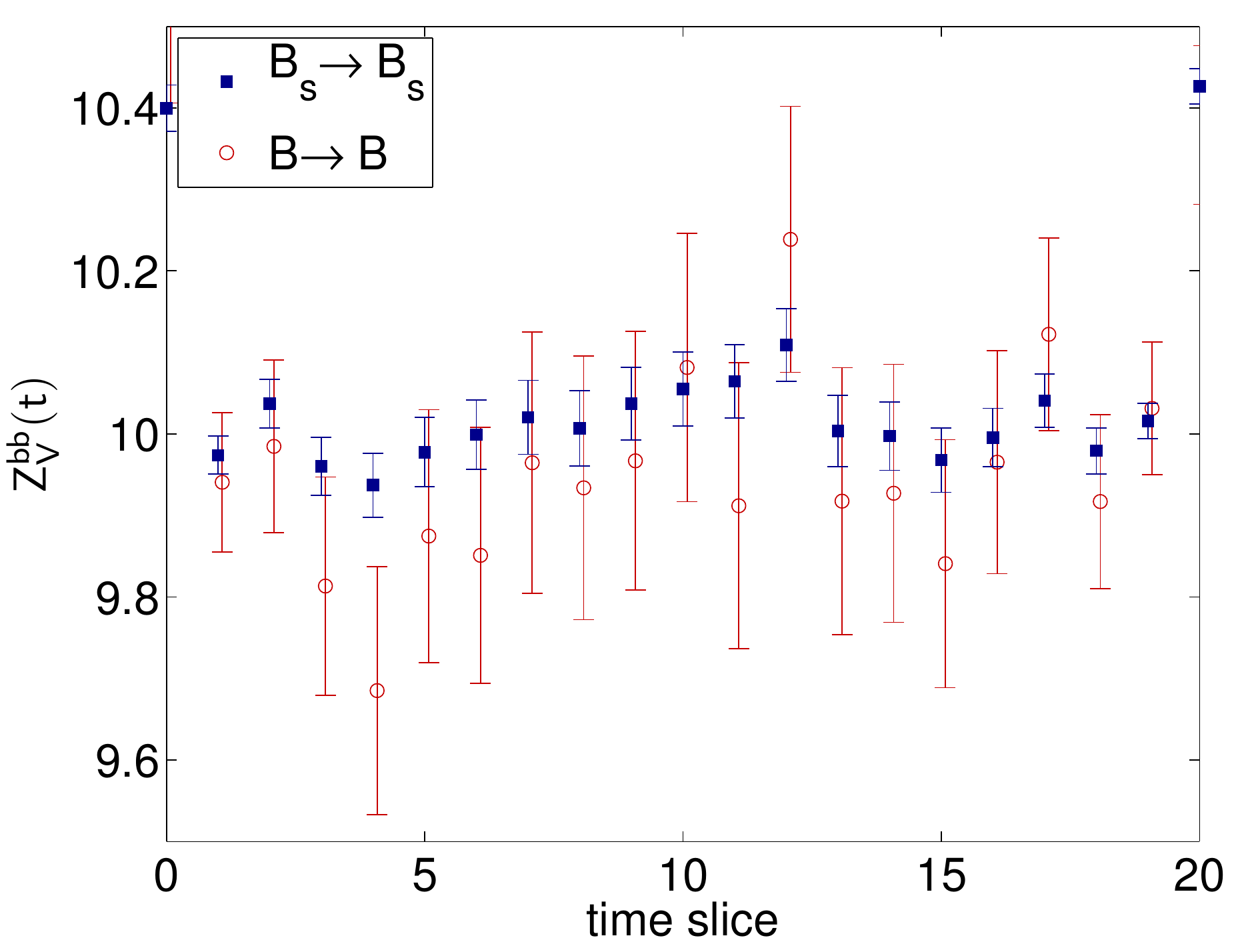}
\caption{Determination of $Z_V^{bb}$ from three-point correlators with unitary {(open red circles)} and strange {(filled blue squares)} spectator quarks on the $24^3$ ensemble with $am_l = 0.005$.  {The $B\to B$ data points are shown with a slight horizontal offset for clarity.}}
\label{Fig:Zvbb_spectator_quark_dep}
\end{figure}

\begin{figure*}[tb]
\flushleft
\includegraphics[scale=0.45]{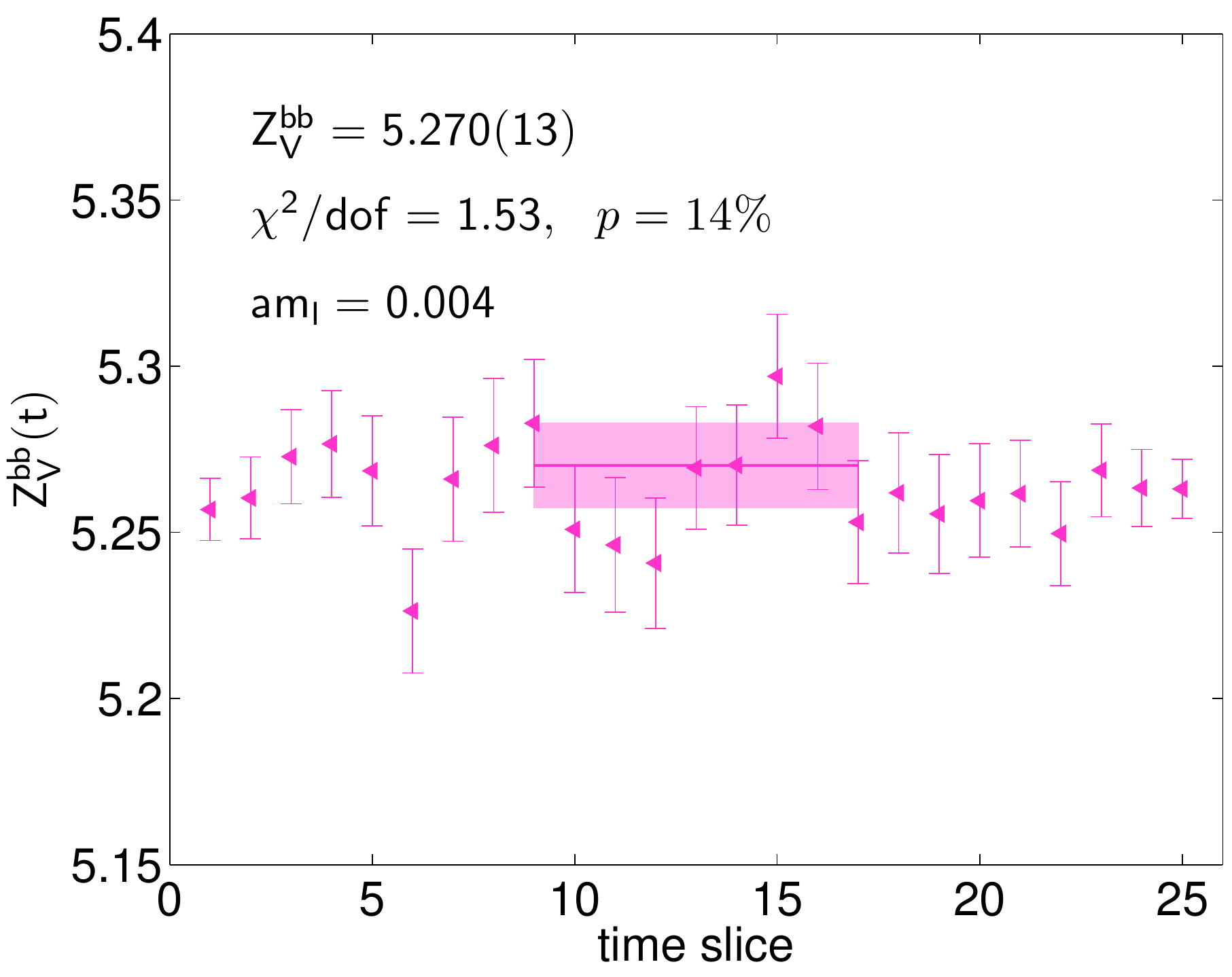}
\includegraphics[scale=0.45]{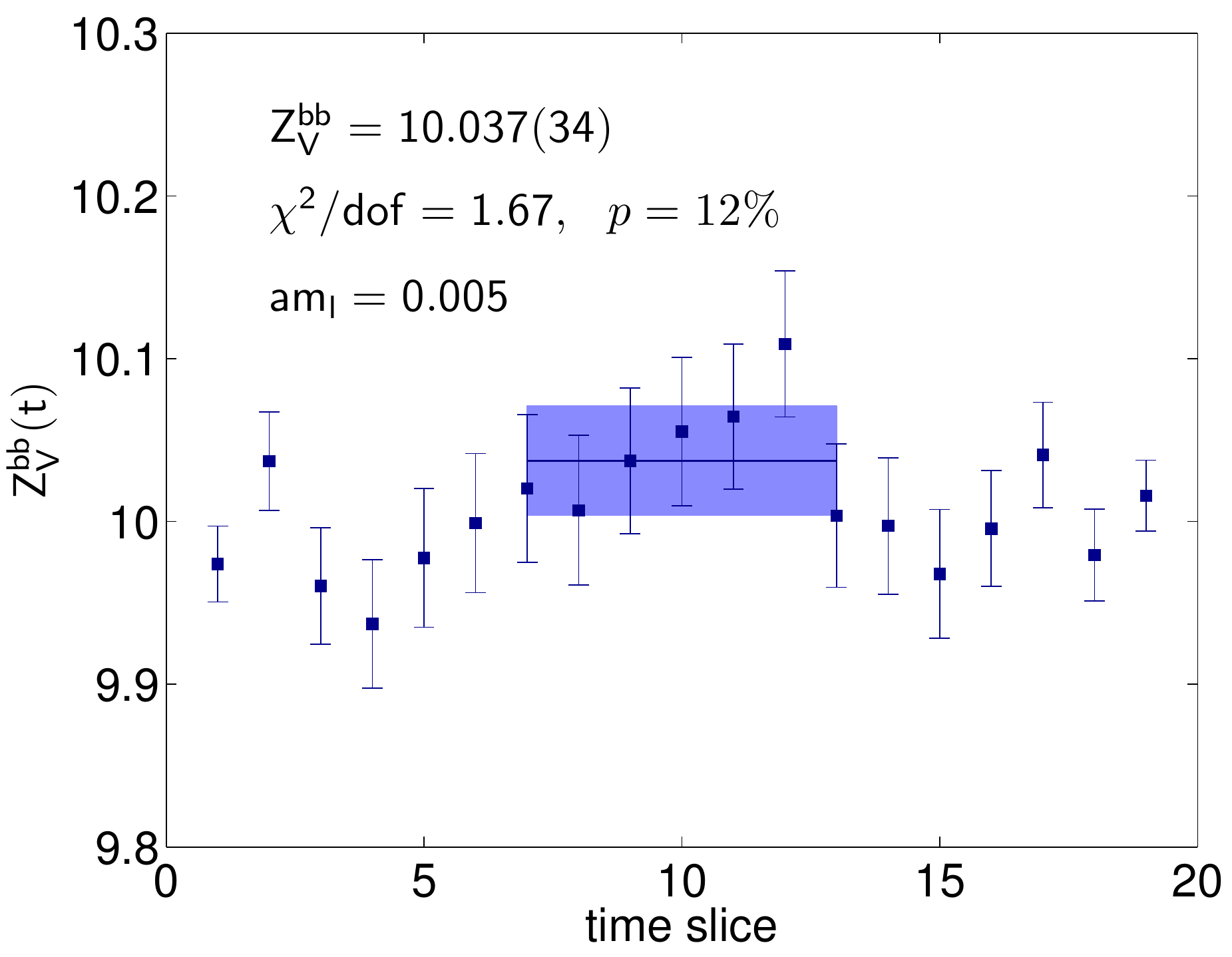}
\includegraphics[scale=0.45]{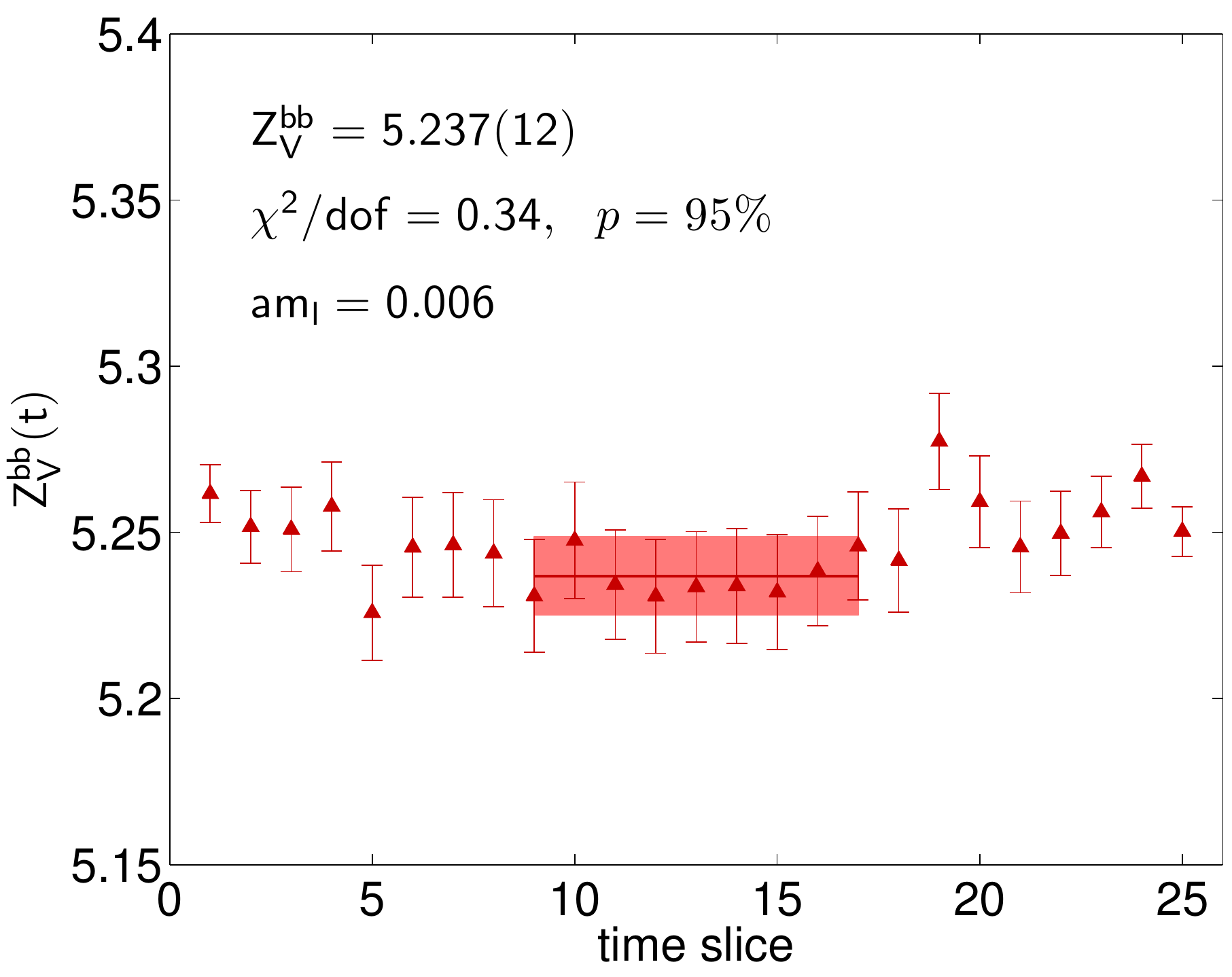}
\includegraphics[scale=0.45]{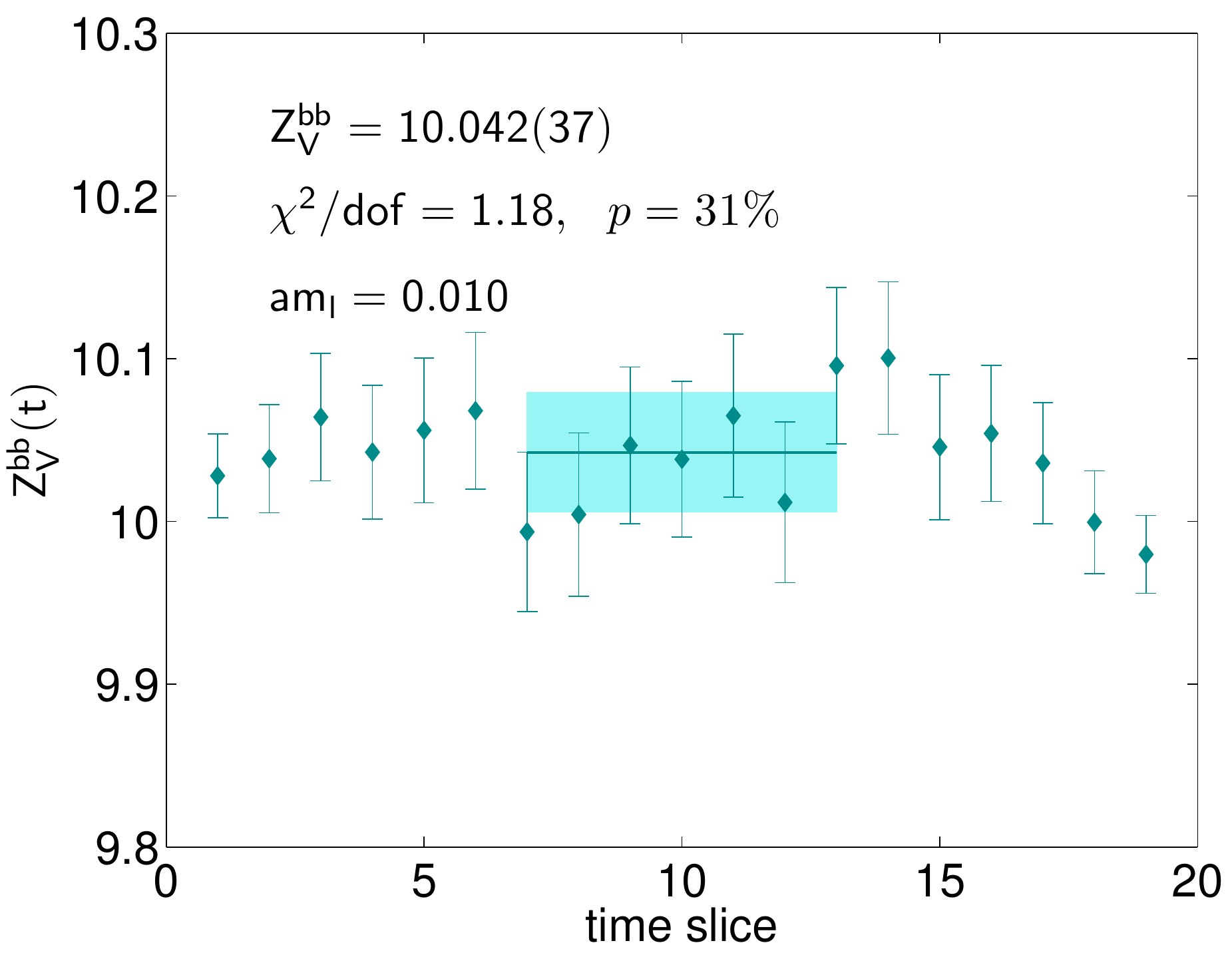}
\includegraphics[scale=0.45]{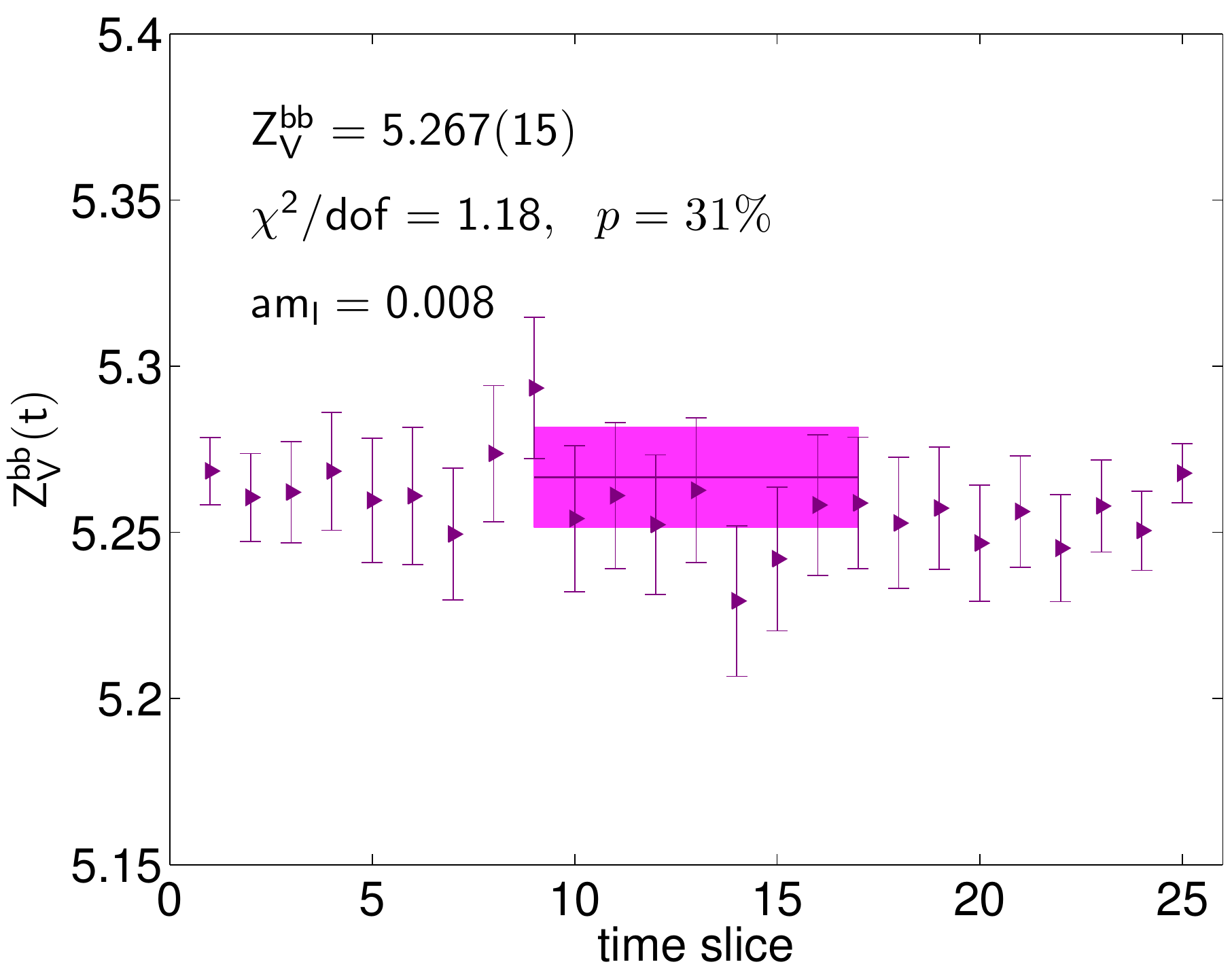}
\caption{Determination of $Z_V^{bb}$ from correlated constant fits to the ratio in Eq.~(\ref{Eq:Zvbb}).  The filled symbols show the data points with jackknife statistical errors, while the horizontal bands show the result of a correlated constant fit to the data on those time slices.  On the left we show the results for our three $32^3$ ensembles and on the right results for both of the the $24^3$ ensembles.}
\label{Fig:ZvbbFits}
\end{figure*}

Figure~\ref{Fig:Zvbb_spectator_quark_dep} shows an example determination of $Z_V^{bb}$ via Eq.~(\ref{Eq:Zvbb}) for two different values of the spectator-quark mass on the $am_l = 0.005$, $24^3$ ensemble; results on other ensembles look similar.
The data display long plateaus with small error bars over almost the entire time range.  We do not observe any spectator-quark mass dependence within statistical uncertainties, but the statistical errors increase with lighter spectator-quark mass.  We therefore determine $Z_V^{bb}$ using a strange spectator quark ($m_q \sim m_s$) in Eq.~(\ref{Eq:Zvbb}).  The values for $Z_V^{bb}$ are extracted by performing a correlated constant fit over time slices [7:13] on the $24^3$ ensembles and [9:17] on the $32^3$.  The fits are shown in Fig.~\ref{Fig:ZvbbFits} and the results with jackknife statistical errors are summarized in Tab.~\ref{Tab:Zvbb}.

\section{Numerical estimate of heavy-quark discretization errors}
\label{App:HQdiscErr}

Here we provide the explicit forms of the relevant operators and mismatch functions for the heavy-quark action in Sec.~\ref{sec:HQDiscAction}, and for the heavy-light current in Sec.~\ref{sec:HQDiscCurrent}.  Then, in Sec.~\ref{sec:HQDiscTotal}, we present numerical estimates of heavy-quark discretization errors in our calculation of the $B_{(s)}$-meson leptonic decay constants on the $24^3$ and $32^3$ ensembles.  For the discretization errors from the current, we compare our estimates from heavy-quark power counting with ones based on the observed sizes of the ${\mathcal O}(a)$ and ${\mathcal O}(\alpha_s a)$ contributions to the decay amplitudes, and find good agreement.

\subsection{\texorpdfstring{${\mathcal O}(a^2)$ errors from the action}{O(a2) errors from the action}}
\label{sec:HQDiscAction}

Oktay and Kronfeld present the complete set of bilinears and four-quark operators that can appear in the Symanzik effective Lagrangian through dimension~7 in Ref.~\cite{Oktay:2008ex}.  At dimension~6, there are two bilinears $\bar{h}\{\bm{\gamma}\cdot\bm{D}, \bm{\alpha}\cdot\bm{E}\}h$ and $\bar{h}\gamma_4(\bm{D}\cdot\bm{E}-\bm{E}\cdot\bm{D})h$ and many four-quark operators.  At tree-level, the mismatch coefficients of all of the four-quark operators vanish.  The tree-level mismatch coefficients of the two bilinears are the same, and are given by:
\begin{equation}
	f_E (m_0a, c_P, \zeta) =  \frac{1}{8 m_E^2 a^2} - \frac{1}{8 m_2^2 a^2}, \label{eq:fE}
\end{equation}
where
\begin{eqnarray}
	\frac{1}{m_2a} & = & \frac{2\zeta^2}{m_0a(2+m_0a)} 
		+ \frac{\zeta}{1+m_0a}, \label{eq:m2_tree} \\
	\frac{1}{4m_E^2a^2} &=&
		\frac{\zeta^2}{[m_0a(2+m_0a)]^2} +
		\frac{\zeta c_P}{m_0a(2+m_0a)} \,. \quad
\end{eqnarray}
The size of the relative error from each of the dimension~6 bilinears is then estimated to be
\begin{equation}
	{\rm error}_E \sim f_E(m_0a, c_P, \zeta) \left( a \Lambda_{\rm QCD} \right) ^2 \,.
\end{equation}
%

\subsection{\texorpdfstring{${\mathcal O}(\alpha_s^2 a, a^2)$ errors from the current}{O(alphas2 a, a2) errors from the current}}
\label{sec:HQDiscCurrent}

Harada {\it et al.} present the complete set of operators needed to improve the vector and axial-vector heavy-light currents to all orders in ${\mathcal O}(a)$ in Ref.~\cite{Harada:2001fi}.  There are six such operators; two for the temporal current and four for the spatial current.  Although their coefficients have been computed numerically at one-loop for the perturbative matching used in this work, the expressions are not known analytically.  We therefore use the tree-level mismatch functions as a guide.  At tree-level, all of the operators have the same mismatch coefficient:
\begin{align}
	f_3^{[0]} (m_0a, c_P, \zeta) &=  \frac{\zeta (1 + m_0a)}{m_0a (2+m_0a) }- \frac{1}{2 m_2a} - d_1 \\
						& =    \frac{\zeta}{m_0a (2+m_0a)} + \frac{\zeta}{(2 + m_0a)} \nonumber\\ &- \frac{\zeta}{2(1 + m_0a)} - \frac{\zeta^2}{m_0a (2 + m_0a)} - d_1\,, 
\end{align}
where at tree-level $d_1^{[0]}$ is defined such that $f_3^{[0]} = 0$.  For the 2-loop mismatch function(s), we multiply the above expression by $\alpha_s^2$ and set $d_1^{[2]} = 0$.  The result, however, approaches infinity in the $m_0a \to 0$ limit.  We therefore instead consider several functions similar to $f_3^{[0]}$ that have the expected parametric dependence on the strong coupling and $\zeta$, as well as the correct asymptotic behavior in both the chiral and static limits.  For our final estimate, we use the simple ansatz
\begin{equation}
	f_3^{[2]} (m_0a, c_P, \zeta) = \alpha_s^2 \zeta  \frac{2}{(2 + m_0a)}  \,,
\end{equation}
where the factor of two in the numerator allows for more than one term of this size in the true mismatch function.  In our numerical simulations, the parameter $\zeta$ is of ${\mathcal O}(1)$, so the small size of $f_3^{[2]}$ is primarily due to the perturbative factor $\alpha_s^2$.  The exact dependence on $m_0a$ in the denominator does not impact the size of $f_3^{[2]}$ significantly, but we conservatively take the function that leads to the largest value of the mismatch function.  The size of the relative error from the ${\mathcal O}(a)$ heavy-light current operators is then estimated to be
\begin{equation}
	{\rm error}_3 \sim f_3^{[2]}(m_0a, c_P, \zeta) \left( a \Lambda_{\rm QCD} \right) \,.
\end{equation}

El~Khadra, Kronfeld, and Mackenzie present the expression for the tree-level ${\mathcal O}(a^2)$-improved heavy-light electroweak current in Eq.~(A.17) of Ref.~\cite{ElKhadra:1996mp}.  At ${\mathcal O}(a^2)$, there are three relevant operators --- $\bar{q} \Gamma \mathbf{D}^2 h$, $\bar{q} \Gamma i \mathbf{\Sigma} \cdot \mathbf{B} h$, and $\bar{q} \Gamma \mathbf{\alpha} \cdot \mathbf{E} h$ --- where $q$ and $h$ denote the light- and heavy-quark fields, respectively.   Their tree-level coefficients are given in Eq.~(A.19) of the same paper, from which the mismatch functions can be inferred:
\begin{align}
	f_{X_1} (m_0a, c_P, \zeta) &= -\frac{1}{2} \left[ d_1^2 - \frac{\zeta}{2(1+m_0a)} \right] \\
	f_{X_2} (m_0a, c_P, \zeta) &= -\frac{1}{2} \left[ d_1^2 - \frac{c_P}{2(1+m_0a)} \right] \\
	f_Y (m_0a, c_P, \zeta) &= -\frac{1}{2} \left[ \frac{(\zeta - c_P)(1+m_0a)}{m_0a(2+m_0a)} - \frac{d_1}{m_2a} \right] \,, 
\end{align}
with
\begin{equation}
	d_1 = (m_0a, c_P, \zeta) = \frac{\zeta (1 + m_0a)}{m_0a (2+m_0a)}- \frac{1}{2 m_2a} \,.
\end{equation}
The sizes of the relative errors from the three operators are then estimated to be
\begin{eqnarray}
	{\rm error}_{X_1} &\sim& f_{X_1}(m_0a, c_P, \zeta) \left( a \Lambda_{\rm QCD} \right) ^2 \\
	{\rm error}_{X_2} &\sim& f_{X_2}(m_0a, c_P, \zeta) \left( a \Lambda_{\rm QCD} \right) ^2 \\
	{\rm error}_{Y} &\sim& f_Y(m_0a, c_P, \zeta) \left( a \Lambda_{\rm QCD} \right) ^2 \,.
\end{eqnarray}
%

\subsection{Numerical estimates}
\label{sec:HQDiscTotal}

Table~\ref{tab:Mismatch_Fcns} presents the numerical values of the mismatch functions at the tuned values of the RHQ parameters given in Table~\ref{tab:RHQParamErr}.  Table~\ref{tab:HQDiscErrs_fB} presents the estimated size of heavy-quark discretization errors in $f_{B_{(s)}}$ from operators of ${\mathcal O}(a^2)$ in the action and of ${\mathcal O}(\alpha_s^2a, a^2)$ in the axial-vector current taking $\Lambda_{\rm QCD} = 500$~MeV.  The last column gives the sum of the errors from the individual operators added in quadrature.

\begin{table}[tb]
	\centering 
	\caption{Mismatch functions for the nonperturbatively-tuned parameters of the RHQ action on the $24^3$ and $32^3$ ensembles given in Table~\ref{tab:RHQParamErr}.  The tree-level coefficients $f_E$, $f_{X_i}$, and $f_Y$ are known exactly.  The two-loop coefficient $f_{3}^{[2]}$ is not known, so we use an ansatz based on the tree-level expression.} \vspace{1mm}
	\label{tab:Mismatch_Fcns}
\begin{tabular}{lcccccc}
	\hline\hline
	& $f_E$ & $f_{X_1}$ & $f_{X_2}$ & $f_{Y}$ & $f_{3}^{[2]}$ \\ \hline
	$a \approx$~0.11 fm & 0.0652  & 0.0803 & 0.1517 & 0.1605 & 0.0659 \\
	$a \approx $~0.086 fm & 0.0864  & 0.0953 & 0.1774 & 0.1900 & 0.0312 \\
	\hline\hline
\end{tabular}
\end{table}

\begin{table*}[tb]
	\centering 
	\caption{Percentage errors from mismatches in the action and current for the bottom quark on the $24^3$ and $32^3$ ensembles.  For this estimate, we calculate the mismatch functions for the nonperturbatively-tuned parameters of the RHQ action from Table~\ref{tab:RHQParamErr}.  We estimate the size of operators using HQET power counting with $\Lambda_{\rm QCD}=500$~MeV.  To obtain the total, we add the individual errors in quadrature, counting contributions $E$ and 3 twice because they each arise from two operators in the Symanzik effective Lagrangian.} \vspace{1mm}
	\label{tab:HQDiscErrs_fB}
\begin{tabular}{l@{\hskip 3mm}c@{\hskip 2mm}c@{\hskip 3mm}ccc @{\hskip 3mm} cc}
	\hline\hline
	& & ${\mathcal O}(a^2)$ error & \multicolumn{3}{c}{${\mathcal O}(a^2)$ errors} {\hskip 1.mm} & ${\mathcal O}(\alpha_s^2 a)$ error  \\[-0.3ex]
	& & from action & \multicolumn{3}{c}{from current} {\hskip 1.mm} & from current  \\	
	& $\alpha_s^{\bar{\rm MS}}(1/a)$ & $E$ & $X_1$ & $X_2$ & $Y$ & 3  & Total (\%) \\\hline
	$a \approx$~0.11 fm & 1/3 &  0.55 & 0.67 & 1.27 & 1.34 & 1.91 & 3.42 \\
	$a \approx $~0.086 fm & 0.22 & 0.42 & 0.46 & 0.85 & 0.91 & 0.68 & 1.75 \\
	\hline\hline
\end{tabular}
\end{table*}

We can also estimate the size of heavy-quark discretization errors from the current, {\it i.e.} those corresponding to operators $X_1, X_2, Y$, and $3$ in Table~\ref{tab:HQDiscErrs_fB}, by looking at the known ${\mathcal O}(a)$ and ${\mathcal O}(\alpha_s a)$ contributions to the decay amplitudes in our data.  We estimate the size of the omitted ${\mathcal O}(a^2)$ contributions by assuming that they are approximately $a \Lambda_{\rm QCD}$ times the size of the tree-level ${\mathcal O}(a)$ contribution, which is 0.75\% on the finer $32^3$ ensembles.  This is of the same size as the estimates of the contributions from ${\mathcal O}_{X_1,X_2,Y}$ in Table~\ref{tab:HQDiscErrs_fB} based on heavy-quark power counting.  We estimate the size of the omitted ${\mathcal O}(\alpha_s ^2a)$ contributions in two ways:  by assuming that they are approximately $\alpha_s^2$ times the size of the tree-level contribution, or that they are $\alpha_s$ times the size of the 1-loop contribution.  On the $32^3$ ensembles, the first approach leads to an estimate of 0.17\%, while the second leads to one of 0.59\%.  These are both similar in magnitude to the estimated contribution from ${\mathcal O}_3$ in Table~\ref{tab:HQDiscErrs_fB}.  Given the overall consistency between the two error estimation approaches, we use the values obtained from heavy-quark power counting in our systematic error budget.  This, in fact, leads to a slightly larger quoted heavy-quark discretization error than what we would obtain if we used the data-driven numbers.


\bibliography{B_meson}
\bibliographystyle{apsrev4-1} 
\end{document}